\documentclass[fleqn,usenatbib]{mnras}
\usepackage[dvipsnames]{xcolor}
\usepackage{newtxtext,newtxmath}
\usepackage{caption}
\usepackage{subcaption}
\usepackage{hyperref}
\usepackage[T1]{fontenc}
\usepackage{enumitem}
\usepackage{stfloats}

\DeclareRobustCommand{\VAN}[3]{#2}
\let\VANthebibliography\thebibliography
\def\thebibliography{\DeclareRobustCommand{\VAN}[3]{##3}\VANthebibliography}

\usepackage{graphicx}	% Including figure files
\usepackage{amsmath}	% Advanced maths commands

\title[Ly$\alpha$ in the EoR from FRESCO and Keck]{Lyman-$\alpha$ Visibility During the Epoch of Reionization: Combining JWST FRESCO Grism Data with Keck Archival Spectroscopy }

\author[E. Leonova et al.]{\parbox{17.5cm}{Ecaterina Leonova,$^{1,2}$\thanks{E-mail: e.leonova@uva.nl}
Rohan P. Naidu,$^{3}$\thanks{NASA Hubble Fellow, Pappalardo Fellow}
Pascal A. Oesch,$^{4,5,6}$
Gabriel Brammer,$^{5,6}$
Jorryt Matthee,$^{7}$
Romain A. Meyer,$^{4}$
Daniel Schaerer,$^{4}$
Mengyuan Xiao,$^{4}$}
\\ \vspace{0.2cm} \\
$^{1}$GRAPPA, Anton Pannekoek Institute for Astronomy and Institute of High-Energy Physics\\
$^{2}$University of Amsterdam, Science Park 904, NL-1098 XH Amsterdam, the Netherlands\\
$^{3}$MIT Kavli Institute for Astrophysics and Space Research, 70 Vassar Street, Cambridge, MA 02139, USA \\
$^{4}$Department of Astronomy, University of Geneva, Chemin Pegasi 51, 1290 Versoix, Switzerland\\
$^{5}$Cosmic Dawn Center (DAWN)\\
$^{6}$Niels Bohr Institute, University of Copenhagen, Jagtvej 128, DK-2200, Copenhagen N, Denmark\\
$^{7}$Institute of Science and Technology Austria (ISTA), Am Campus 1, 3400 Klosterneuburg, Austria\\
}

\date{Accepted XXX. Received YYY; in original form ZZZ}

\pubyear{2025}

\begin{document}
\label{firstpage}
\pagerange{\pageref{firstpage}--\pageref{lastpage}}
\maketitle

% Abstract of the paper
\begin{abstract}

%abstract=250words
The visibility of Lyman-$\alpha$ emission at $z>7$ provides crucial insights into the reionization process and the role of galaxies in shaping the ionized intergalactic medium. Using JWST FRESCO data, we investigate the environments of Lyman-$\alpha$ emitters (LAEs) in the GOODS-N and GOODS-S fields by identifying [OIII] emitters and analyzing their large-scale distribution. Using the FRESCO redshifts, we recover eight new LAEs from archival Keck/MOSFIRE observations at $z=7.0-7.7$, including a potential AGN candidate at $z \sim 7.2$. Complemented by six literature LAEs, our sample consists of 14 LAEs in total, all of which are [OIII] emitters except for one very faint source not detected by FRESCO.
We define seven groups of [OIII] emitters centered around the brightest LAEs and find that these bright LAEs do not reside in more overdense environments than the average galaxy population. The overdensity parameters for LAEs and [OIII] emitters without Lyman-$\alpha$, calculated for sources with $\mathrm{M_{UV}<-19.5}$ to ensure completeness, are similar, indicating that overdensities alone cannot fully explain Lyman-$\alpha$ visibility. While LAEs have slightly higher recent star formation (SFR$_{10}$/SFR$_{50} \approx 1.3\times$) and [OIII] EW ($\approx1.5\times$), they show no significant differences from [OIII] emitters in UV slope ($\beta$), UV magnitude ($\mathrm{M_{UV}}$), or stellar mass ($\log_{\mathrm{M}_{\star}}$). Our results suggest that other factors may contribute to the observability of Lyman-$\alpha$ emission. Future spectroscopic surveys with broader wavelength coverage and more complete sampling will be crucial for refining our understanding of reionization.

\end{abstract}


%%%%%%%%%%%%%%%%%%%%%%%%%%%%%%%%%%%%%%%%%%%%%%%%%%

%%%%%%%%%%%%%%%%% BODY OF PAPER %%%%%%%%%%%%%%%%%%

\section{Introduction}

Understanding the last major phase transition of the Universe—the Epoch of Reionization (EoR)—is a key goal of extragalactic astronomy. The EoR begins with the first luminous sources that start to ionize their surrounding intergalactic medium (IGM).  Early galaxies and stars likely played the driving role in ionizing the hydrogen, leading to the universe's transition from a predominantly neutral to a fully ionized state by $z\sim5.5$ \citep{Barkana2001, Dayal2018, Robertson2022}.

One of the most effective ways to study the EoR is by observing Lyman-$\alpha$ emitters (LAEs) – galaxies that emit strong Lyman-$\alpha$ radiation at 1215.67\AA   \citep{MiraldaEscude98, McQuinn2007, Dayal11, Mason2018a, Lu2024, Napolitano2024, Prieto2025, Witten2024, Witstok2025}. Due to its resonant nature, Lyman-$\alpha$ provides critical information about the neutral hydrogen fraction in the IGM \citep[e.g.,][]{Dijkstra2014, Hutter17, Mason2018b, Hoag2019, Ouchi20, Jones2025, Kageura2025}. Hydrodynamical simulations also provide important theoretical insight into Lyman-$\alpha$ emission in the early Universe \citep{Smith21, Bhagwat2025, Neyer2025}.

However, detecting LAEs becomes progressively more challenging at higher redshifts, especially beyond $z > 7$, when a larger fraction of hydrogen in the universe is still neutral \citep{Stark2010, Pentericci2011, Kageura2025, Tang2025}. Neutral hydrogen scatters Lyman-$\alpha$ photons, which results in suppression of their visibility at high redshifts \citep[e.g.,][]{MiraldaEscude98, Dijkstra2014}. The Lyman-$\alpha$ emission from these galaxies is scattered and absorbed by neutral hydrogen in the intergalactic medium, reducing its visibility at high redshift. A dramatic decline in the number of galaxies showing Lyman-$\alpha$ emission was observed between $z=6-7$, suggesting a rapid evolution of the neutral hydrogen fraction in the IGM \citep{Stark2010, Stark2017, Pentericci2011, Mason2018b, Napolitano2025}. Despite these difficulties, detections of LAEs at even higher redshifts have been reported in several recent studies \citep{Song16, Zitrin2015, Tilvi2020, Jung2020, Roberts-Borsani2023, Saxena23, Chen2024, Tang2024}, with the record being at $z=13.0$ \citep{Witstok2024}. 

What makes a galaxy a LAE at high redshift? High-$z$ LAEs are thought to be located within ionized bubbles created by neighbouring galaxies, allowing Lyman-$\alpha$ photons to escape before being redshifted and absorbed by the neutral intergalactic medium \citep[e.g.,][]{Mason2018a, Matthee2018, Napolitano2024, Chen2024, Witstok2024, Tang2024, Napolitano2025}. These overdense regions contain many galaxies, each contributing to the process of ionization, creating larger ionized regions that allow Lyman-$\alpha$ photons to escape \citep[e.g.,][]{Castellano2018, Jung2022, Larson2022, Leonova2022, Endsley2022, Saxena23, Tang2023, Tang2024, Witten2024, Chen2025, Hashemi2025, Whitler2025}. However, alternative scenarios suggest that individual luminous sources, particularly bright galaxies or AGN, may drive localized ionized bubbles sufficiently large to facilitate Lyman-$\alpha$ transmission \citep[e.g.,][]{Matthee2018, Meyer2021}

On the other hand, Lyman-$\alpha$ visibility is also shaped by intrinsic galaxy properties, including star formation history, ionizing photon production, and energetic sources such as AGNs. Notably, the Lyman-$\alpha$ escape fraction correlates strongly with dust attenuation \citep[e.g.,][]{Atek2009, Hayes2011, Martin2025}, suggesting that dust geometry and UV absorption play critical roles in determining whether Lyman-$\alpha$ photons reach the observer. Since dust-free or low-dust systems exhibit enhanced Lyman-$\alpha$ transmission, systematic variations in dust properties across our LAE and non-LAE samples could substantially influence observability independent of environmental factors. Bursty star formation episodes can enhance ionization, potentially creating clearer pathways for Lyman-$\alpha$ photons to escape \citep{Matthee2017, Begley2024, Boyett2024}. Other studies suggest that Lyman-$\alpha$ visibility may also be linked to galaxy mergers. \citet{Witten2024} showed that galaxies with frequent mergers can experience periods of intense star formation, leading to strong Lyman-$\alpha$ emission. Additionally, AGN activity could contribute to ionization, further influencing Lyman-$\alpha$ transmission and its detectability at high redshifts \citep{Giallongo2015, YungL2021, Dayal2024}. Disentangling the role of these intrinsic properties from environmental factors is a crucial step towards constraining the physics, timeline, and protagonists of reionization \citep[e.g.,][]{Finkelstein2019, Naidu20, Matthee2022, Asthana2024, Lu25}.

% Its advanced infrared capabilities allow us to detect and characterise galaxies at redshifts well beyond $z = 7$ with high precision, making it particularly effective for studying reionization \citep{Naidu2022, Bunker2024, Finkelstein2025a}. 

The launch of the James Webb Space Telescope (JWST) has transformed the study of the Epoch of Reionization (EoR) \citep{Ellis2025, Stark2026}. The Lyman-$\alpha$ emission lines from EoR galaxies are redshifted into the infrared, and JWST's sensitivity enables the detection of even faint and distant LAEs, providing key data about the state of the universe during the EoR. 

Critical to the study of galaxy environments, JWST enables detailed spectroscopic analysis of distant galaxies, providing unprecedented insights into their physical conditions and emission properties \citep[e.g.,][]{Covelo-Paz2024, Herard-Demanche2025, Matharu2024, Matthee2024, Nelson2024, Xiao2024, Witten2025}.
Building on these capabilities, wide-area NIRCam grism surveys now allow us to measure the three-dimensional positions of complete galaxy samples for the first time, enabling the identification of large-scale overdensities and ionized bubbles in the reionization-era universe \citep{Chen2024, Witstok2024, Tang2024, Helton2024, Torralba-Torregrosa2024, Naidu2024, Kashino2025}. For example, the [O\,\textsc{iii}] emission line at 5008~\AA\ is a powerful probe of the ionization state and gas-phase metallicity of galaxies, while Balmer recombination lines such as H$\alpha$ and H$\beta$ more directly trace ongoing star formation. Together, these diagnostics provide insights into the ionizing photon output of galaxies and their contribution to reionizing the IGM \citep{Matthee2023, Begley2024, Endsley21_oiii, Endsley2024, Meyer2024}.
Additionally, rest-optical lines like [OIII] enable the measurement of systemic redshifts, which in turn provide much more reliable measurements on reionization with the use of the Lyman-$\alpha$ emission lines \citep[e.g.,][]{Chen2024, Napolitano2024, Saxena2024, Tang2024}. By studying LAEs alongside [OIII] emitters, we can examine whether LAEs preferentially reside in over-dense regions of star-forming galaxies, which may contribute to the formation of large ionized bubbles that facilitate Lyman-$\alpha$ escape.

% \citet{Lu2024} introduced techniques to detect the edges of these bubbles and estimate their sizes by analyzing Lyman-$\alpha$ transmission.

In this paper, we investigate the environments of LAEs during the Epoch of Reionization, focusing on identifying overdensities of [OIII] emitters around LAEs using data from the JWST FRESCO program \citep{Oesch2023, Meyer2024}. Our goal is to understand the mechanisms driving reionization, particularly through the analysis of LAEs and their surrounding environments. In particular, we use the new systemic redshift measurements obtained from the FRESCO grism program in order to search for new Lyman-$\alpha$ emission lines from the large archive of Keck/MOSFIRE data taken over the last decade in the GOODS fields. Using these new LAEs together with JWST/NIRSpec LAEs reported in the literature, we will examine the properties of LAEs at  $z=7-8$ in GOODS-N and GOODS-S, and focus on the spatial distribution and clustering of [OIII] emitters around LAEs.

In Section~\ref{sec:data}, we describe the observational data used in this work, including a new Keck archival search for LAEs. In Section~\ref{sec:results}, we present our analysis and results. Sections~\ref{sec:discussion} and~\ref{sec:summary} summarize our findings. Throughout this paper, we adopt a flat $\Lambda$CDM cosmology with parameters from the \textit{Planck} Collaboration \citep{Planck2018}: $H_{0} = 67.4~\mathrm{km~s^{-1}~Mpc^{-1}}$, $\Omega_{\mathrm{m}} = 0.315$, and $\Omega_{\Lambda} = 0.685$. All magnitudes are given in the AB system \citep{OkeGunn1983}.

%------------------------

%---------------------------------------

\section{Data}
\label{sec:data}
Understanding the environments of LAEs requires high-quality spectroscopic and imaging data. In this section, we describe the dataset used in this study, which allows us to identify and analyze the distribution of [OIII] emitters around LAEs in GOODS-N and GOODS-S fields.

\subsection{FRESCO [OIII] emitters}

We use NIRCam/grism data from  FRESCO \citep[First Reionization Epoch Spectroscopically Complete Observations;][]{Oesch2023}  to select [OIII] emitters in the GOODS-N and GOODS-S fields, focusing on galaxies at $z>7$. FRESCO covers 62 arcmin$^2$ in each of these fields, providing a total survey area of 124 arcmin$^2$. 

The NIRCam/grism observations of FRESCO were performed with the F444W filter, exposing for approximately 2 hours per pointing. These observations, with a resolution of R $\sim$ 1600, cover the wavelength range of 3.8–5.0 $\mu$m, capturing spectra for all galaxies within the NIRCam field of view. The setup allows for the measurement of critical emission lines across different cosmic epochs, including [OIII] and H$\beta$ for galaxies at redshifts $\sim$ 6.8–9. In particular, we use the [OIII] emitter sample of 137 galaxies that was previously discussed and published in \citet{Meyer2024}. Note that in GOODS-N, our galaxy IDs differ from those in \citet{Meyer2024} as we use an earlier version of the ID catalog.

In addition to the spectroscopic data, FRESCO provides deep, high-resolution imaging in the F182M, F210M, and F444W filters, reaching a depth of $\sim$  28.2 mag (5$\sigma$ in 0.32'' diameter apertures). This imaging complements the spectral data by facilitating stellar mass estimates and calibrating slit-loss corrections for NIRSpec/MSA spectra in overlapping fields. 

Additionally, the FRESCO field is partially covered by the JWST Advanced Deep Extragalactic Survey (JADES) \citep{Eisenstein2023}. JADES offers much deeper imaging and broader wavelength coverage compared to FRESCO, with an average of 130 hours of exposure time spread over 9 NIRCam filters, covering the spectral range from 0.6 to 5.3 $\mu$m, and also provides deep NIRSpec spectroscopy for large galaxy samples in this field \citep[][]{DEugenio24}. 

This extensive dataset enables the identification of LAEs over a wide redshift range ($6.6 \lesssim z \lesssim 9.3$) \citep{Saxena23, Tang2024}. In this work we focus on the interval $7 < z < 8$, which corresponds to the range where Lyman-$\alpha$ falls within the MOSFIRE $Y$ band, providing valuable context and complementing the sample of new LAEs discovered using the FRESCO data.

\subsection{Lyman-$\alpha$ emitters}

\begin{table*}
\centering
\begin{tabular}{ccccccccccc}
\hline\hline
ID & RA & DEC & z$_\mathrm{Ly\alpha}$ & z$_\mathrm{[OIII]}$ & $\Delta v$ [km s$^{-1}$] &F$\mathrm{_{Ly\alpha}}$[10$^{-17}$ erg s$^{-1}$ cm$^{-2}$] & EW$\mathrm{_{Ly\alpha}}$[\AA] &Instrument & References \\\hline\hline\\[-5pt]

\multicolumn{10}{c}{\textbf{GN LAEs}}\\[5pt]

GN\_27795 & 189.13498 & 62.29190 & 7.038 & 7.017 & $792^{+36}_{-37}$ &$3.10 \pm 0.91$ &  $151 \pm 17$ & Keck MOSFIRE\href{#note2}{\textsuperscript{1}} & \\
GN\_26051 & 189.17981 & 62.28239 & 7.087 & 7.084 & $122\pm53$\href{#refa}{\textsuperscript{a}}  & - & $52\pm 9$\href{#refa}{\textsuperscript{a}}&JWST NIRSPEC & \href{#refa}{a} \\
GN\_22679 & 189.20377 & 62.26843 & 7.090 & 7.090 & $233\pm53$\href{#refa}{\textsuperscript{a}} & $1.61 \pm 0.09$ \href{#refa}{\textsuperscript{a}}&$143\pm 8$\href{#refa}{\textsuperscript{a}}& JWST NIRSPEC& \href{#refa}{a} \\

GN\_4394  & 189.08349 & 62.20258 & 7.197 & 7.186 & $398^{+34}_{-29}$ &$3.80 \pm 0.48 $&$221 \pm 15 $& Keck MOSFIRE\href{#note4}{\textsuperscript{3}} &  \\
GN\_26844 & 189.22512 & 62.28629 & 7.211 & 7.205 & $168^{+41}_{-36}$ &$0.79 \pm 0.02$  & $54 \pm 5$ & Keck MOSFIRE\href{#note1}{\textsuperscript{2}} &  \\
GN\_8563  & 189.08405 & 62.22204 & 7.206 & 7.200 & $236^{+45}_{-43}$ &$1.74 \pm 0.22 $ &  $51 \pm 2 $ &Keck MOSFIRE \href{#note5}{\textsuperscript{4}} &  \\

GN\_29192 & 189.15779 & 62.30234 & 7.507 & 7.498 & $314^{+43}_{-42} $ &$2.07 \pm 0.09 $  & $72 \pm 4 $ & Keck MOSFIRE\href{#note2}{\textsuperscript{1}} & \href{#refb}{b} \\
GN\_29193 & 189.15798 & 62.30240 & 7.507 & 7.502 & $171^{+41}_{-42}$ &$1.06 \pm 0.02 $  & $27 \pm 1$ & Keck MOSFIRE\href{#note1}{\textsuperscript{2}} & \\

GN\_19441 & 189.33307 & 62.25722 & 7.600 & 7.595 & $161^{+40}_{-37}$ &$1.90 \pm 0.05$  & $46 \pm 3 $ & Keck MOSFIRE\href{#note6}{\textsuperscript{6}} & \href{#refb}{b}\\
GN\_9680  & 189.28867 & 62.22651 & 7.636 & 7.629 & $235^{+35}_{-38}$ &$1.86 \pm 0.04$ &  $143 \pm 6$ & Keck MOSFIRE\href{#note7}{\textsuperscript{7}} & \\
GN\_14535 & 189.20307 & 62.24249 & 7.657 & 7.648 & $315^{+40}_{-44} $ &$1.56 \pm 0.03$ &  $86 \pm 10  $ & Keck MOSFIRE\href{#note3}{\textsuperscript{5}} &  \\
GN\_25604 & 189.30014 & 62.28034 & 7.700 & 7.694 & $226^{+37}_{-38}$ &$2.86 \pm 0.08 $&$241 \pm 47$ & Keck MOSFIRE\href{#note2}{\textsuperscript{1}} & \\\hline\hline\\[-5pt]

\multicolumn{10}{c}{\textbf{GS LAEs}}\\[5pt]

GS\_1744    & 53.16958 & -27.73806 & 7.242 & 7.242     & $269\pm94$\href{#refa}{\textsuperscript{a}}  & $0.48 \pm 0.07$ \href{#refa}{\textsuperscript{a}}& $26 \pm 3$ \href{#refa}{\textsuperscript{a}} & JWST NIRSPEC & \href{#refa}{a}\\
JADES-13682 & 53.16745 & -27.77203 & 7.275 & - & $217\pm93$\href{#refa}{\textsuperscript{a}} & $0.23 \pm 0.03$ \href{#refa}{\textsuperscript{c}}& $259 \pm 54$ \href{#refa}{\textsuperscript{a}}&JWST NIRSPEC & \href{#refa}{a}, \href{#refc}{c}\\
GS\_6644    & 53.13347 & -27.76039 & 7.660 & 7.661 & $277\pm89$\href{#refa}{\textsuperscript{a}} & $0.55 \pm 0.09$ \href{#refa}{\textsuperscript{d}}& $33 \pm 3 $ \href{#refa}{\textsuperscript{a}}&JWST NIRSPEC & \href{#refa}{a}, \href{#refd}{d}\\
GS\_28631   & 53.08733 & -27.86028 & 7.962 & 7.958 & $327\pm65$\href{#refa}{\textsuperscript{e}} & $0.24 \pm 0.08$\href{#refa}{\textsuperscript{e}}& $16\pm8$\href{#refa}{\textsuperscript{e}}&JWST NIRSPEC & \href{#refe}{e}\\

\hline\hline
\end{tabular}

\vspace{10pt}
\textbf{Notes:} 

 [\hypertarget{note1}{1}] N107M, 2014A, PI: Finkelstein;
[\hypertarget{note2}{2}] N100M, 2013A, PI: Finkelstein;
[\hypertarget{note3}{3}] U076M, 2014A, PI: Faber;
[\hypertarget{note4}{4}] N118, 2022A, PI: Casey, Nunez, Prusinski, Steidel;
[\hypertarget{note5}{5}] U092M, 2016A, PI: Faber, Guo, Yesuf, Koo, Barro;
[\hypertarget{note6}{6}] U085M, 2015A, PI: Faber;
[\hypertarget{note7}{7}] C194M, 2013A, PI: Konidaris

\vspace{10pt}
\textbf{References:} 
[\hypertarget{refa}{a}] \cite{Tang2024};
[\hypertarget{refb}{b}] \cite{Jung2020};
[\hypertarget{refc}{c}] \cite{Saxena23};
[\hypertarget{refd}{d}] \cite{Song16};
[\hypertarget{refe}{e}] \cite{Roberts-Borsani2023}

\caption{Table of Lyman-$\alpha$ Emitters in GOODS-S and GOODS-N fields, including coordinates, redshifts, velocity offsets, instruments used, and relevant references. The sample is restricted to sources with a signal-to-noise ratio (SNR) greater than 5 in the Lyman-$\alpha$ emission, ensuring reliable detection.}
\label{tab:LAE}
\end{table*}

To investigate the overdensities around LAEs in the GOODS-N and GOODS-S fields, we will examine all known LAEs at redshifts $7 < z < 8$ within the FRESCO field both from ground-based Keck data \citep[e.g.,][]{Jung2022} as well as from new NIRSpec observations \citep[e.g.,][]{Saxena2024}.
Additionally, we search for new LAEs in archival ground-based Keck/MOSFIRE data. 

Since the GOODS fields have been studied extensively over the past decades, in particular in the search for high-redshift galaxy candidates, these fields have been covered with a large amount of Keck/MOSFIRE spectroscopy in the NIR. The main purpose of these observations was often to target the redshifted Lyman-$\alpha$ line. While reliably detecting faint Lyman-$\alpha$  can be challenging due to the NIR sky, we can now leverage FRESCO spectroscopic data, to accurately determine the redshifts of [OIII] emitters to discover new LAEs.
Knowing the [OIII] redshifts of galaxies is crucial, as it allows us to revisit ground-based archival data to search for Lyman-$\alpha$ lines at these redshifts.

We utilize archival Keck MOSFIRE data made available via the grizli-Keck archive\footnote{\url{https://grizli-cutout.herokuapp.com/mosfire?mode=table}} that has previously been used in \citet{Valentino22} and includes reductions of MOSFIRE data up to 2022.
Among the 71 FRESCO [OIII] emitters at $z \sim 7-8$ in GOODS-N, 24 have existing Keck/MOSFIRE data in the Lyman-$\alpha$ wavelength range. Of these, 10 exhibit a Lyman-$\alpha$ line, allowing us to identify 8 new LAEs in the GOODS-N field (Table \ref{tab:LAE}). These include GN$\_$27795 (N107M, 2014A, PI: Finkelstein), GN$\_$26844 (N100M, 2013A, PI: Finkelstein), GN$\_$4394 (U076M, 2014A, PI: Faber), GN$\_$8563 (N118, 2022A, PI: Casey, Nunez, Prusinski, Steidel), GN$\_$29193 (N100M, 2013A, PI: Finkelstein), GN$\_$14535 (U092M, 2016A, PI: Faber, Guo, Yesuf, Koo, Barro), GN$\_$9680 (C194M, 2013A, PI: Konidaris), and GN$\_$25604 (N107M, 2014A, PI: Finkelstein). In addition to these, we include four previously known LAEs in GOODS-N (GN$\_$26051, GN$\_$22679, GN$\_$29192, and GN$\_$19441) identified in earlier studies \citep{ Jung2020, Saxena23, Tang2024}.

In the GOODS-S field, 13 out of 40 FRESCO [OIII] emitters have Keck/MOSFIRE coverage in the Lyman-$\alpha$ wavelength range. Among these, we detect Lyman-$\alpha$ emission in three galaxies: GS$\_$1744, GS$\_$6644, and GS$\_$28631, all previously identified in the literature \citep{Song16, Roberts-Borsani2023, Saxena23, Tang2024}. We also include the previously confirmed LAE JADES-13682 in our analysis \citep{Tang2024}.

We extracted the Lyman-$\alpha$ emission spectra from the 1D and 2D spectra files associated with each target in our sample. For the flux calibration, we used slit stars observed within the same mask as the potential LAEs. We determined the best stellar fit using model stellar spectra from \citet{Castelli2003}, along with Gaia \citep{Gaia2023} photometric magnitudes and astrophysical parameters (effective temperature, surface gravity, and metallicity). We then flux- and slit-corrected the galaxy spectra using the standard star spectra. For each source, we selected LAEs by requiring a signal-to-noise ratio (\(\sigma\)) greater than 5 in the Lyman-$\alpha$ line to ensure reliable detection. Each target file was loaded and aligned with its redshift-corrected wavelength grid based on the Lyman-$\alpha$ rest-frame wavelength of 1215.67 \AA. To analyze the Lyman-$\alpha$ emission in our sample of LAEs, we applied a Bayesian approach to model the observed 1D spectra. We modeled the Lyman-$\alpha$ line profile with a Gaussian function, using the known [OIII] redshift to inform our initial estimate for the line center. Since the redshift is determined from the [OIII] line, we have an approximate expected wavelength for the Lyman-$\alpha$ emission.

To determine the velocity offset (\(\Delta v\)) between the Lyman-$\alpha$ and [OIII] emission lines, we use the formula:

\begin{equation}
    \Delta v = c\frac{(z \mathrm{_{Ly\alpha}} - z\mathrm{_{[OIII]}})}{(1 + (z\mathrm{_{Ly\alpha}} + z\mathrm{_{[OIII]}})/2)},
\end{equation}

where z$_\mathrm{Lya}$ is a Lyman-$\alpha$ redshift, z$_\mathrm{[OIII]}$ is [OIII] redshift, and c is the speed of light. We applied a Gaussian fitting approach combined with bootstrapping to estimate uncertainties. For each source, we perturbed the [OIII] redshift by random variations within 60 km/s to simulate observational uncertainties. For each perturbed redshift, we ran an MCMC sampler to obtain the Lyman-$\alpha$ redshift. This process was repeated 1000 times to generate a distribution of velocity offsets. From this sample, we calculated the mean velocity offset and its uncertainty, reported as the 16th and 84th percentiles, which account for both intrinsic and observational variations in the velocity offset measurements. For fitting the Lyman-$\alpha$ line parameters, we used a Markov Chain Monte Carlo (MCMC) sampler to explore the posterior distribution and maximize the likelihood function, accounting for uncertainties and potential asymmetry in parameter distributions.

\begin{figure*}
     \centering
    \begin{subfigure}[t]{0.46\textwidth}
         \centering
         \includegraphics[width=1.\textwidth]{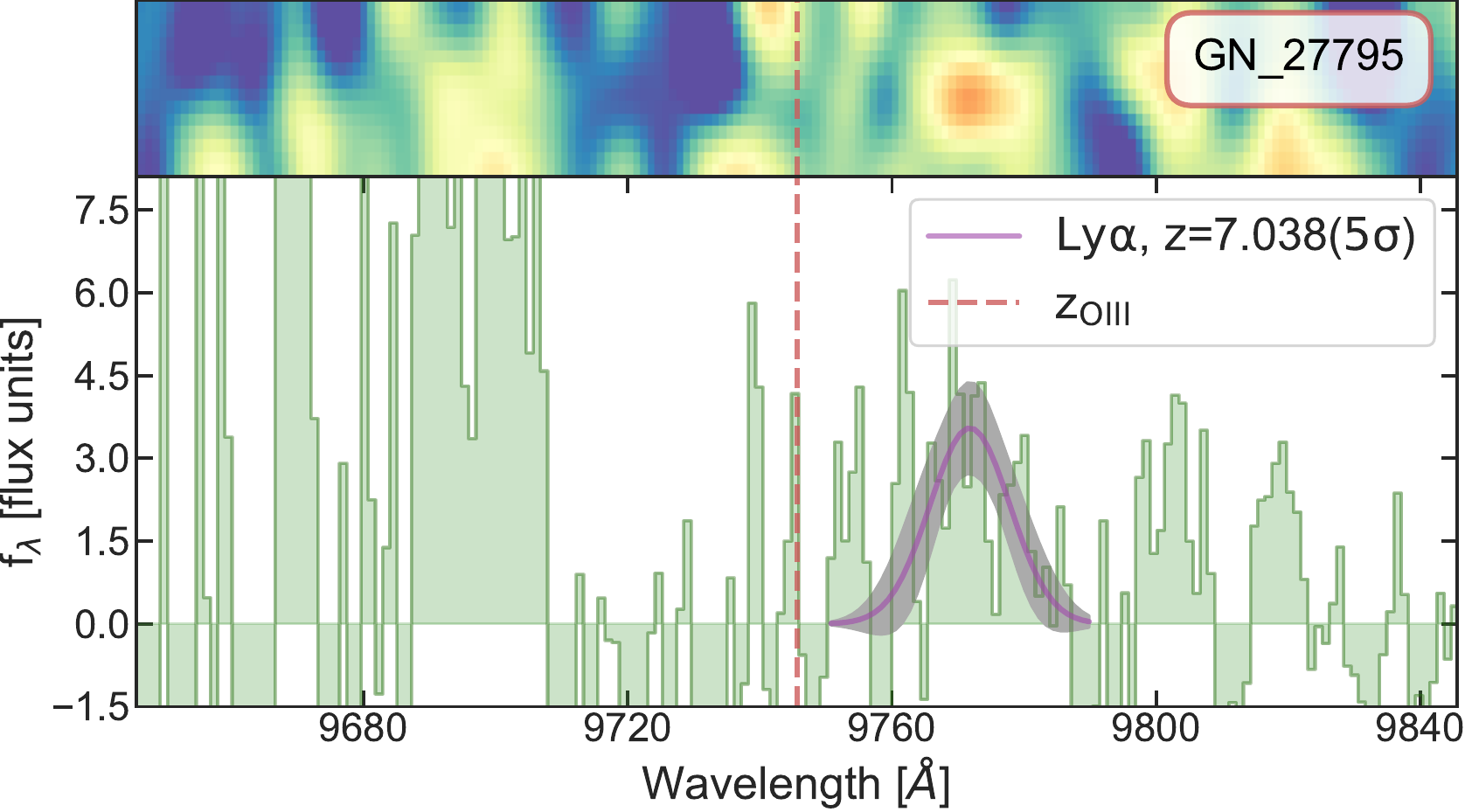}
         \includegraphics[width=1.\textwidth]{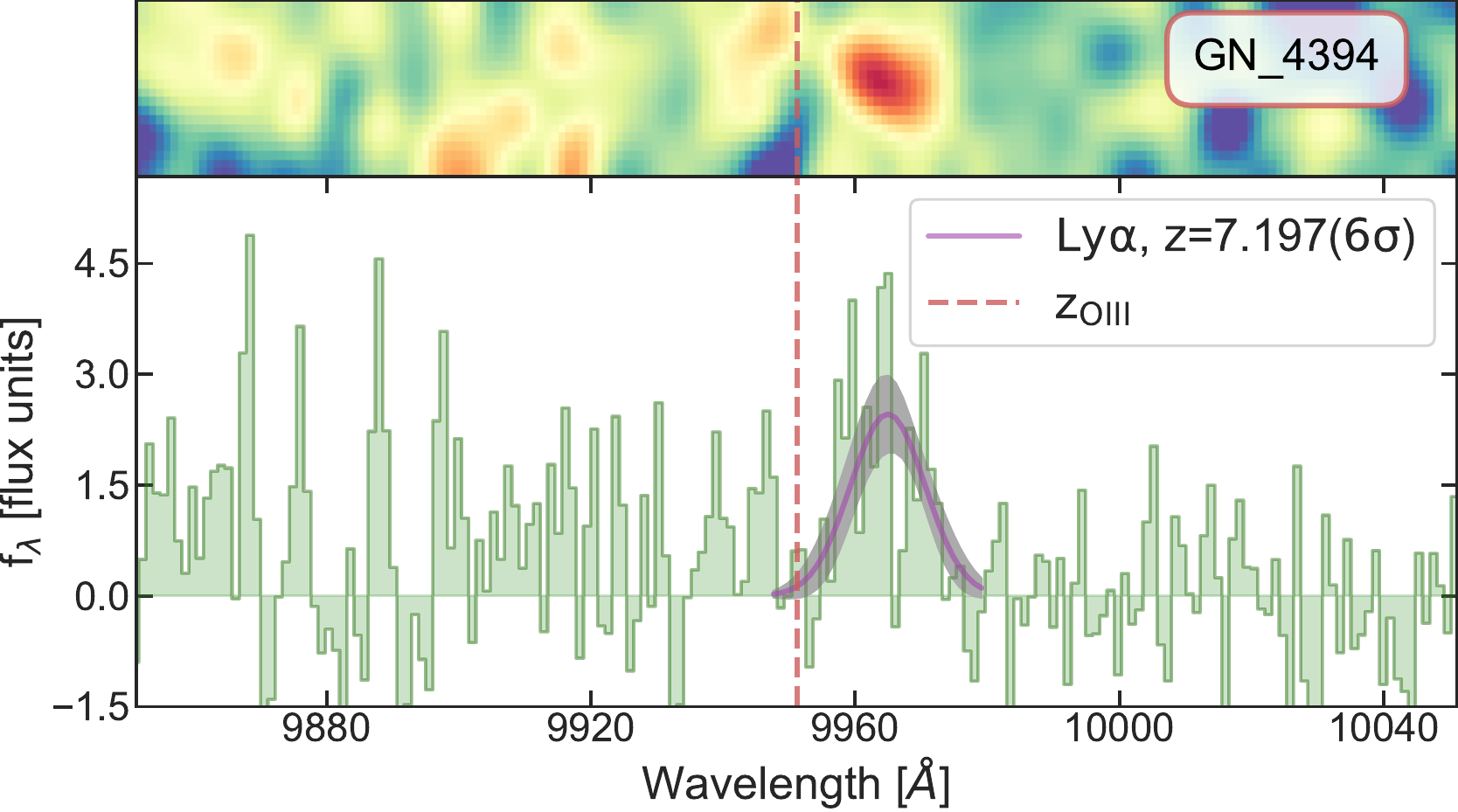}
         
         \includegraphics[width=1.\textwidth]{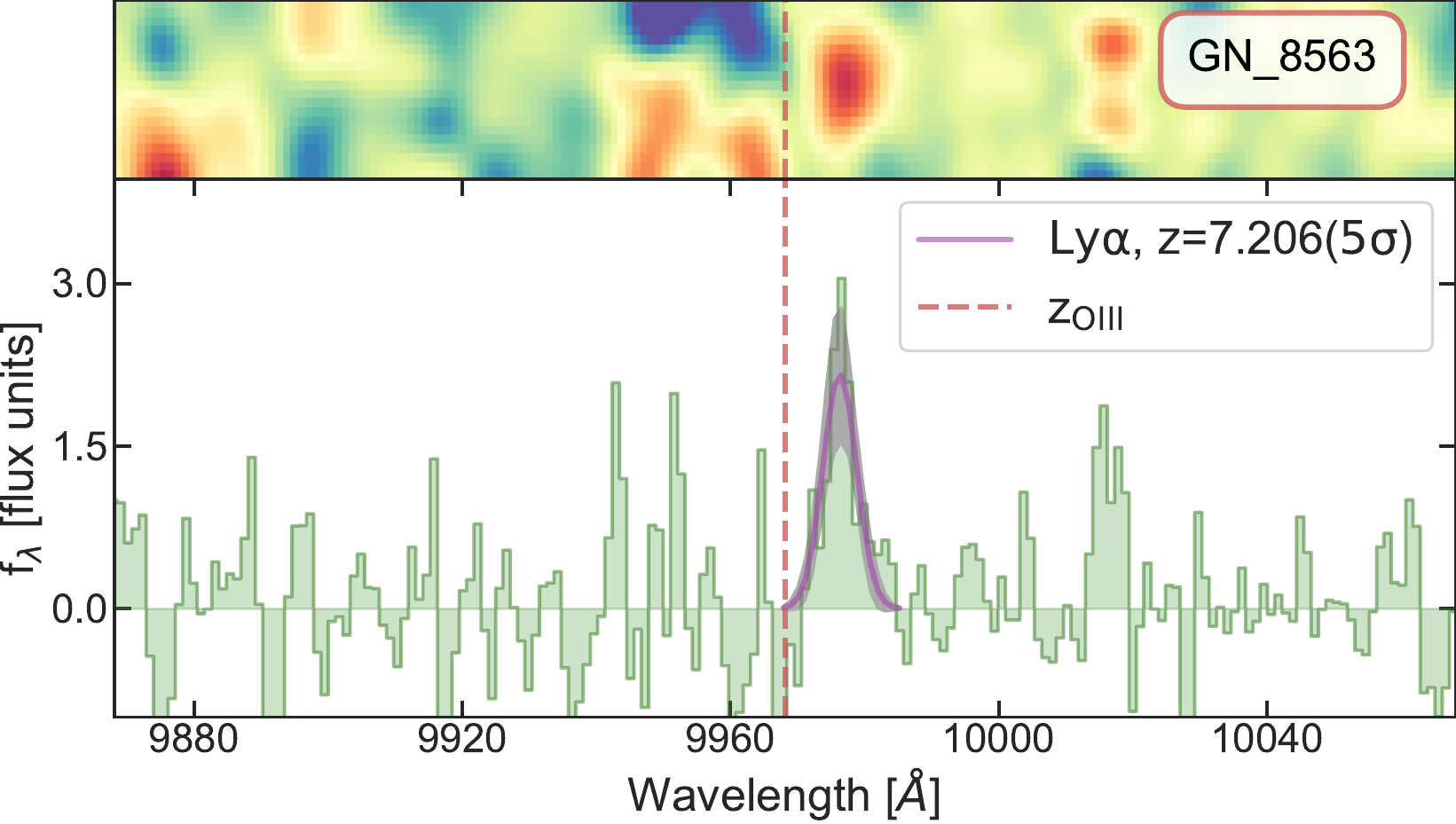}
         \includegraphics[width=1.\textwidth]{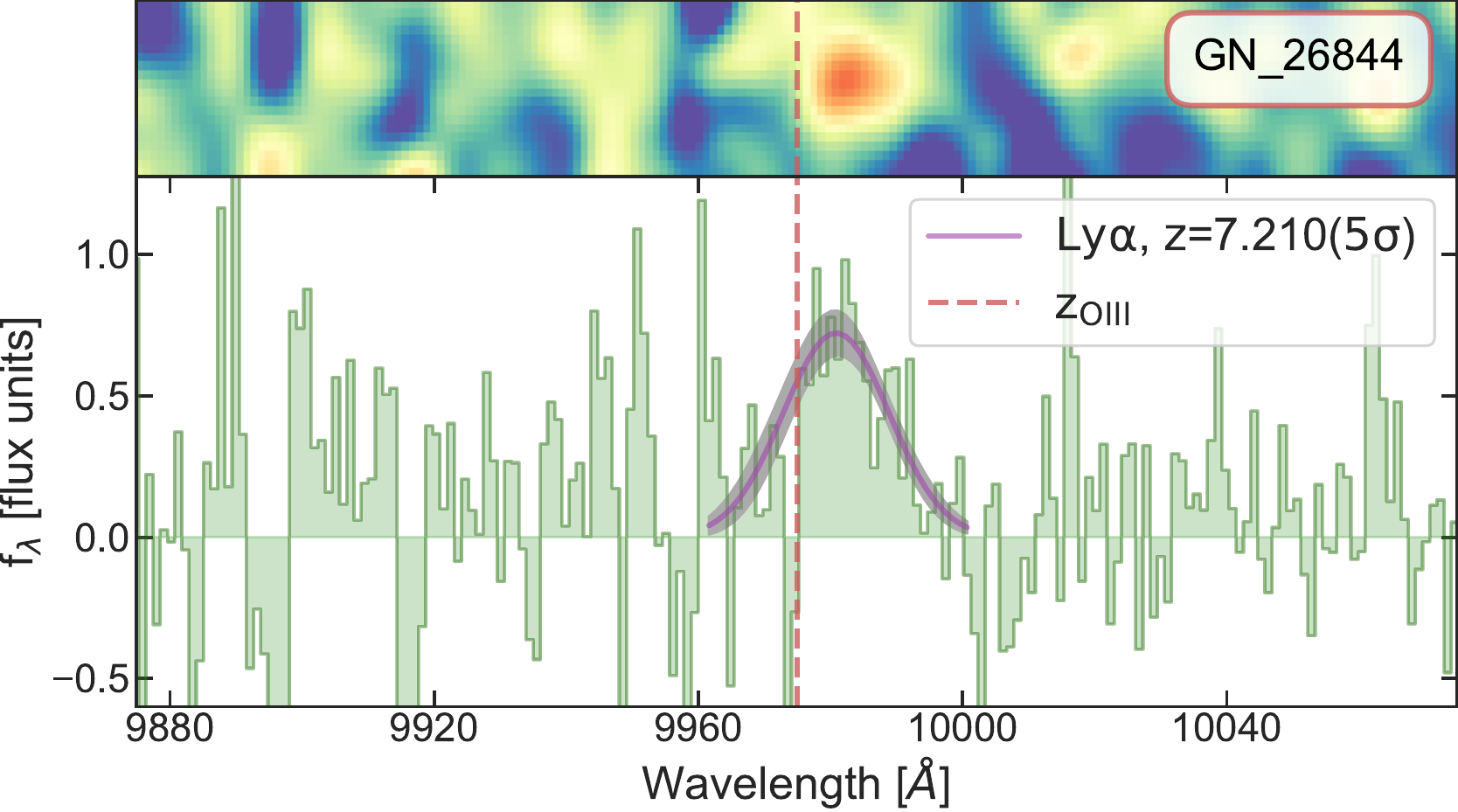}
         \includegraphics[width=1.\textwidth]{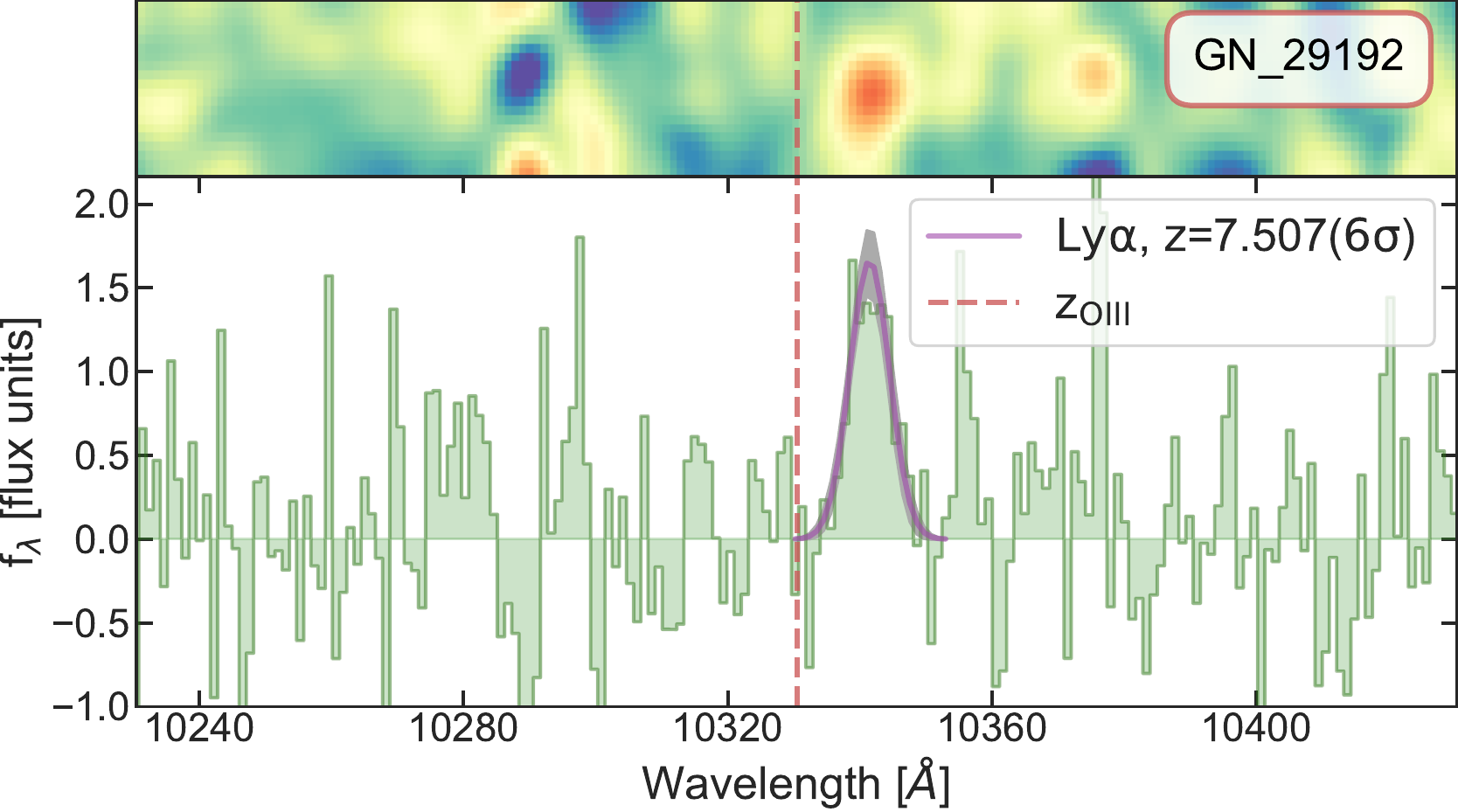}
        
         %\vspace{-0.8 cm}
    \end{subfigure}
    \hspace{0.05\textwidth}
     \begin{subfigure}[t]{0.46\textwidth}
         \centering
            \includegraphics[width=1.\textwidth]{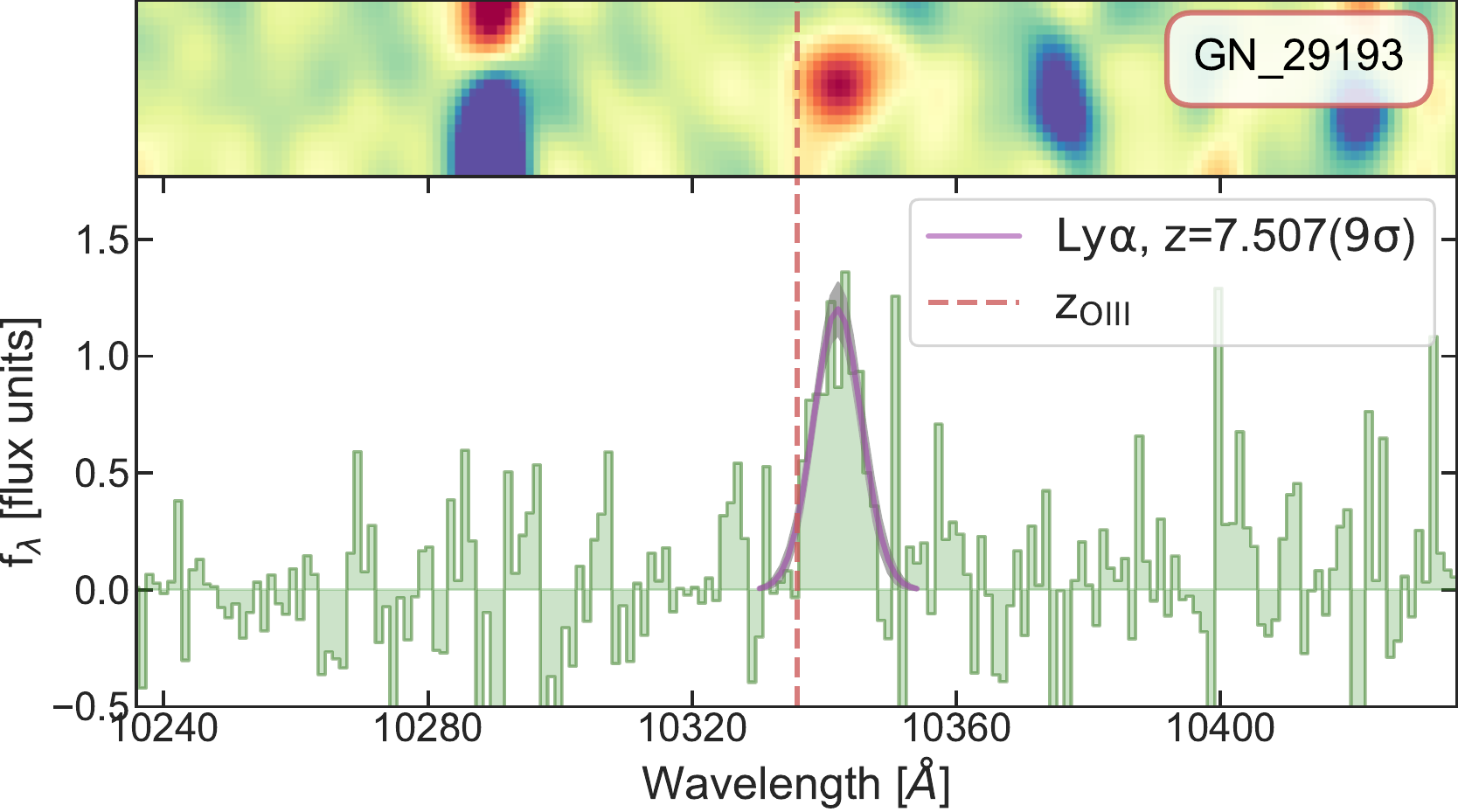}
        
            \includegraphics[width=1.\textwidth]{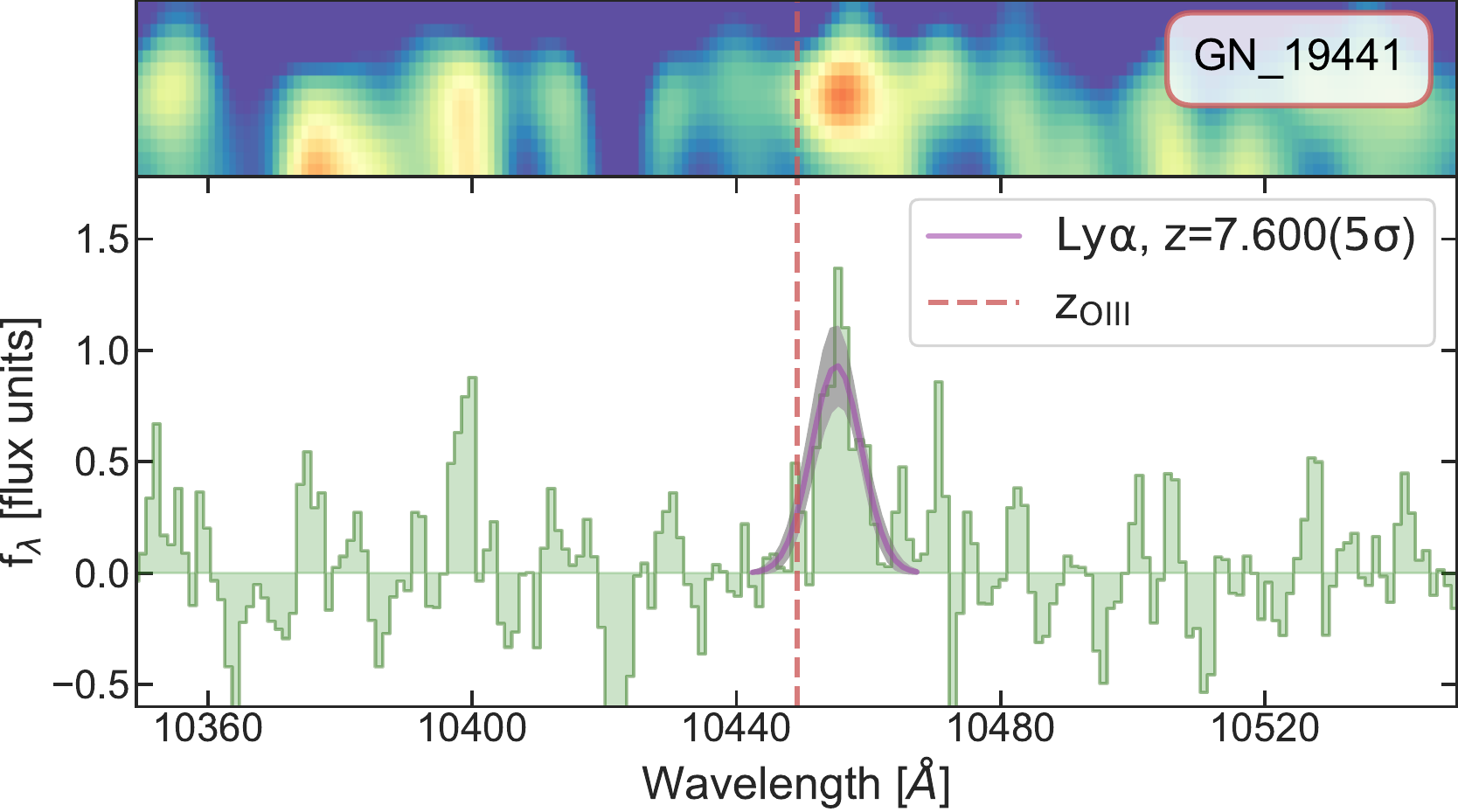}
            \includegraphics[width=1.\textwidth]{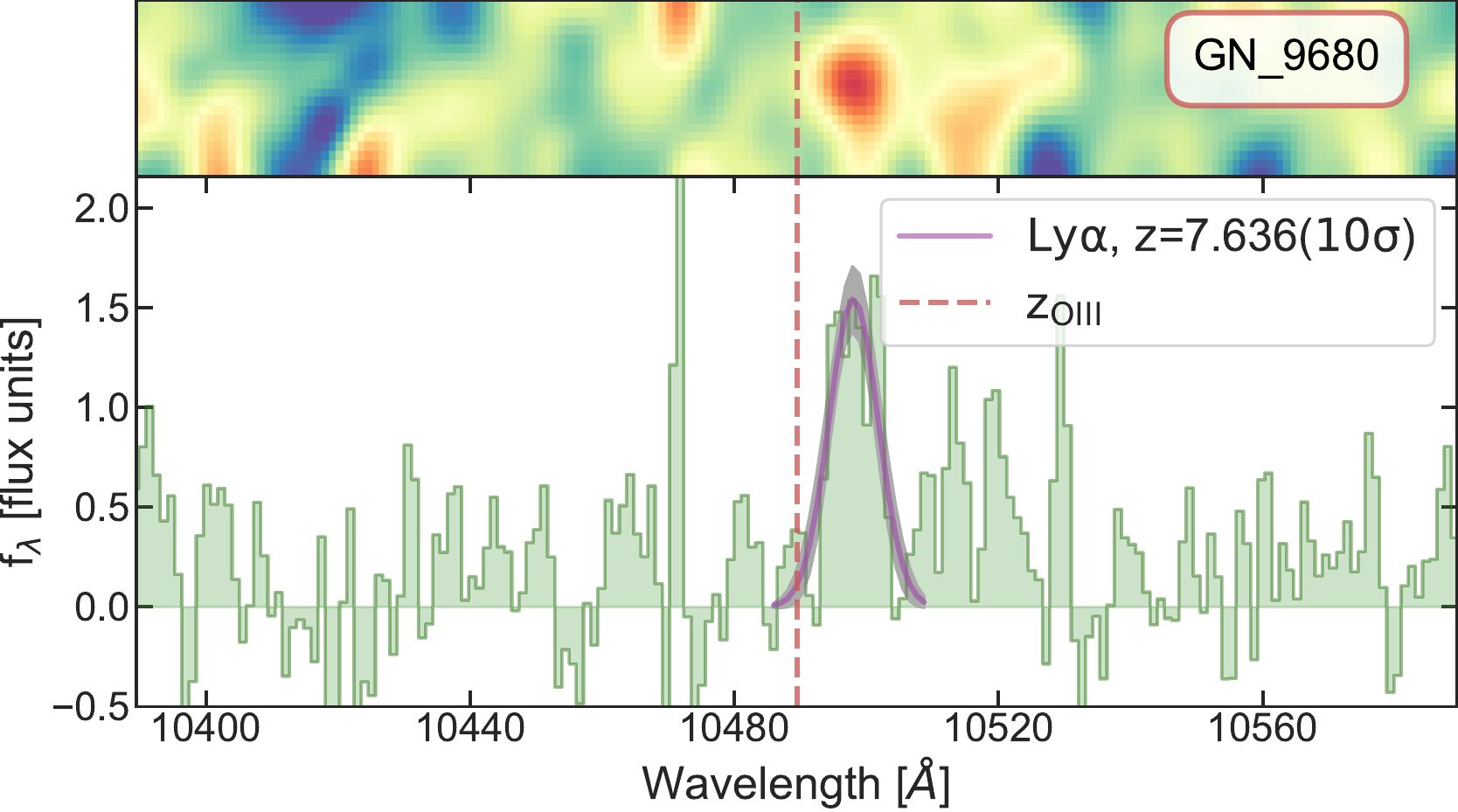}
            \includegraphics[width=1.\textwidth]{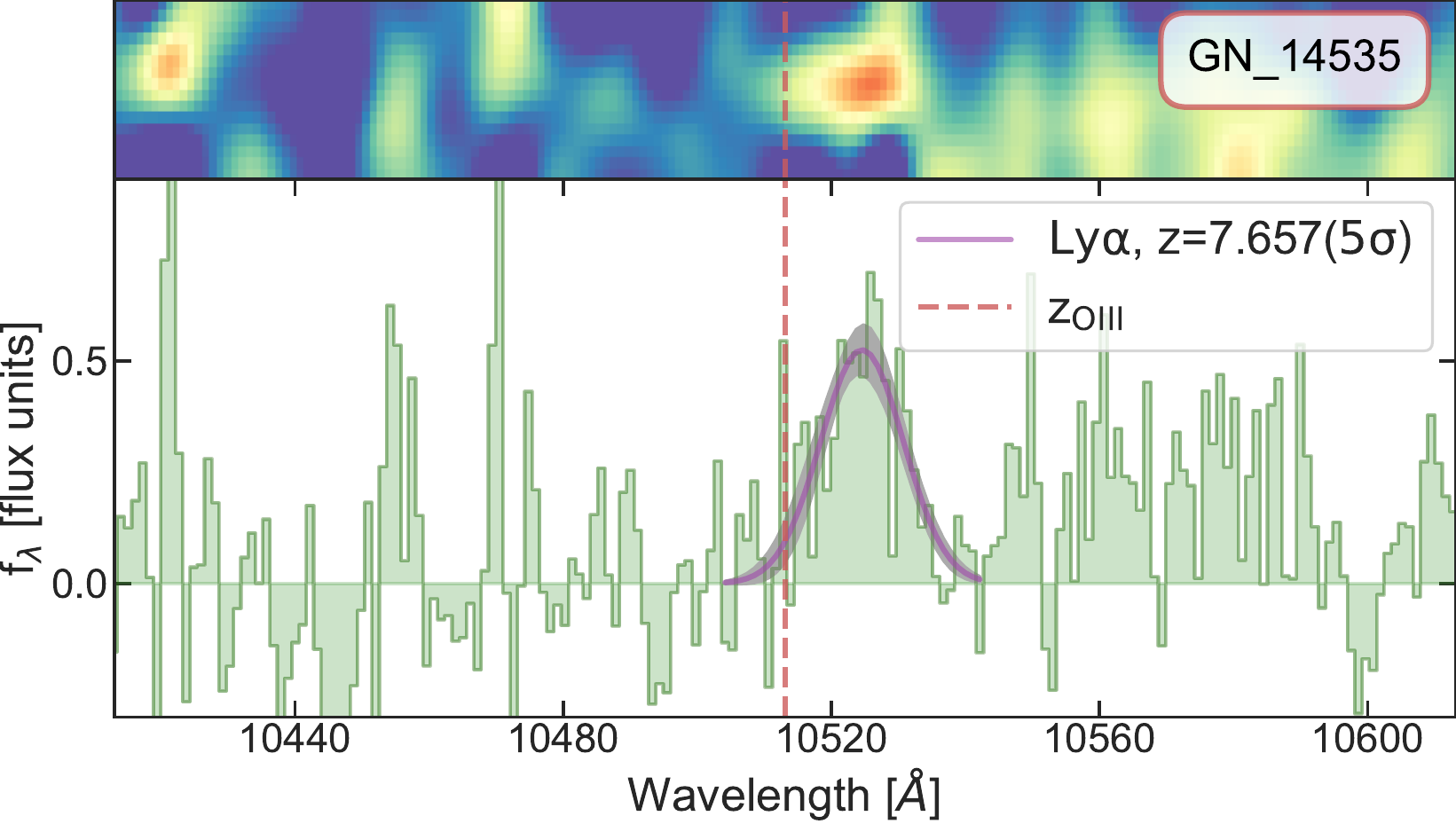}
            \includegraphics[width=1.\textwidth]{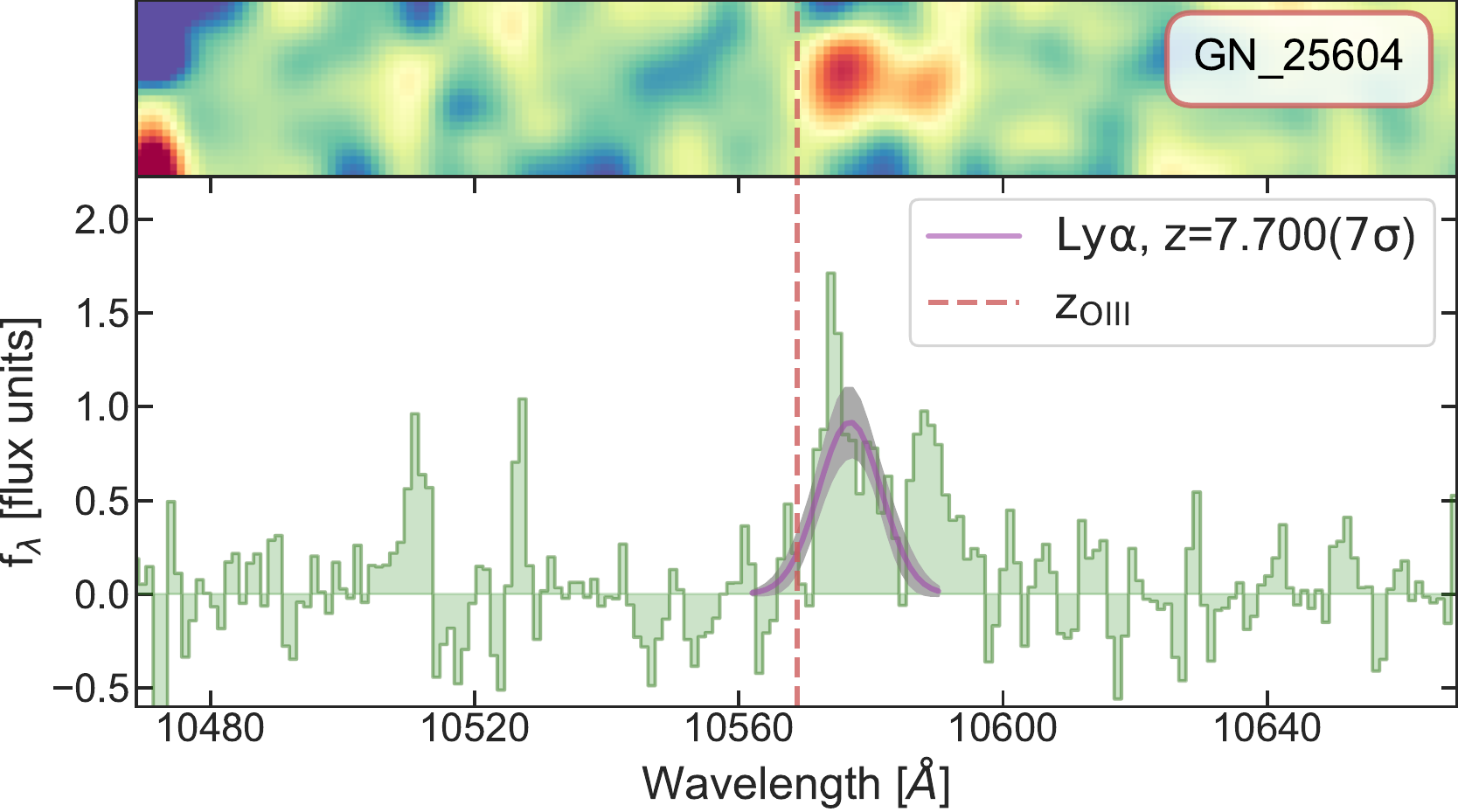}

        \end{subfigure}
    
        \caption{Spectra of Lyman-$\alpha$ emission for LAEs in the GOODS-N field. The top panel presents the 2D signal-to-noise spectra obtained using Keck MOSFIRE (Table \ref{tab:LAE}
), with each plot centered around the Lyman-$\alpha$ emission line. The dotted vertical line indicates the expected rest-frame Lyman-$\alpha$ position based on [OIII] redshift measurements. The bottom panel shows the corresponding 1D spectra, fitted with Gaussian profiles to model Lyman-$\alpha$ emission line.}
    \label{fig:LAE_sp}
\end{figure*}

The 2D spectra shown in Fig. \ref{fig:LAE_sp} were normalized, clipped, and smoothed using a Gaussian filter to emphasize the Lyman-$\alpha$ line region for each target. For visualization, we used a dual-panel format, where the smoothed and resampled 1D spectra are displayed below the processed 2D image, both centered around the Lyman-$\alpha$ emission line (Figure \ref{fig:LAE_sp}).

In Table \ref{tab:LAE} we list the LAEs identified in the GOODS-S and GOODS-N fields. The z$\mathrm{_{Ly\alpha}}$, z$\mathrm{_{[OIII]}}$ and Lyman-$\alpha$ flux (F$_\mathrm{Ly\alpha}$) in erg s$^{-1}$ cm$^{-2}$ for each target are provided, along with the corresponding velocity offset ($\Delta v$) in km s$^{-1}$ which reflects the difference between the Lyman-$\alpha$ and [OIII] redshifts. 

Out of 71 [OIII] emitters in GOODS-N and 40 [OIII] emitters in GOODS-S, we find that the fraction of LAEs (only including LAEs with EW>25 \AA) among galaxies with  $\mathrm{M_{UV} \leq -19.5}$ at $7 < z < 8$ is $0.28 \pm 0.07$ in GOODS-N and $0.08 \pm 0.05$ in GOODS-S. Combined, the overall LAE fraction in the GOODS fields is $0.19 \pm 0.05$. 

It is important to note that the Lyman-$\alpha$ line was not detected in the Keck data for some [OIII] emitters, and certain emitters were outside the coverage of Keck or JWST observations within the Lyman-$\alpha$ wavelength range. Therefore, the LAE fractions reported above should be considered as lower limits, since we cannot definitively rule out the possibility that these [OIII] emitters are also LAEs

%-------------------------------------

%---------------------------------------

\section{Results}
\label{sec:results}

In this section, we analyze the spatial distribution, clustering properties, and intrinsic characteristics of [OIII] emitters and LAEs at $7<z<8$. We first identify and categorize groups of LAEs in GOODS-N and GOODS-S to explore their large-scale environments. We then investigate the relationship between Lyman-$\alpha$ visibility and galaxy overdensities, comparing the environments of LAEs to those of non-LAE [OIII] emitters. Additionally, we examine the intrinsic properties of these galaxies, including their UV magnitudes, $\beta$ slopes, stellar masses, and [OIII] equivalent widths, to assess the factors influencing Lyman-$\alpha$ emission.

By combining environmental and intrinsic analyses, we aim to understand how both factors contribute to Lyman-$\alpha$ visibility and which plays a more dominant role.

\subsection {Groups of LAEs}
\label{sec:groups}

\begin{figure*}
\centering
\includegraphics[width=\linewidth]{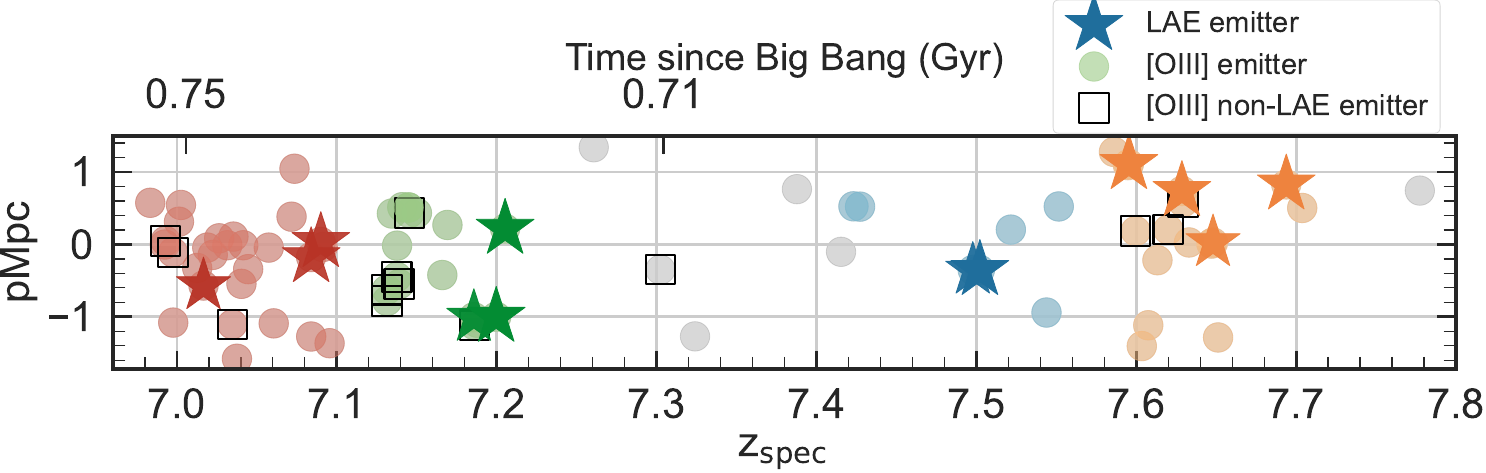}
\includegraphics[width=\linewidth, trim=0pt 180pt 0pt 200pt, clip]{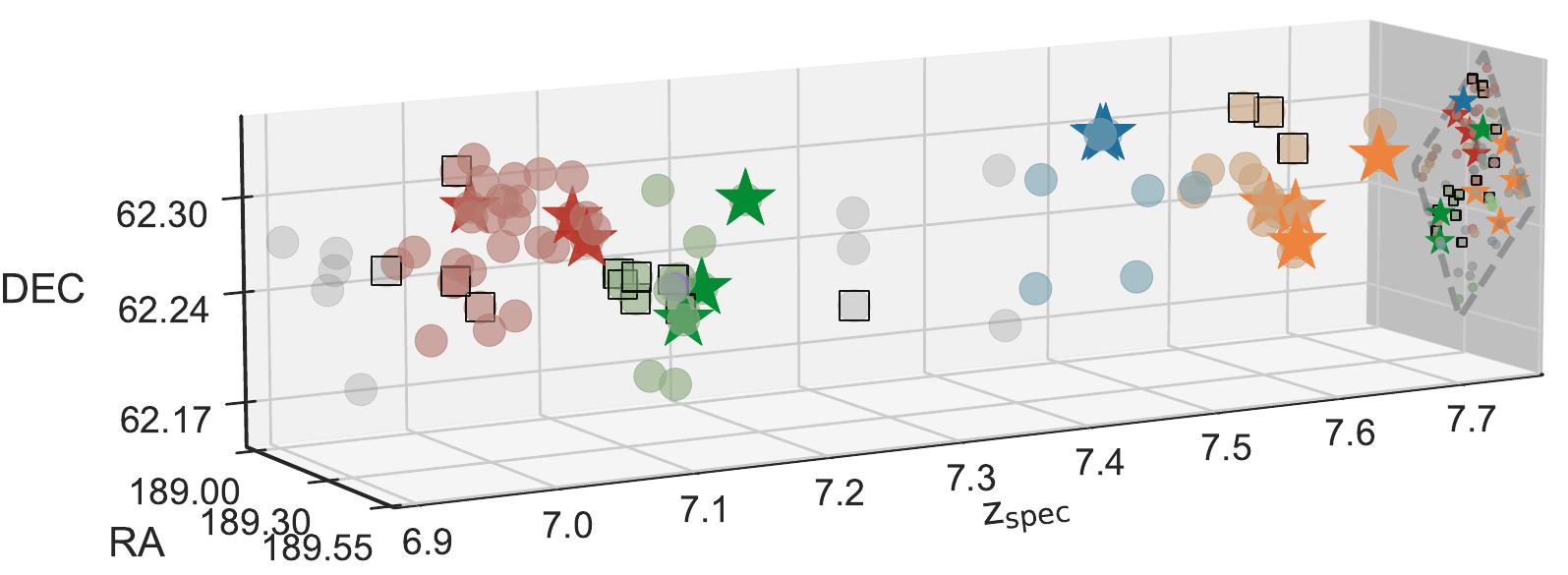}
\caption{The spatial distribution of Lyman-$\alpha$ and [OIII] emitters in GOODS-N is shown through 2D (top panel) and 3D (bottom panel) projections. The galaxies in the FRESCO field in GOODS-N are organized into four groups, each spanning 30 cMpc, with each group centered around the brightest LAE or one of the brightest galaxies in the region. Circles represent [OIII] emitters, stars indicate LAEs, and black squares highlight [OIII] emitters with spectra in the Keck archive at the Lyman alpha wavelength that were not detected as LAEs. The color scheme highlights the LAEs and their companions within each group: red for GN\_z7\_0, green for GN\_z7\_2, blue for GN\_z7\_5, and orange for GN\_z7\_6.}
\label{fig:GN2d}
\end{figure*}

\begin{figure*}
\centering
\includegraphics[width=0.9\linewidth]{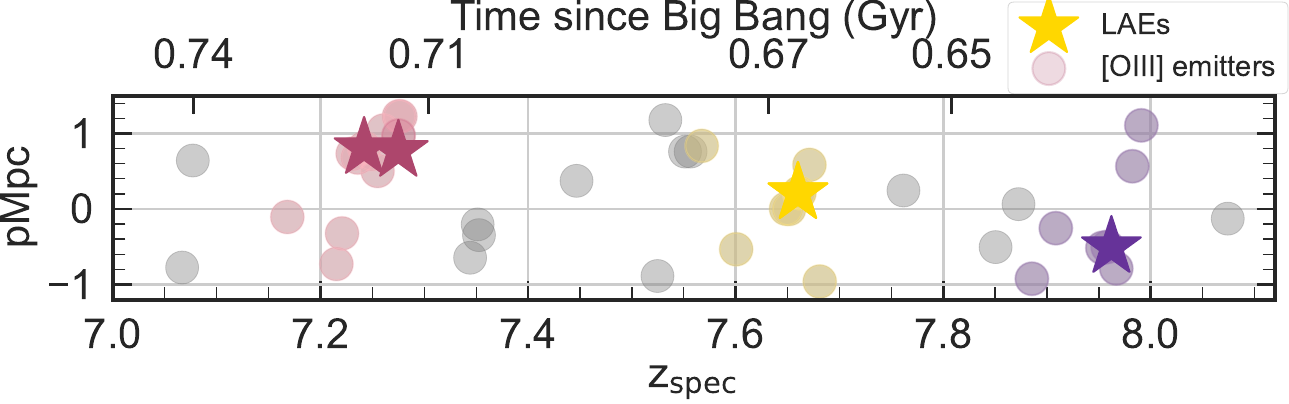}
\includegraphics[width=0.9\linewidth, trim=0pt 180pt 0pt 200pt, clip]{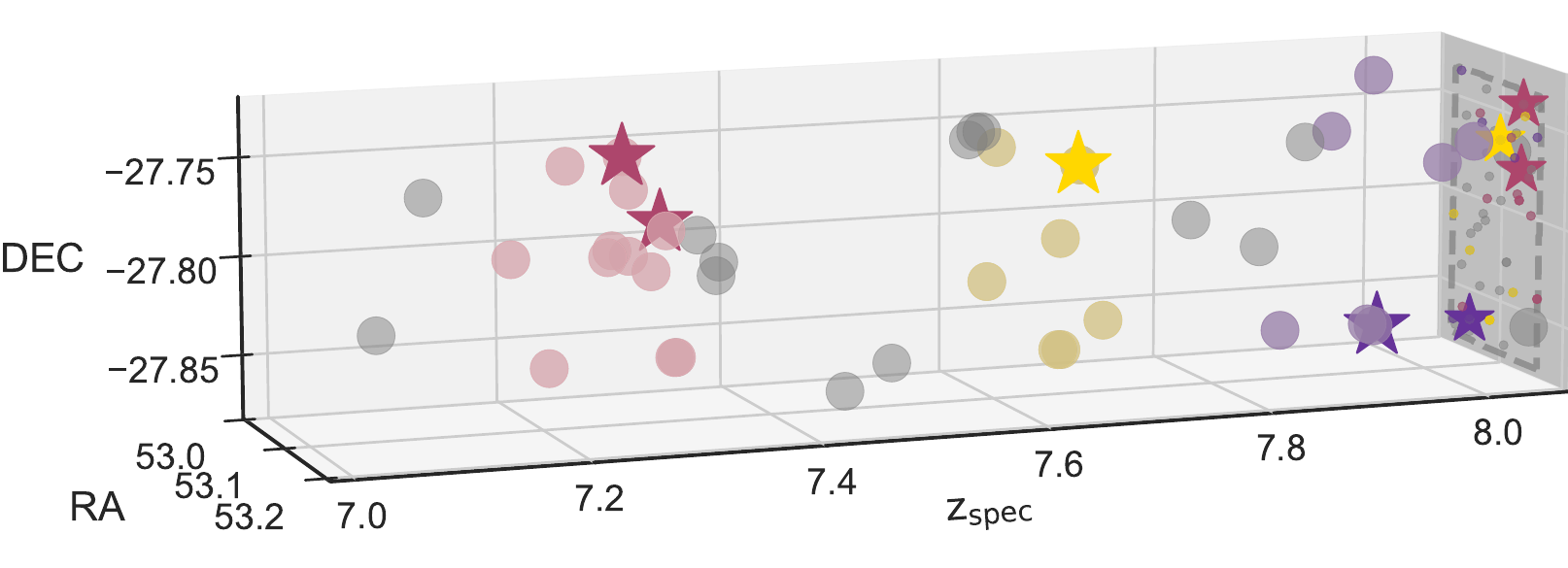}

\caption{The spatial distribution of Lyman-$\alpha$ and [OIII] emitters in GOODS-S is shown through 2D (top panel) and 3D (bottom panel) projections. The galaxies in the FRESCO field in GOODS-S are organized into three groups, each spanning 30 cMpc, with each group centered around the brightest LAE or one of the brightest galaxies in the region. Circles represent the [OIII] emitters, while stars indicate the LAEs. The color scheme highlights the LAEs and their companions within each group: pink for GS\_z7\_2, yellow for GS\_z7\_7, and purple for GS\_z7\_8.}
\label{fig:GS2d}
\end{figure*}

To investigate the environment of LAEs at 
$7<z<8$, we categorized the galaxies into four groups in GOODS-N (Figure \ref{fig:GN2d}) and three groups in GOODS-S (Figure \ref{fig:GS2d}). Each group spans 30 cMpc and is centered around the brightest LAE or one of the brightest in the region. This approach minimizes overlap between groups, allowing us to analyze distinct spatial regions independently.

In many studies at $7<z<8$ the typical size of ionized bubbles surrounding LAEs is estimated to be 0.5-1 pMpc \citep{Hayes2023, Umeda04, Witstok2024, Napolitano2024}. However, LAEs are expected to form within larger-scale overdensities, where multiple ionized regions may merge into extended structures. By using a 30 cMpc grouping scale, we aim to capture the broader cosmic environment of LAEs. Additionally, a 30 cMpc scale ensures that we analyze distinct spatial regions while minimizing overlap, helping us distinguish environmental differences between groups.

 Each group and its corresponding LAE are represented by different colors for visual clarity. The naming of the groups was based on the redshift of the central LAE in each region: GN\_z7\_0, GN\_z7\_2, GN\_z7\_5, GN\_z7\_6, GS\_z7\_2, GS\_z7\_7, and GS\_z8. This classification allows for a systematic investigation of the local environment and clustering properties of LAEs in the GOODS-N and GOODS-S fields, particularly focusing on the ionized regions around them at redshifts 7 to 8.
 The individual [OIII] and H$\beta$ grism spectra and tables of all sources in each group are presented in the appendix.

\subsubsection{GOODS-N}

In the GOODS-N field, the first group, \textbf{GN\_z7\_0} (Figure \ref{fig:group1}, Table \ref{tab:1f}), centered at redshift $z \sim 7$, consists of \textbf{30 [OIII] emitters}, including \textbf{3 LAEs}: GN\_27795, GN\_26051, and GN\_22679. On Figure \ref{fig:GN2d}, this group is represented by red circles, and the LAEs are marked as red stars. This is the most populated group in our study. It includes several luminous galaxies with $\mathrm{M_{UV}} < -20$ and massive galaxies, GN\_14967 and GN\_30464, with $\log_{\mathrm{M}_{\star}} >9.25$.  GN\_26051 and GN\_22679 were first identified in \citet{Tang2024}, where the large overdensity at this redshift around these sources was discovered.

The second group, \textbf{GN\_z7\_2} (Figure \ref{fig:group2}, Table \ref{tab:2f}), is situated at z$\sim$7.2 and contains \textbf{17 [OIII] emitters}, depicted as green circles in Figure \ref{fig:GN2d}. Among these, \textbf{3 are LAEs}—GN\_4394, GN\_8563, and GN\_26844—highlighted as green stars. This group houses exceptionally luminous galaxies, and also features extraordinarily bright AGN, such as GN\_5688 (GNz7q; \citealt{Fujimoto2022}).

Another candidate, GN\_14101, previously identified in \citep{Jung2020} as z8\_GND\_22233, was reported to have an emission line at 1.015 $\mu$m with S/N = 7.1. At a redshift of 
z=7.145, this corresponds to the NV line. This raises the possibility that GN\_14101 could also host AGN activity, bringing the total number of AGNs in this group to three, including one AGN that is also an LAE. An interesting feature of the group GN\_z7\_2 is the presence of GN\_26844, an LAE, which is accompanied by another LAE, GN\_4394, that stands out due to its exceptionally large Lyman-$\alpha$ equivalent width of 221 \AA. This unusually high EW of Lyman-$\alpha$, combined with a broad H$\beta$ line with FWHM = 1069 $\pm$ 239 km/s \citep{xiao2025}, raises the possibility that GN\_4394 may also be an AGN. Notably, it is located just 0.167 pMpc from GN\_5688. The potential presence of an AGN in this context highlights the need for deeper investigation into the interactions between LAEs and AGNs, especially regarding their contributions to ionization and the overall evolution of the group.

The third group, \textbf{GN\_z7\_5} (Figure \ref{fig:group3}, Table \ref{tab:3f}), at $z \sim 7.5$, includes \textbf{7 companions} (blue circles), and \textbf{two LAEs}, GN\_29192 and GN\_29193, marked by blue stars. These two LAEs are very close to each other, suggesting a potential merger, which might explain the Lyman-$\alpha$ emission.

Finally, the group \textbf{GN\_z7\_6} (Figure \ref{fig:group4}, Table \ref{tab:4f}), consisting of \textbf{16 sources} (orange circles), includes \textbf{4 LAEs}: GN\_19441, GN\_14535, GN\_25604, and GN\_9680, marked by orange stars.

Interestingly, there is a significant clustering of bright galaxies in the GN\_z7\_0 group, which is not observed between the GN\_z7\_2 and GN\_z7\_5 groups. Bright galaxies appear to reside in the most overdense region and may serve as tracers of large-scale structure.

\subsubsection{GOODS-S}

The first group in the GOODS-S field, \textbf{GS\_z7\_2} (Figure \ref{fig:group5}, Table \ref{tab:group5}), with a redshift of $z \sim 7.2$, consists of \textbf{12 [OIII] emitters}, represented in Figure \ref{fig:GS2d} as pink circles, including 2 LAEs: GN\_1744 and JADES-13682. One of the LAEs, JADES-13682, identified in \citep{Saxena23}, is very faint and is not detected as an [OIII] emitter in FRESCO. The overdensity around this LAE and the presence of bright companions in its vicinity may explain the visibility of  Lyman-$\alpha$ line and for such a faint source. The overdensity around JADES-13682 was also confirmed in \citet{Helton2024, Endsley2024, Tang2024, Witstok2024,Chen2025}.

Another group in GOODS-S, \textbf{GS\_z7\_7} (Figure \ref{fig:group6}, Table \ref{tab:group6}), marked as yellow, located around the bright LAE GS\_6644, previously identified by \citet{Song16}, consists of only \textbf{7 companions} within 30 cMpc. One of the companions, GS\_6645, is in very close proximity to LAE GS\_6644. Their close location and matching redshift suggest that these two galaxies might be undergoing a merger process.

The final group in GOODS-S, \textbf{GS\_z8} (Figure \ref{fig:group7}, Table \ref{tab:group7}), is centered around LAE GS\_28631, previously identified by \citet{Roberts-Borsani2023} at $z \sim 8$. This group, marked as purple, includes \textbf{7 [OIII] emitters}.

The local environment in GOODS-S shows slightly different clustering properties compared to GOODS-N, potentially indicative of varied large-scale structure at $z \sim 7-8$. The density of bright [OIII] emitters ($\mathrm{M_{UV}}<-19.5$) in GOODS-S are lower compared to that in GOODS-N, suggesting that fainter sources dominate the population in this field.

% completeness of GS?

\subsection{Overdensities around LAEs} 
\label{sec:overd}

The spatial distribution of galaxies can play a crucial role in shaping their visibility and the formation of ionized bubbles during reionization. LAEs are often found in overdense regions, where neighboring galaxies may help create ionized bubbles that facilitate Lyman-$\alpha$ escape \citep{Castellano2018, Jung2022, Larson2022,Leonova2022, Endsley2022, Saxena23, Tang2023, Tang2024, Witten2024}. To test this hypothesis, we compare the overdensities around LAEs with those of [OIII] emitters that do not exhibit detected Lyman-$\alpha$ emission.

To estimate the number of galaxies expected within a given area, we use the UV luminosity function (LF) derived in \citep{Bouwens2021} at redshifts $z \sim 2-10$:

\begin{equation}
    \begin{cases}
 M_{\rm{UV}}^*=-21.03 -0.04(z-6) \\
 \phi^*/\rm{Mpc}^{-3}=0.4\times 10^{-3} \times
 10^{-0.33(z-6) -0.24(z-6)^2} \\
 \alpha=-1.94-0.11(z-6) \\
    \end{cases}
    \label{bouwcr}
\end{equation}

To compute the expected number of detections within the certain volume, we integrate the UV LF as follows:
 
  \begin{equation}
     N_\mathrm{exp}=\int \Omega \frac{dV_{\rm{com}}}{dz}  dz\int\phi(M) dM
     \label{eqexp}
 \end{equation}

We define the overdensity parameter $\delta$ as the difference between the number of companions and LAEs within 30 cMpc and the expected number of galaxies, normalized by the expected number of galaxies for each group:

\begin{equation}
    \mathrm{\delta_{overd}}=\mathrm{\frac{N_{det}-N_{exp}}{N_{exp}}}
\end{equation}

where $\mathrm{N_{det}}$ is the number of sources detected by the JWST FRESCO survey within the specified distance 1-2 pMpc (or $\Delta z\sim 0.03-0.05$) from the central LAE, and $\mathrm{N_{exp}}$ is the expected number of sources derived from the UV luminosity function (LF) at the relevant redshift. Only sources with $\mathrm{M_{UV}<-19.5}$ are considered for the overdensity calculation for completeness \citep{Meyer2024}.

A key question in understanding Lyman-$\alpha$ visibility is whether overdense regions, where large ionized bubbles could enhance the transmission of Lyman-$\alpha$ photons, are the primary factor influencing the visibility of LAEs. To test this, we compare the overdensities of LAEs with [OIII]-non LAEs. If overdensity alone determined Lyman-$\alpha$ visibility, we would expect LAEs to reside in significantly higher-density regions.

\begin{figure*}
\centering
\includegraphics[width=\linewidth]{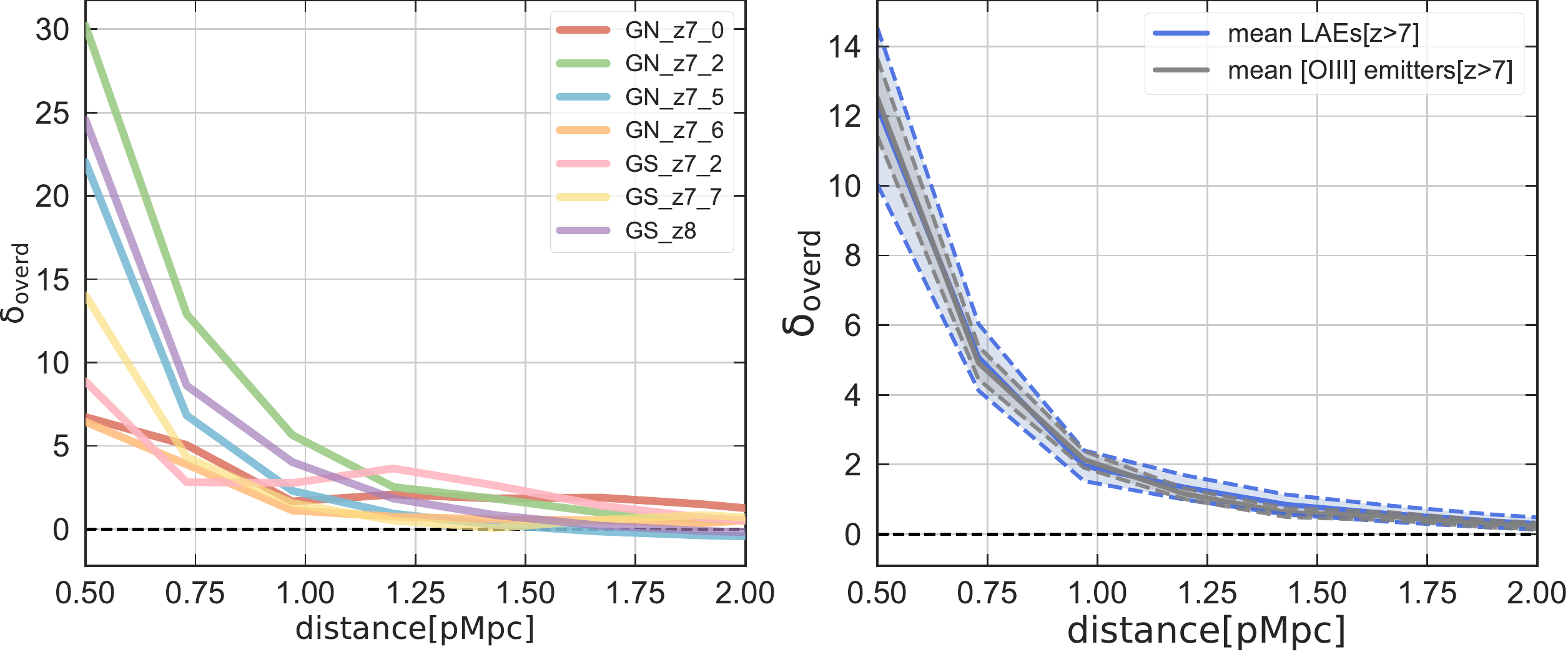}
\caption{\textbf{Left panel}: The overdensities around the 7 central LAEs in each group within the GOODS-N and GOODS-S fields. Each curve is color-coded according to the same group scheme as in Figures \ref{fig:GN2d} and \ref{fig:GS2d}: red for GN\_z7\_0, green for GN\_z7\_2, blue for GN\_z7\_5, orange for GN\_z7\_6, pink for GS\_z7\_2, yellow for GS\_z7\_7, and purple for GS\_z7\_8. The overdensity parameter is calculated for sources with $\mathrm{M_{UV}<-19.5}$ to ensure completeness. \textbf{Right panel}: The mean overdensity for LAEs across all groups is represented by the blue curve, with the shaded region denoting the 16th to 84th percentile range. The gray line shows the overdensity for [OIII]-non LAEs in the GOODS-N and GOODS-S fields at $7 < z < 8$. }
\label{fig:finalplot}
\end{figure*}

The left panel of Figure \ref{fig:finalplot} shows the overdensity parameter for each group, with different colors corresponding to different regions. The right panel presents the mean overdensity across all seven groups, shown as a blue curve, with the shaded region representing the 16th–84th percentile range from bootstrap resamplings of individual galaxies. For comparison, the gray dashed curve represents the mean overdensity around all [OIII]-non LAEs in the GOODS-N and GOODS-S fields at $7<z<8$.

Surprisingly, we find that the mean overdensity parameters are similar for both LAEs and non-LAEs. This suggests that while LAEs have large $\delta$, overdensity alone does not determine whether a galaxy exhibits Lyman-$\alpha$ emission. Other factors, such as the presence of sufficiently large ionized bubbles or galaxy properties affecting Lyman-$\alpha$ escape, likely play a role in shaping their visibility.

%---------------------------------------
%---------------------------------------

\subsection{Properties of [OIII] emitters and LAEs}
\label{sec:prop}

In the previous subsection, we examined the role of large-scale environment in determining Lyman-$\alpha$ visibility by comparing the overdensities of LAEs and non-LAEs. 
As we have found that overdensities do not correlate with the presence of Lyman-$\alpha$ emission, other physical properties of galaxies must play an important role.
In this section, we compare the intrinsic properties of LAEs and other [OIII] emitter, including their $\beta$ slope, UV magnitude, $\log_{\mathrm{M}_{\star}}$, EW, and star-formation history, and quantify differences with two-sample K–S tests \citep{KolmogorovSmirnovTest,Virtanen2020SciPy}. These quantities are derived using the \texttt{prospector} SED-fitting code \citep{Johnson21} following the choices outlined in \citet{Naidu2024}.

\begin{figure*}
     \centering
    \begin{subfigure}[t]{0.45\textwidth}
         \centering

         \includegraphics[width=1.\textwidth]{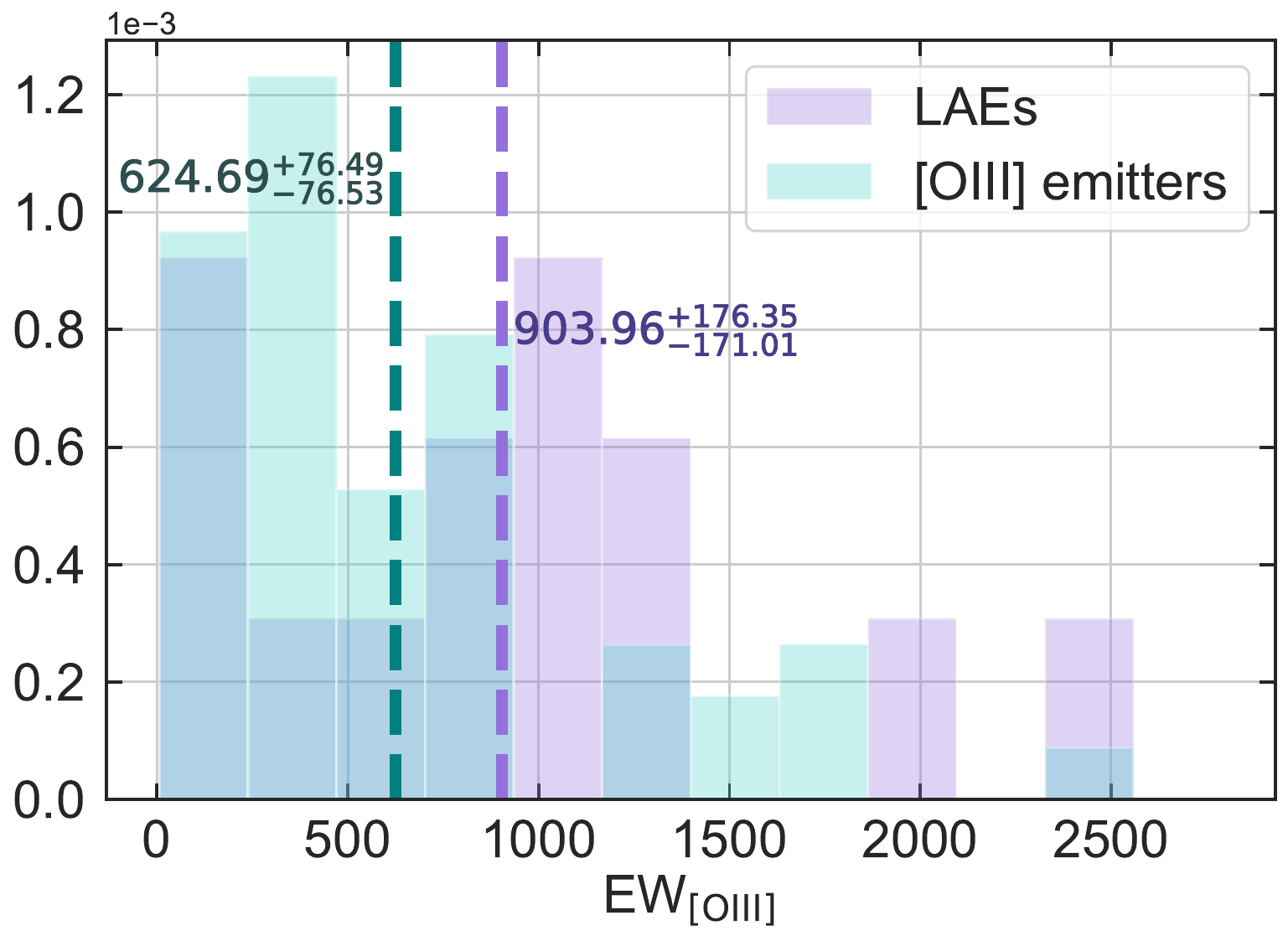}
         
         \includegraphics[width=1.\textwidth]{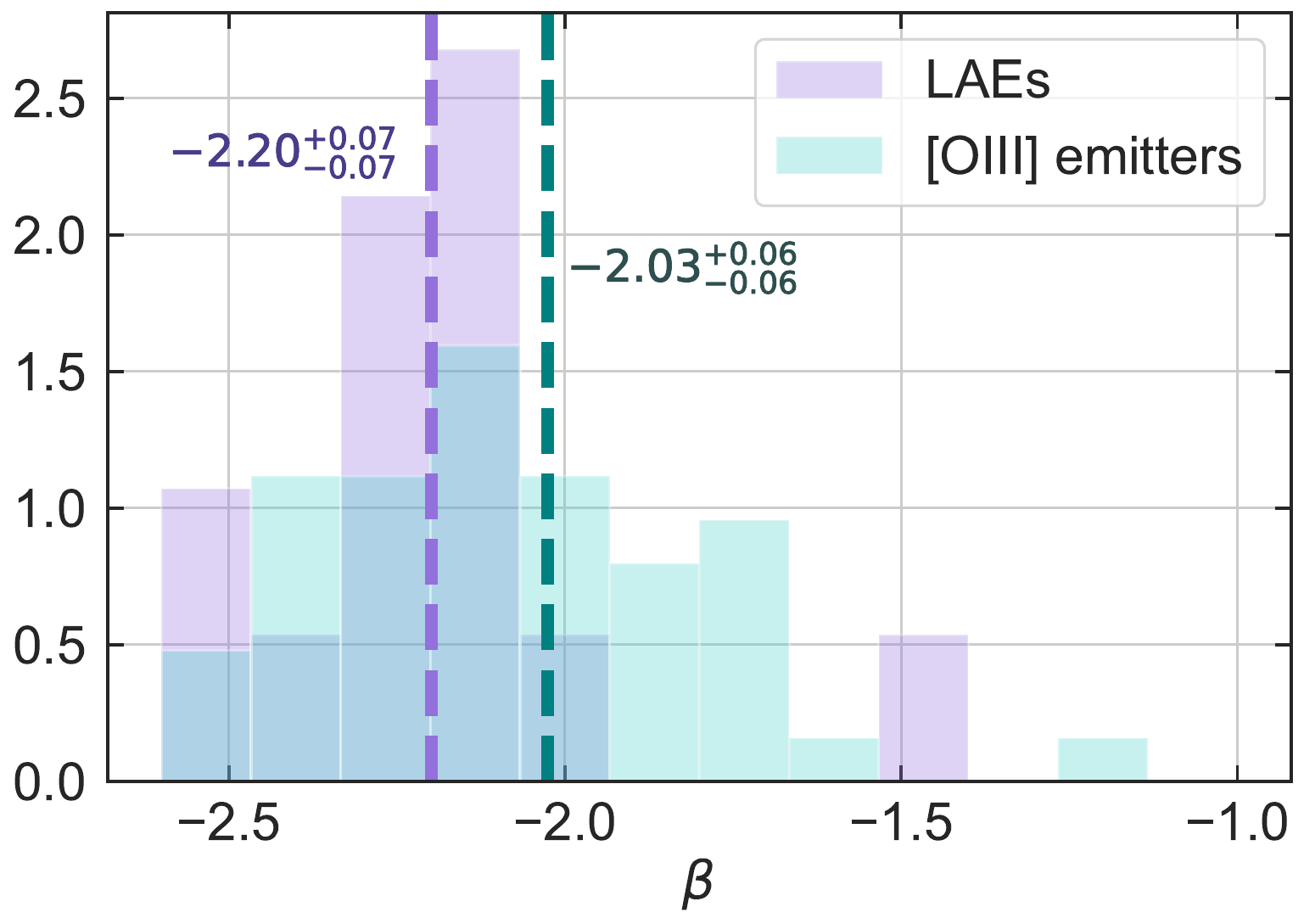}
         
        \includegraphics[width=1.\textwidth]{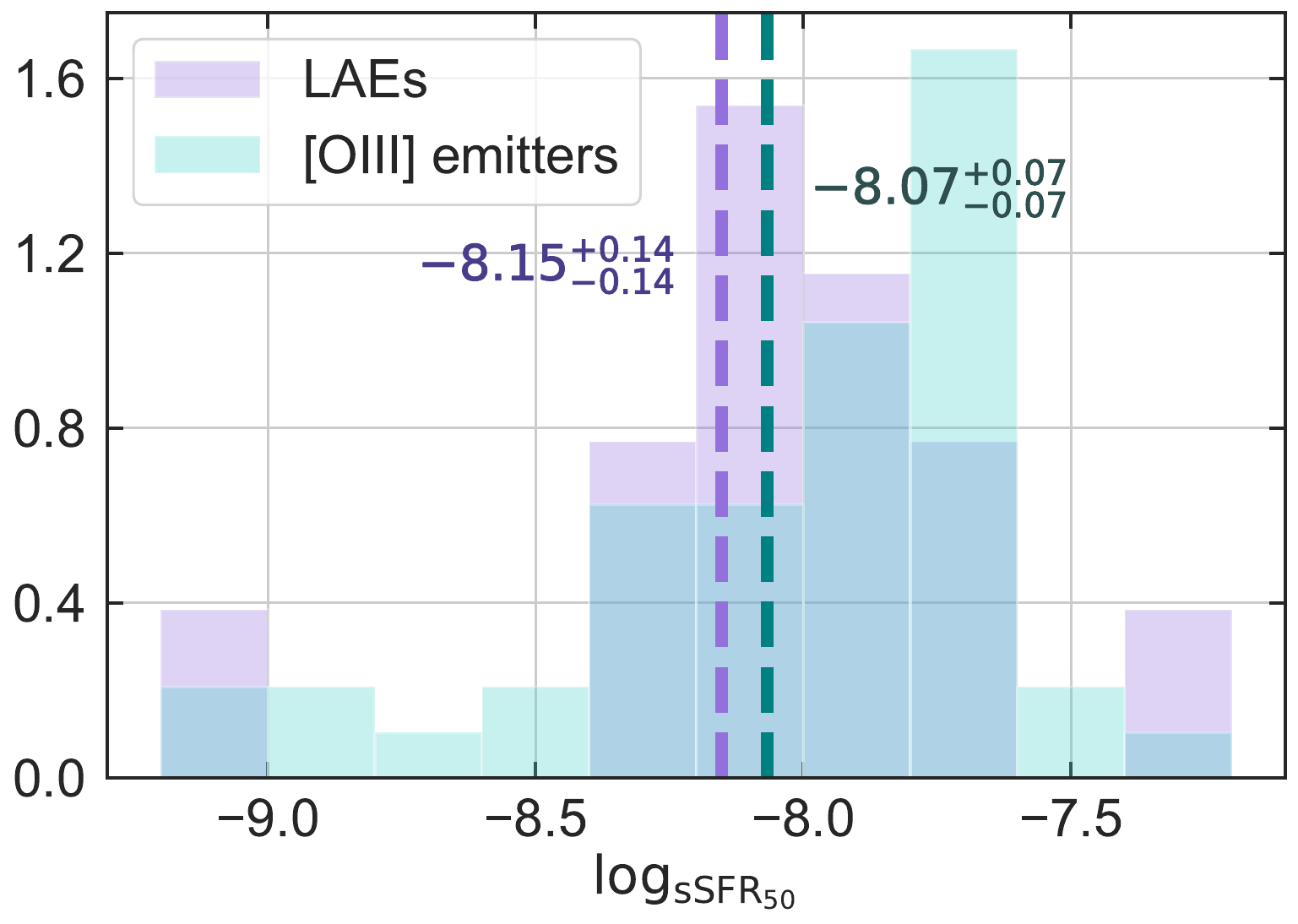}
     \end{subfigure}
     \qquad
     \begin{subfigure}[t]{0.45\textwidth}
         \centering
         %\vspace{-3 cm}
         \includegraphics[width=1.\textwidth]{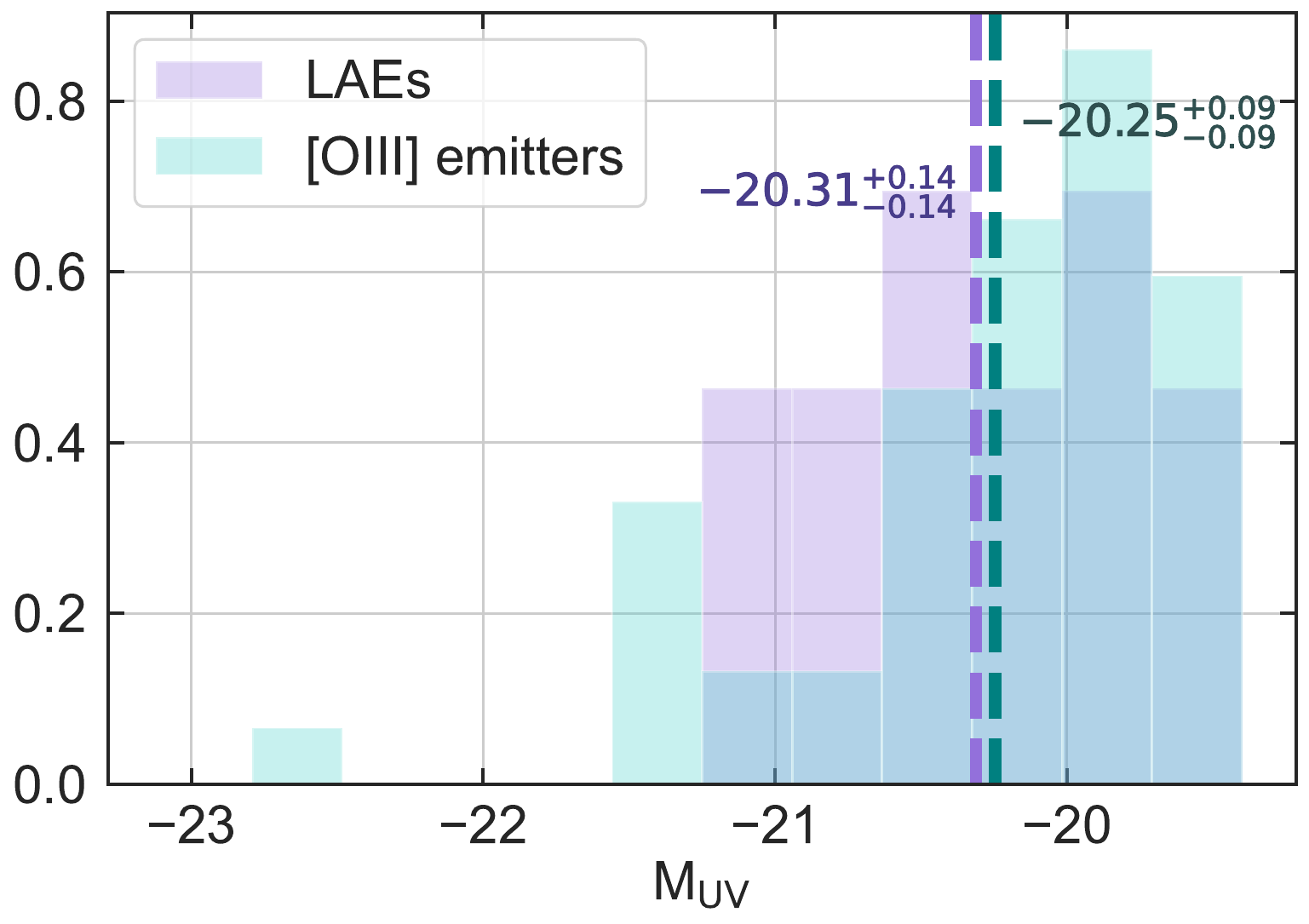}
         %\vspace{-3. cm}
         \includegraphics[width=1.\textwidth]{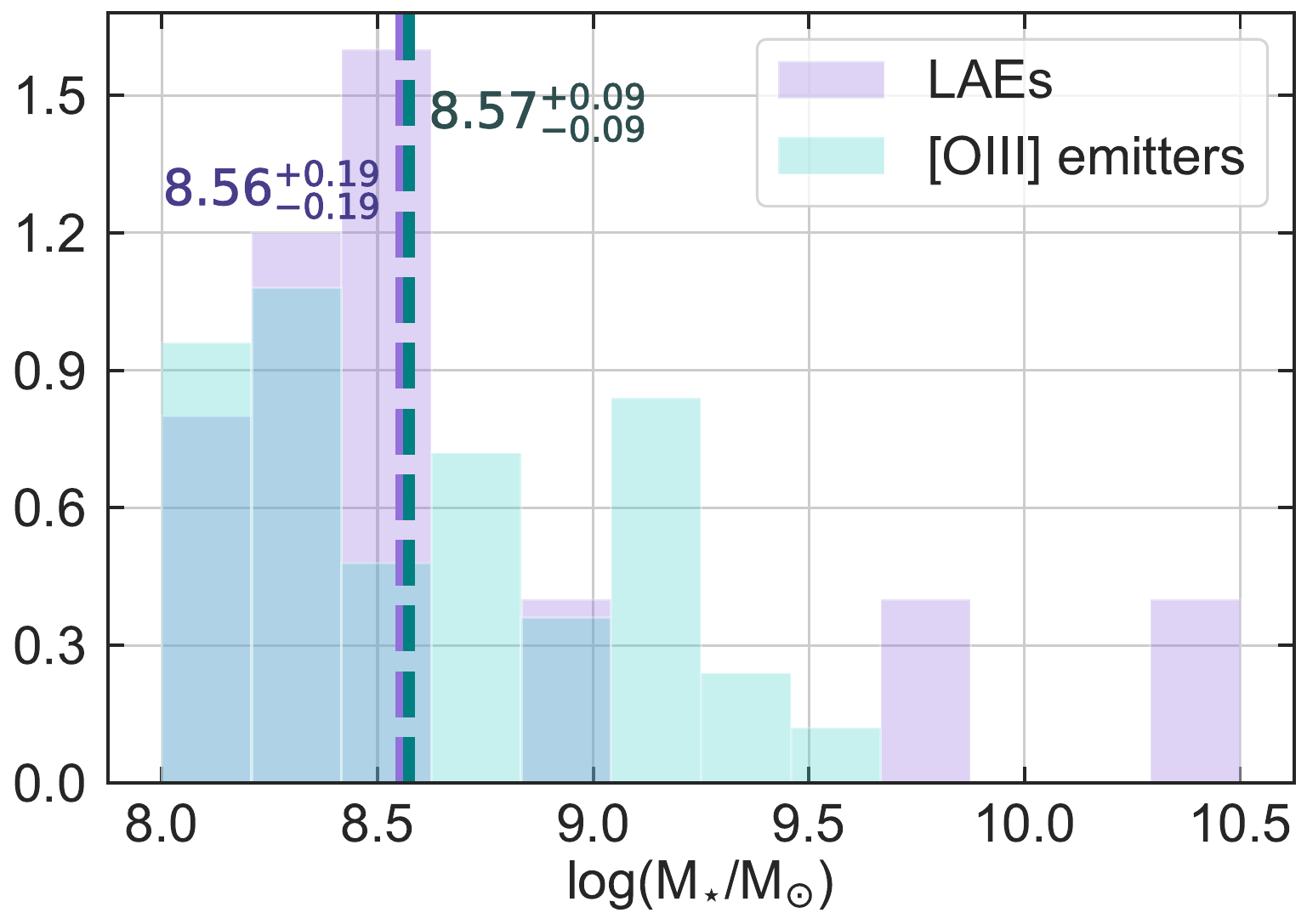}

         \includegraphics[width=1.\textwidth]{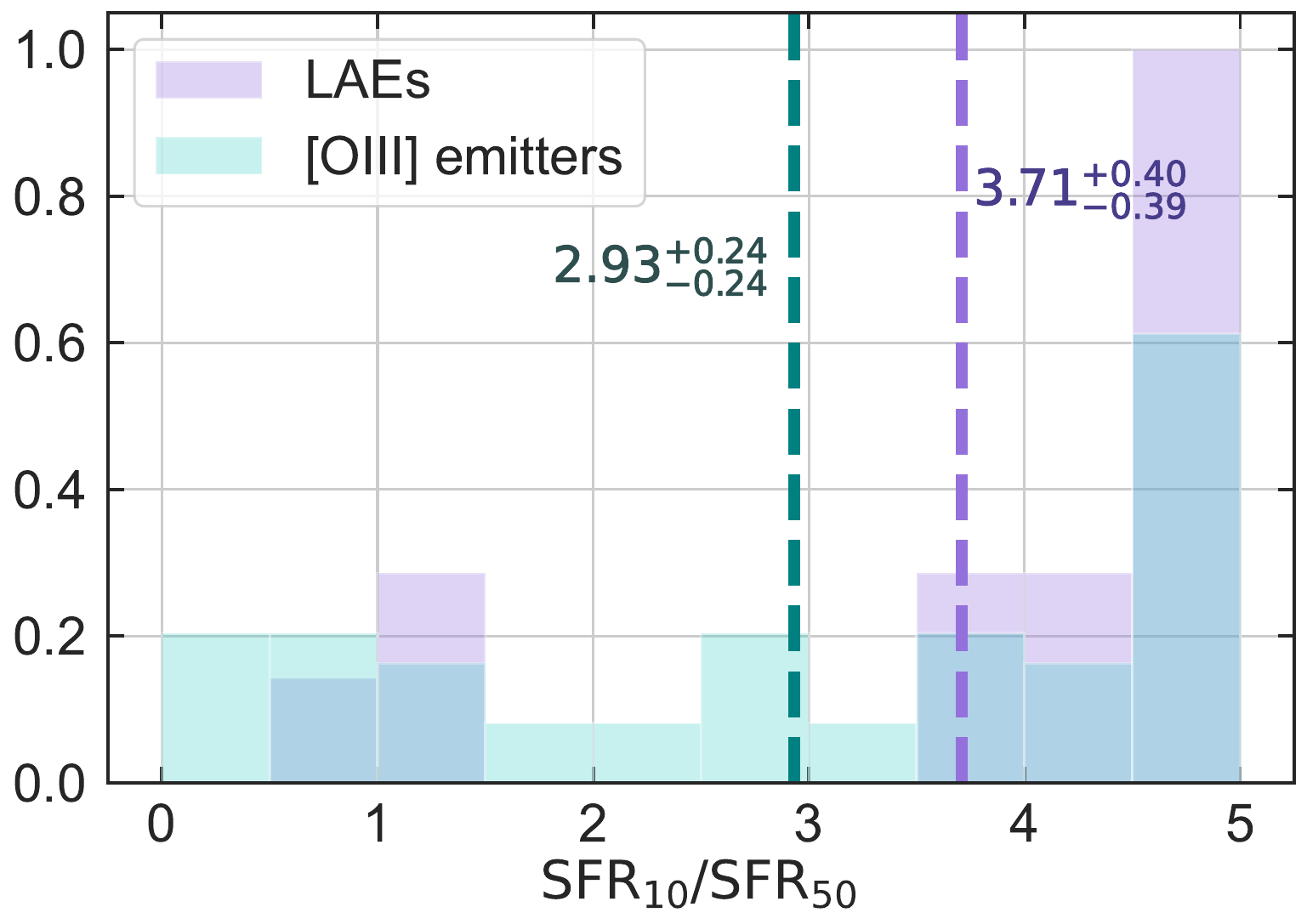}
         %\vspace{-0.8 cm}
     \end{subfigure}
    
        \caption{Properties of [OIII] emitters (green bars) and LAEs (purple bars): This comparison includes several key intrinsic properties, such as the [OIII] EW, $\mathrm{M_{UV}}$, $\beta$ slope, stellar mass ($\log_{\mathrm{M}_{\star}}$), log$\mathrm{sSFR_{10}}$, and SFR${10}$/SFR${50}$). The green and purple dashed lines represent the mean values for [OIII] and Lyman-$\alpha$ emitters, respectively. The numbers near each line indicate the mean value for each property, with error bars showing the 16th to 84th percentile range. }
    \label{fig:properties}
\end{figure*}

Figure \ref{fig:properties} presents these comparisons, with 14 LAEs shown as purple bars and 49 non-LAE [OIII] emitters as green bars. The mean values for each property are marked by dashed lines. To ensure a fair comparison, we only include sources with magnitudes brighter than $-19.5$.

The mean EW of [OIII] for LAEs is higher than that of non-LAE [OIII] emitters ($D=0.40$, $p=0.049$), which may indicate a more active star-forming environment or potentially stronger ionizing radiation associated with LAEs. Additionally, the mean star formation rate ratio (SFR$_{10}$/SFR$_{50}$) for LAEs is also slightly elevated compared to [OIII] emitters ($D=0.30$, $p=0.25$), suggesting that LAEs may be undergoing more recent or enhanced star formation over the considered timescales.

% Despite these slight differences, intrinsic properties such as the $\beta$ slope, UV magnitude ($\mathrm{M_{UV}}$), and $\log_{\mathrm{M}_{\star}}$
%  appear similar between the two populations. This may suggest that, although LAEs tend to show signs of higher star formation activity and potentially stronger ionizing radiation, these factors alone are unlikely to fully account for the presence of Lyman-$\alpha$ emission. 

 The UV slopes also shows difference (bluer $\beta$ for LAEs; $D=0.39$, $p=0.06$), suggestive of lower dust content among LAEs. The bluer $\beta$ slopes of LAEs point to lower dust content, consistent with the known connection between dust attenuation and Lyman-$\alpha$ escape \citep[e.g.][]{Atek2009, Hayes2011}. The lack of significant differences in $\mathrm{M_{UV}}$ ($D=0.24$, $p=0.52$) and $\log_{\mathrm{M}_{\star}}$ ($D=0.21$, $p=0.64$) indicates that luminosity and stellar mass alone are unlikely to fully account for the presence of Lyman-$\alpha$ emission.

%---------------------------------------

\section{Discussion}
\label{sec:discussion}
The study of LAEs at high redshifts (7 < z < 8) in the GOODS-N and GOODS-S fields has provided insights into their spatial distribution, intrinsic properties, and environmental contexts. We compare LAEs with [OIII] emitters that do not show Lyman-$\alpha$ emission to understand how environmental and intrinsic factors affect Lyman-$\alpha$ visibility \citep[e.g.][]{Tang2024, Witstok2024, Chen2025}. 

Our analysis indicates that LAEs and [OIII] emitters without detected Lyman-$\alpha$ emission exhibit similar mean overdensity parameters, suggesting that in GOODS-N and GOODS-S fields environmental factors do not determine whether a galaxy appears as an LAE. It is also important to note that for some [OIII] emitters, we did not detect the Lyman-$\alpha$ line in the Keck data, and some were not observed by Keck or JWST within the Lyman-$\alpha$ wavelength range, meaning we cannot definitively conclude that these [OIII] emitters are not LAEs. These sources are therefore treated as non-LAE [OIII] emitters in our analysis.

While our LAEs do not appear significantly more likely to reside in overdense regions compared to non-LAEs, this may partly reflect the adopted Lyman-$\alpha$ EW threshold used to define the sample. In this work, we include all sources with detected Lyman-$\alpha$ emission, spanning a wide range of rest-frame EWs from $\sim$25~\AA\ to $\sim$250~\AA. In contrast, \citet{Chen2025} focus on the most extreme emitters (EW~$>100$~\AA) and report that these LAEs are preferentially located in overdense regions, likely tracing large ionized bubbles. Indeed, several of our strongest emitters (e.g. GN\_27795, GN\_4394, GN\_25604, and JADES-13682) fall within this regime, but the inclusion of low- and moderate-EW systems likely makes any link between Lyman-$\alpha$ strength and environment less apparent. Adopting a higher EW threshold in our analysis could therefore reveal a stronger dependence of Lyman-$\alpha$ visibility on environment, since very high-EW emission is expected to occur predominantly in ionized regions with reduced IGM attenuation.

The intrinsic properties of LAEs and other [OIII] emitters provide additional insight into the conditions that influence Lyman-$\alpha$ visibility. Our comparison shows that the EW of [OIII] for LAEs is higher on average than for non-LAE [OIII] emitters, and the mean SFR$_{10}$/SFR$_{50}$ ratio is also slightly elevated, suggesting that LAEs may experience more recent or enhanced star formation activity. In contrast, $\mathrm{M_{UV}}$ ($D=0.24$, $p=0.52$) and $\log_{\mathrm{M}_{\star}}$ ($D=0.21$, $p=0.64$) show no significant differences between the two populations. The $\beta$ slopes are bluer for LAEs ($D=0.39$, $p=0.06$), consistent with lower dust content and enhanced Lyman-$\alpha$ escape. Overall, these results indicate that Lyman-$\alpha$ visibility is not governed by a single intrinsic property, but likely arises from a combination of factors including dust attenuation, ionization conditions, and local IGM transmission.

The spatial distribution of bright galaxies provides clues about large-scale structure during reionization. In some cases (e.g. GN\_z7\_0), we find significant clustering of luminous sources, consistent with bright galaxies tracing overdense regions \citep[e.g.][]{Endsley2024, Naidu2024, Helton2024}. However, our LAE comparison shows no strong correlation between Lyman-$\alpha$ visibility and overdensity in GOODS-N/S, indicating that the presence of LAEs is not defined by environment alone. This may partly reflect the range of Lyman-$\alpha$ EWs included in our sample, as studies focusing on the strongest emitters (EW~$>100$~\AA) have found a clearer association with overdense regions \citep[e.g.][]{Chen2025}. Bright galaxies may contribute to the formation of larger ionized regions \citep[e.g.][]{Tang2024, Witstok2024}, but Lyman-$\alpha$ visibility likely also depends on more local conditions within and around individual galaxies. Thus, bright-galaxy clustering traces the underlying structure, while LAE visibility reflects additional small-scale effects.

Possible galaxy mergers also appear to provide a hint that they may be an important factor affecting Lyman-$\alpha$ visibility. Several close galaxy pairs, including GN\_29192 and GN\_29193, GS\_6644 and GS\_6645, GS\_12667 and GS\_12669, GS$\_$23161 and GS$\_$23162, suggest the possibility of ongoing or recent interactions. While some of these pairs are detected as LAEs, others are not confirmed as such, though they could potentially be LAEs. The presence of LAEs in these pairs likely contributes to their detection, but even non-LAEs may still play a role. These non-LAE galaxies might ionize the region around them, creating an environment that allows Lyman-$\alpha$ photons from neighbouring galaxies to pass through the ionized medium, depending on the line of sight. This highlights the possibility that mergers could enhance Lyman-$\alpha$ emission through triggered star formation or AGN activity, affecting the surrounding ionization and contributing to the observed properties of LAEs in dynamically active environments.

The contribution of AGN to Lyman-$\alpha$ visibility is another aspect that requires further exploration. In the group \textbf{GN\_z7\_2}, which includes the AGN GN\_5688 \citep{Fujimoto2022}, we identify a potential AGN, GN\_14101, that could provide additional ionizing radiation, contributing to local reionization. The presence of potential AGNs in these regions raises important questions about their role in ionized bubble formation and their potential contribution to facilitating Lyman-$\alpha$ escape at high redshift. It is also worth noting that GN\_26844, which is an LAE, has a companion LAE, GN\_4394, with a very large equivalent width of 221~\AA. The high EW and stellar mass ($\log_{\mathrm{M}_{\star}}\sim9$) of GN\_4394 suggest that it could also potentially host an AGN, adding complexity to the dynamics of the \textbf{GN\_z7\_2} group.

Finally, Lyman-$\alpha$ visibility may depend not only on the physical properties of galaxies and their environments but also on observational effects such as viewing angles and the geometry of ionized regions \citep[e.g.][]{Rosdahl2018, Katz2020, Garel2021}. The lack of a clear distinction in overdensity measurements between LAEs and non-LAE [OIII] emitters suggests that the detectability of Lyman-$\alpha$ emission may be influenced by the orientation of ionized bubbles relative to the observer. If Lyman-$\alpha$ photons are more likely to escape along certain sightlines, this could explain why some galaxies in dense environments do not appear as LAEs despite having similar intrinsic properties.

Overall, our findings highlight the complexity of Lyman-$\alpha$ visibility at $7<z<8$. The combination of environmental and intrinsic factors, together with observational effects such as viewing angles, shapes the conditions under which Lyman-$\alpha$ photons can escape and be detected. While bright galaxies are often found in clustered regions that may contribute to large-scale reionization \citep[e.g.][]{Castellano2018, Endsley2022, Stark2017, Leonova2022, Larson2022, Tang2024, Witstok2024}, our results indicate that Lyman-$\alpha$ visibility is not solely determined by local overdensity. Instead, other factors, including mergers, AGN activity, or ionized bubble geometry — could influence Lyman-$\alpha$ visibility. In addition, ultra-faint galaxies, undetected by the FRESCO survey, may contribute to reionization by producing small ionized regions that allow Lyman-$\alpha$ photons to escape.

\section{Summary}
\label{sec:summary}
In this study, we investigated the spatial distribution and intrinsic properties of LAEs and [OIII] emitters at redshifts $7 < z < 8$ using FRESCO NIRCam/grism data from the GOODS-N and GOODS-S fields. Our goal was to determine the relative contributions of environmental and intrinsic factors in regulating Lyman-$\alpha$ visibility and to assess the role of these galaxies in the epoch of reionization. Our main results are:
\begin{itemize}\itemsep1em%[label=\arabic*)]

   \item[$-$] Using Keck archival data we discover 8 new LAEs in the GOODS-N field, expanding the known sample of high-redshift LAEs. 
    
    \item[$-$] In \ref{sec:groups}, we investigate the environment of LAEs at $7<z<8$ by categorizing galaxies into four groups in GOODS-N and three in GOODS-S, each centered around the brightest LAE. Our results show that GOODS-N contains significant clustering of galaxies in the GN\_z7\_0 group, and potential mergers, such as in GN\_z7\_5. The GN\_z7\_6 group is also notable for containing four LAEs. In GOODS-S, the environment is marked by fewer bright [OIII] emitters, with notable objects like the faint LAE JADES-13682 in GS\_z7\_2, and a potential merger in GS\_z7\_7. Additionally, the role of AGNs in Lyman-$\alpha$ visibility is hinted by the presence of AGN GN\_5688 and potential AGNs in the GN\_z7\_2 group, including GN\_14101 and GN\_4394.

    \item[$-$] In \ref{sec:overd} we analyze the overdensity properties of LAEs compared to [OIII] emitters without detected Lyman-$\alpha$ emission by calculating the overdensity parameter using the UV luminosity function. We explore whether LAEs' presence in overdense regions, where ionized bubbles enhance photon transmission, primarily explains their Lyman-$\alpha$ visibility. We find that the mean overdensity parameters for both groups are similar, suggesting that while LAEs are located in overdense regions, overdensity alone does not determine Lyman-$\alpha$ emission, indicating that other factors are likely involved in shaping Lyman-$\alpha$ escape.

   \item[$-$] In \ref{sec:prop} we examine physical properties of LAEs and non-LAE [OIII] emitters, such as UV magnitudes, $\beta$ slopes, stellar masses, and EW$_{\mathrm{[OIII]}}$. This comparison allows us to assess whether factors such as dust attenuation, ionizing photon production, or galaxy assembly history influence the emergence of Lyman-$\alpha$ emission. The slightly higher mean EW$_{\mathrm{[OIII]}}$ and SFR$_{10}$/SFR$_{50}$ for LAEs could suggest more intense star formation and stronger ionizing radiation compared to non-LAE [OIII] emitters. However, in general all intrinsic properties, including the $\beta$ slope, UV magnitude, and stellar mass are similar between LAEs and non-LAEs. This suggests that while LAEs may exhibit higher active star formation and stronger ionization, these factors alone do not fully explain the presence of Lyman-$\alpha$ emission.

    \item[$-$] Our analysis shows that LAEs in GOODS–N and GOODS–S are not preferentially located in overdense regions, suggesting that large-scale environment alone does not determine Lyman-$\alpha$ visibility. However, we note a few cases with nearby companions, which may indicate small-scale interactions (e.g. mergers) that facilitate Lyman-$\alpha$ escape through enhanced star formation or AGN-driven ionization.

\end{itemize}

Our understanding of Lyman-$\alpha$ visibility at $7<z<8$ will improve with more complete galaxy samples that include fainter sources and a confirmed classification of [OIII] emitters as LAEs. Future spectroscopic surveys in the FRESCO field (such as GO-9214; PIs: Mason \& Stark), with broader wavelength coverage, will be essential for making these determinations. Identifying which [OIII] emitters are LAEs, along with detecting fainter sources, will provide deeper insight into the factors governing Lyman-$\alpha$ transmission, such as local ionization conditions, viewing angles, large-scale environment, and the role of low-luminosity galaxies.\\\\

%%%%%%%%%%%%%%%%%%%%%%%%%%%%%%%%%%%%%%%%%%%%%%%%%%
\section*{Acknowledgements}
%%%%%%%%%%%%%%%%%%%%%%%%%%%%%%%%%%%%%%%%%%%%%%%%%%

The work presented in this paper is based on observations made with the NASA/ESA/CSA James Webb Space Telescope. The data were obtained from the Mikulski Archive for Space Telescopes at the Space Telescope Science Institute, which is operated by the Association of Universities for Research in Astronomy, Inc., under NASA contract NAS 5-03127 for JWST. These observations are associated with program \#1895.

This research has made use of the Keck Observatory Archive (KOA), which is operated by the W. M. Keck Observatory and the NASA Exoplanet Science Institute (NExScI), under contract with the National Aeronautics and Space Administration.

EL is grateful for financial support from the Nedelandse Organisatie voor Wetenschappelijk Onderzoek (NWO) through the NWO/Groot grant  "Gravitational waves: The new cosmic messengers".
This work has received funding from the Swiss State Secretariat for Education, Research and Innovation (SERI) under contract number MB22.00072, as well as from the Swiss National Science Foundation (SNSF) through project grant 200020\_207349.
The Cosmic Dawn Center (DAWN) is funded by the Danish National Research Foundation under grant DNRF140.

Telescope facilities: JWST (NIRCam, NIRSpec), Keck (MOSFIRE)

% Several publicly available software tools have facilitated this work. We extend our thanks to the authors of the following softwares:
% {\tt IPython} \citep{Perez2007},
% {\tt jupyter} \citep{jupyter},
% {\tt astropy} \citep{Astropy2013,Astropy2018},
% {\tt matplotlib} \citep{Hunter2007},
% {\tt numpy} \citep{numpy},
% {\tt photutils} \citep{Bradley2022},
% {\tt scipy} \citep{Virtanen2020},
% {\tt EAZY} \citep{Brammer2008},
% {\tt BAGPIPES} \citep{Carnall2018},
% {\tt GalfitM} \citep{Haussler2013,Vika2015},
% {\tt grizli} \citep{Brammer2018},
% {\tt SExtractor} \citep{Bertin1996},
% {\tt pypher} \citep{Boucaud2016},
% {\tt extinction} \citep{Fitzpatrick2007}, 
% {\tt GLACiAR2} \citep{Carrasco2018,Leethochawalit2022}.

%%%%%%%%%%%%%%%%%%%%%%%%%%%%%%%%%%%%%%%%%%%%%%%%%%
\section*{Data Availability}
%%%%%%%%%%%%%%%%%%%%%%%%%%%%%%%%%%%%%%%%%%%%%%%%%%

The JWST raw data products used in this work are available via the Mikulski Archive for Space Telescopes (\url{https://mast.stsci.edu}). 
The [OIII] emitter catalogs from \citet{Meyer2024} are available at \url{https://github.com/rameyer/fresco/}.
The Keck spectra can be retrieved via \url{https://grizli-cutout.herokuapp.com/}.
Additional data presented in this work will be made available by the authors upon request.

%-----------------------%-----------------------

%-----------------------%-----------------------%-----------------------%-----------------------

%---------------------------------------

%---------------------------------------

%---------------------------------------
\fboxsep=10mm%padding thickness
\fboxrule=4pt%border thickness

%---------------------------------------

%---------------------------------------

%---------------------------------------

%---------------------------------------

%%%%%%%%%%%%%%%%%%%% REFERENCES %%%%%%%%%%%%%%%%%%

\bibliographystyle{mnras}
\bibliography{GOODSBubblePaper} 

%%%%%%%%%%%%%%%%%%%%%%%%%%%%%%%%%%%%%%%%%%%%%%%%%%

%%%%%%%%%%%%%%%%% APPENDICES %%%%%%%%%%%%%%%%%%%%%
 
\appendix

%---------------------------------------
\section{Appendix}

\captionsetup{justification=centering}

\begin{table*}
%\centering
\begin{tabular}{cccccccccc}

\hline\hline

  ID & RA & DEC & z$_\mathrm{Lya}$ \href{#refc}{\textsuperscript{c}}& z$_\mathrm{[OIII]}$\href{#refd}{\textsuperscript{d}}  & M$_\mathrm{{UV}}$ & $\log_{\mathrm{M}_{\star}}$

  & EW$_\mathrm{{[OIII]}}$[\AA] \href{#refe}{\textsuperscript{e}}  \\\hline\hline\\[-5pt]

\multicolumn{9}{c}{\textbf{GN\_z7\_0 group}}\\[5pt]
27795\href{#refa}{\textsuperscript{a}}  & 189.13498 & 62.29190 & 7.038 & 7.017 & -19.6 $^{+0.1}_{-0.1}$ & 7.59 $^{+0.15}_{-0.08}$ & 1263 $^{+125}_{-121}$ \\
26051\href{#refb}{\textsuperscript{b}} \href{#ref1}{\textsuperscript{1}} & 189.17981 & 62.28239 & 7.087 & 7.084 & -19.5 $^{+0.1}_{-0.1}$ & 7.60 $^{+0.24}_{-0.07}$ & 2371 $^{+222}_{-260}$ \\
22679  \href{#refb}{\textsuperscript{b}} \href{#ref1}{\textsuperscript{1}} & 189.20377 & 62.26843 & 7.09 & 7.090 & -19.4 $^{+0.1}_{-0.1}$ & 7.51 $^{+0.12}_{-0.04}$ & 1839 $^{+155}_{-222}$ &\\
\hline\hline\\
24082 & 189.07370 & 62.22345 & - & 7.035 & -20.5 $^{+0.0}_{-0.1}$ & 8.18 $^{+0.24}_{-0.06}$ & 298 $^{+45}_{-34}$ \\
17367 & 189.20531 & 62.25077 & - & 6.993 & -20.4 $^{+0.1}_{-0.0}$ & 8.57 $^{+0.10}_{-0.20}$ & 820 $^{+89}_{-66}$ \\
14967 & 189.01924 & 62.24353 & - & 7.038 & -20.2 $^{+0.0}_{-0.1}$ & 9.25 $^{+0.15}_{-0.28}$ & 563 $^{+97}_{-40}$ \\
30464 & 189.26217 & 62.31565 & - & 7.003 & -20.0 $^{+0.1}_{-0.1}$ & 9.26 $^{+0.37}_{-0.99}$ & 373 $^{+599}_{-73}$ \\
5412 & 189.16078 & 62.28308 & - & 7.045 & -20.0 $^{+0.1}_{-0.0}$ & 8.28 $^{+0.22}_{-0.37}$ & 289 $^{+54}_{-48}$ \\
27094 & 189.18337 & 62.28774 & - & 7.022 & -19.9 $^{+0.1}_{-0.1}$ & 8.50 $^{+0.39}_{-0.48}$ & 398 $^{+79}_{79}$ \\
30164 & 189.24393 & 62.31213 & - & 7.072 & -19.8 $^{+0.1}_{-0.1}$ & 7.98 $^{+0.37}_{-0.18}$ & 715 $^{+134}_{-130}$ \\
30668 & 189.18736 & 62.31884 & - & 6.998 & -19.9 $^{+0.1}_{-0.1}$ & 8.05 $^{+0.46}_{-0.25}$ & 396 $^{+115}_{-94}$ \\
16993 & 189.19970 & 62.24981 & - & 6.992 & -19.8 $^{+0.1}_{-0.0}$ & 8.14 $^{+0.28}_{-0.26}$ & 393 $^{+64}_{-57}$ \\
30187 & 189.21231 & 62.31240 & - & 7.035 & -19.8 $^{+0.1}_{-0.1}$ & 7.93 $^{+0.38}_{-0.19}$ & 440 $^{+112}_{-78}$ \\
169 & 189.23573 & 62.32835 & - & 7.001 & -19.5 $^{+0.1}_{-0.1}$ & 8.15 $^{+0.41}_{-0.61}$ & 492 $^{+157}_{-133}$ \\
26059 & 189.19570 & 62.28243 & - & 7.091 & -19.4 $^{+0.1}_{-0.1}$ & 8.71 $^{+0.51}_{-0.65}$ & 356 $^{+116}_{-57}$ \\
21799 & 189.13759 & 62.26521 & - & 7.040 & -19.4 $^{+0.1}_{-0.1}$ & 8.64 $^{+0.38}_{-0.47}$ & 363 $^{+60}_{60}$ \\
20218 & 189.05374 & 62.25970 & - & 7.084 & -19.3 $^{+0.1}_{-0.1}$ & 7.57 $^{+0.18}_{-0.11}$ & 1626 $^{+185}_{-247}$ \\
5365 & 189.31999 & 62.28323 & - & 7.074 & -19.2 $^{+0.1}_{-0.1}$ & 7.87 $^{+0.62}_{-0.44}$ & 780 $^{+360}_{-206}$ \\
28812 & 189.20960 & 62.29928 & - & 7.026 & -18.8 $^{+0.2}_{-0.1}$ & 8.20 $^{+0.29}_{-0.35}$ & 373 $^{+84}_{84}$ \\
10162 & 189.26546 & 62.26544 & - & 6.983 & -19.1 $^{+0.2}_{-0.1}$ & 8.44 $^{+0.50}_{-0.61}$ & 1702 $^{+764}_{-325}$ \\
7409 & 189.19492 & 62.21700 & - & 7.019 & -19.0 $^{+0.0}_{-0.1}$ & 7.70 $^{+0.14}_{-0.30}$ & 1259 $^{+161}_{-116}$ \\
25360 & 189.31580 & 62.27925 & - & 6.943 & -19.0 $^{+0.3}_{-0.2}$ & 8.54 $^{+0.35}_{-0.38}$ & 520 $^{+87}_{87}$ \\
7092 & 189.07495 & 62.21567 & - & 7.061 & -19.0 $^{+0.2}_{-0.2}$ & 7.60 $^{+0.24}_{-0.09}$ & 989 $^{+183}_{-253}$ \\
9825 & 189.22239 & 62.26662 & - & 6.948 & -19.0 $^{+0.1}_{-0.1}$ & 7.57 $^{+0.14}_{-0.20}$ & 1086 $^{+274}_{-203}$ \\
18770 & 189.04297 & 62.25504 & - & 7.096 & -18.9 $^{+0.3}_{-0.2}$ & 8.35 $^{+0.43}_{-0.67}$ & 447 $^{+236}_{-113}$ \\
30222 & 189.19487 & 62.31272 & - & 7.057 & -18.7 $^{+0.2}_{-0.2}$ & 8.51 $^{+0.35}_{-0.32}$ & 571 $^{+126}_{126}$ \\
27069 & 189.16280 & 62.28755 & - & 7.013 & -18.6 $^{+0.3}_{-0.2}$ & 8.91 $^{+0.17}_{-0.23}$ & 247 $^{+53}_{53}$ \\
28523 & 189.19861 & 62.29704 & - & 7.032 & -18.4 $^{+0.3}_{-0.2}$ & 8.95 $^{+0.18}_{-0.19}$ & 358 $^{+77}_{-48}$ \\
28524 & 189.19861 & 62.29704 & - & 7.042 & -19.6 $^{+0.2}_{-0.1}$ & 9.27 $^{+0.28}_{-0.45}$ & 272 $^{+42}_{42}$ \\
28378 & 189.07686 & 62.20544 & - & 6.998 & -17.7 $^{+0.5}_{-0.5}$ & 8.29 $^{+0.46}_{-0.72}$ & 984 $^{+515}_{-258}$ \\

\hline\hline
\end{tabular}
\caption{Summary of the GN$\_$z7$\_$0 group.}\label{table1}

\label{tab:1f}
\vspace{1ex}
\footnotesize{\raggedright Notes:  \textsuperscript{a,b} LAEs ,
\textsuperscript{c} spectroscopic redshift from Lyman-$\alpha$ emission line, 
\textsuperscript{d} systemic redshift, 
\textsuperscript{e} [OIII] equivalent width (Naidu et al. in preparation)}
\footnotesize{\raggedright\par}
\footnotesize{\raggedright References:  {\textsuperscript{1}} \cite{Tang2024} \par}
\end{table*}

%-----------------------

\begin{figure*}
\centering
\begin{subfigure}[t]{1\textwidth}
\centering
\includegraphics[width=0.46\textwidth]{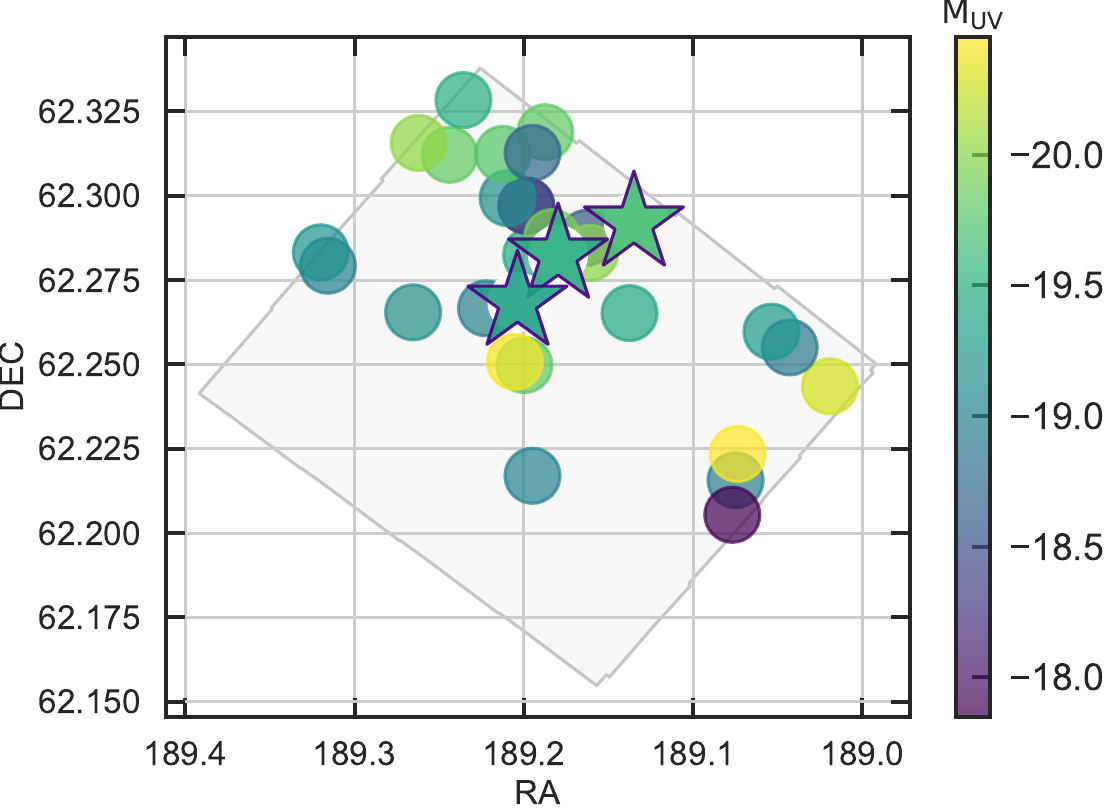}
\includegraphics[width=0.42\textwidth,height=0.42\textwidth]{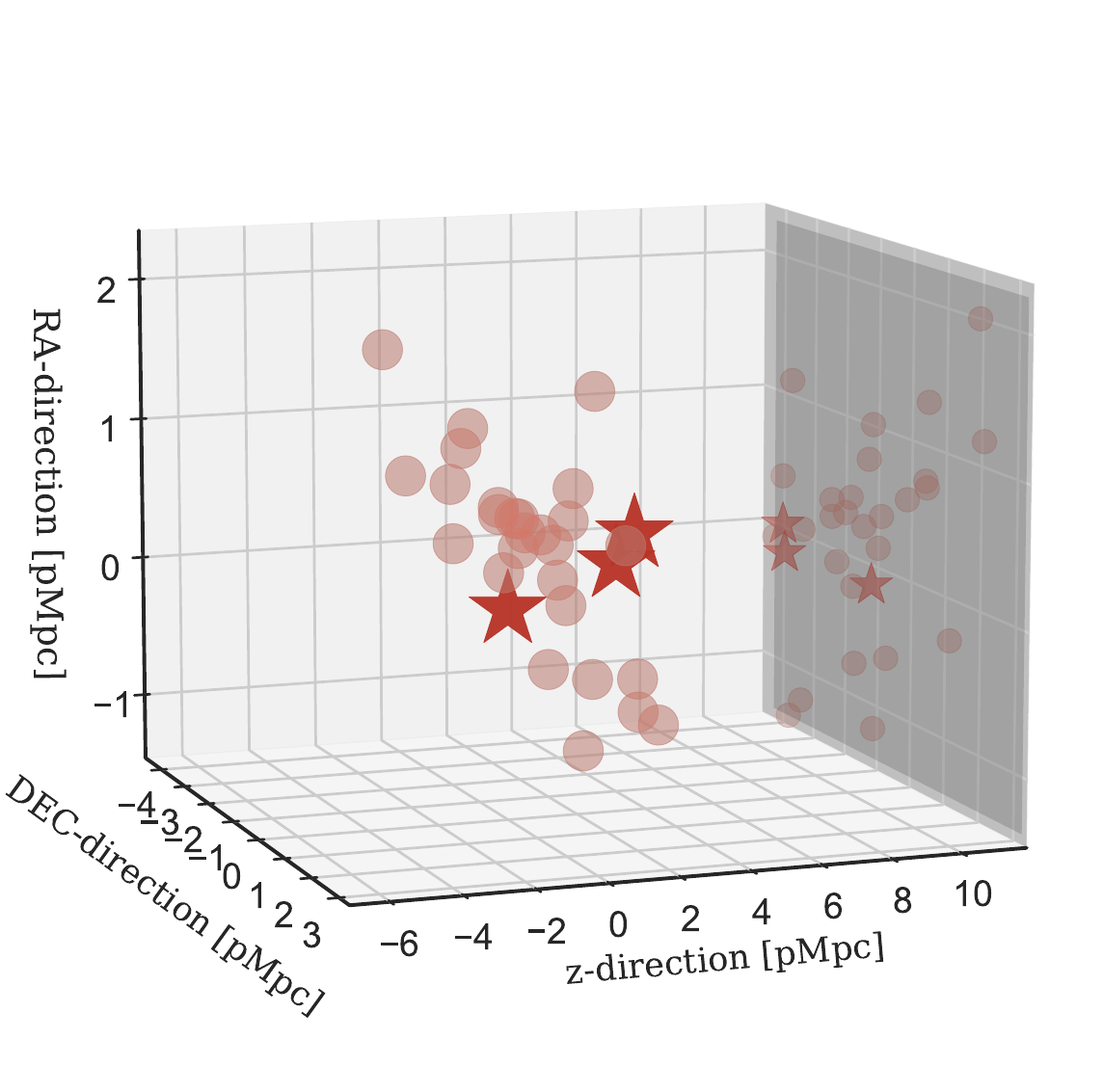}
\end{subfigure}
\end{figure*}
\begin{figure*}
\ContinuedFloat
\begin{subfigure}[t]{0.47\textwidth}
\centering
\includegraphics[width=1.0\textwidth]{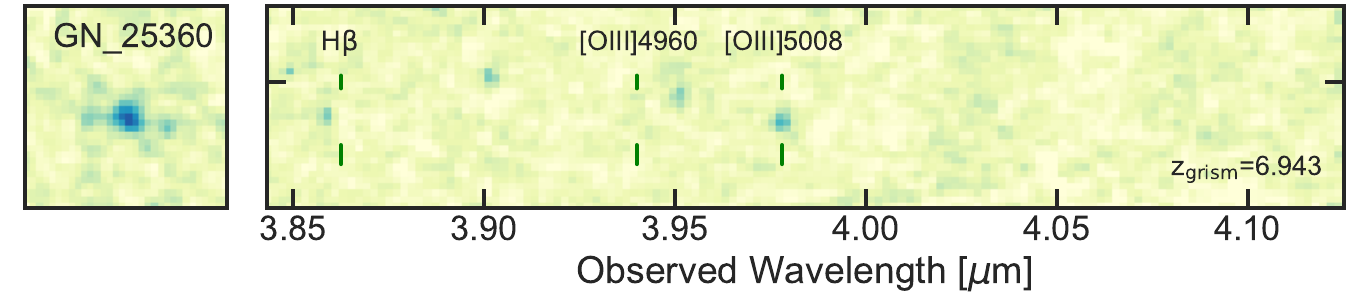}
\vspace{-0.49 cm}
\vspace*{-0.49 cm}
\includegraphics[width=1.0\textwidth]{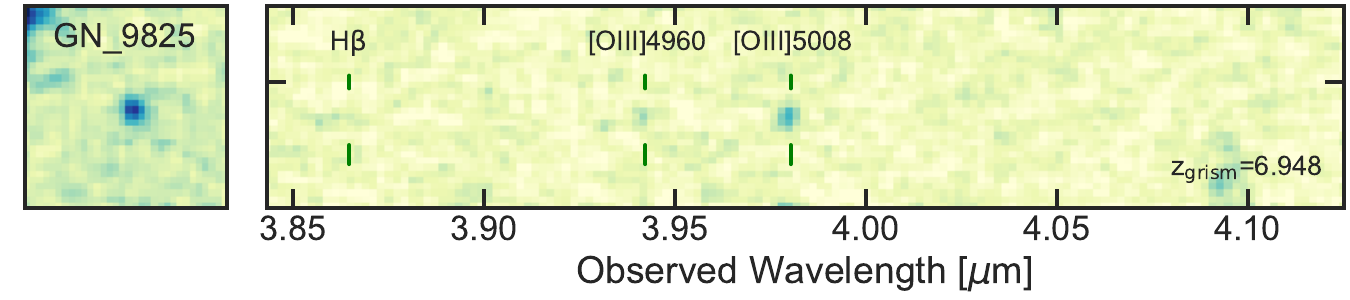}
\vspace{-0.49 cm}
\vspace*{-0.49 cm}
\includegraphics[width=1.0\textwidth]{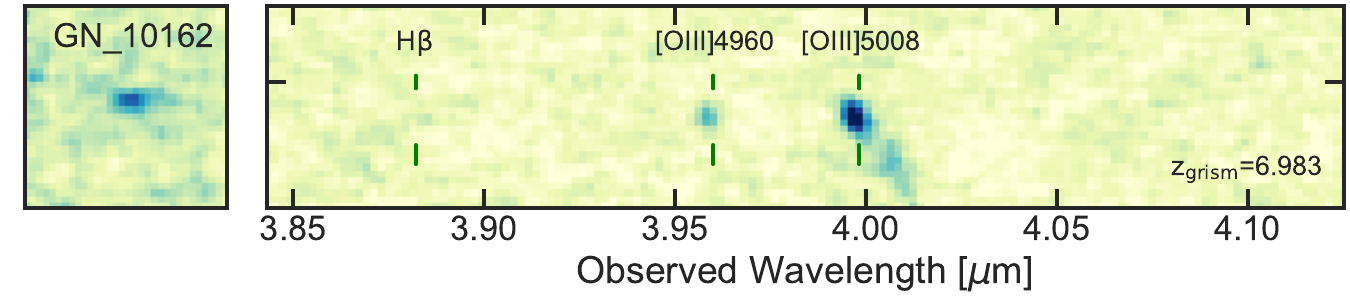}
\vspace{-0.49 cm}
\vspace*{-0.49 cm}
\includegraphics[width=1.0\textwidth]{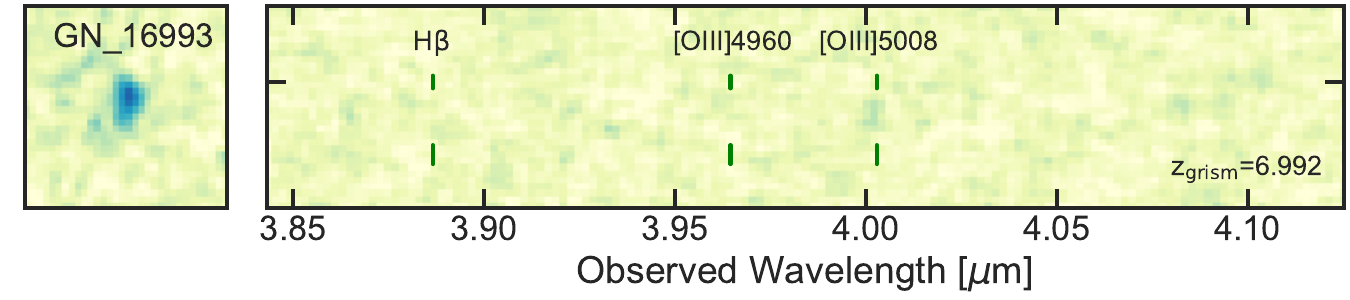}
\vspace{-0.49 cm}
\vspace*{-0.49 cm}
\includegraphics[width=1.0\textwidth]{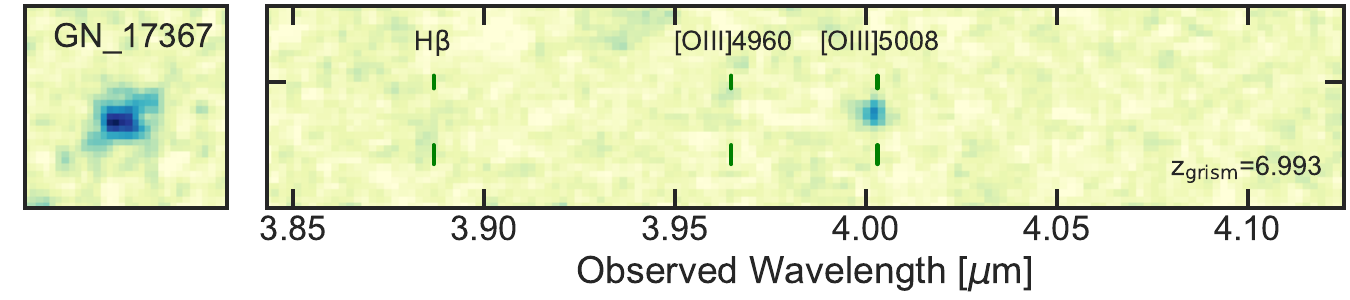}
\vspace{-0.49 cm}
\vspace*{-0.49 cm}
\includegraphics[width=1.0\textwidth]{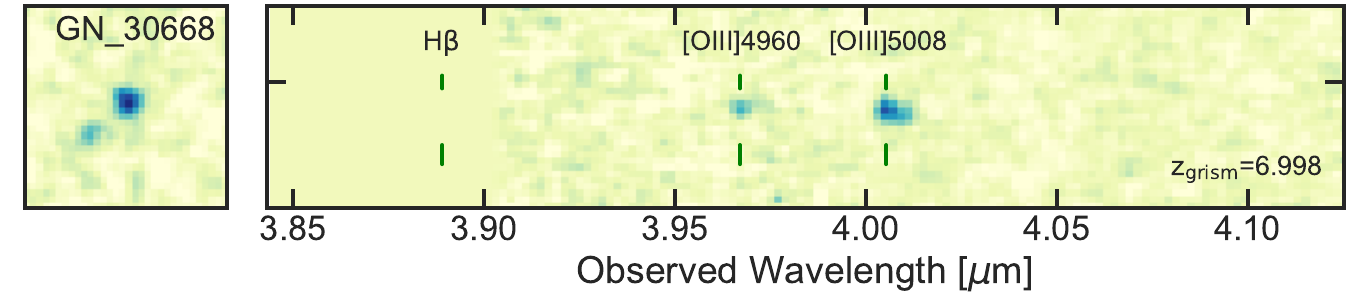}
\vspace{-0.49 cm}
\vspace*{-0.49 cm}
\includegraphics[width=1.0\textwidth]{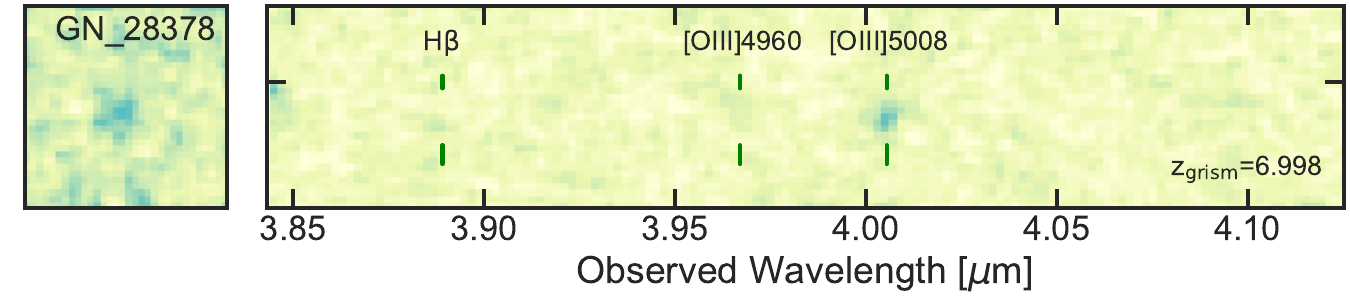}
\vspace{-0.49 cm}
\vspace*{-0.49 cm}
\includegraphics[width=1.0\textwidth]{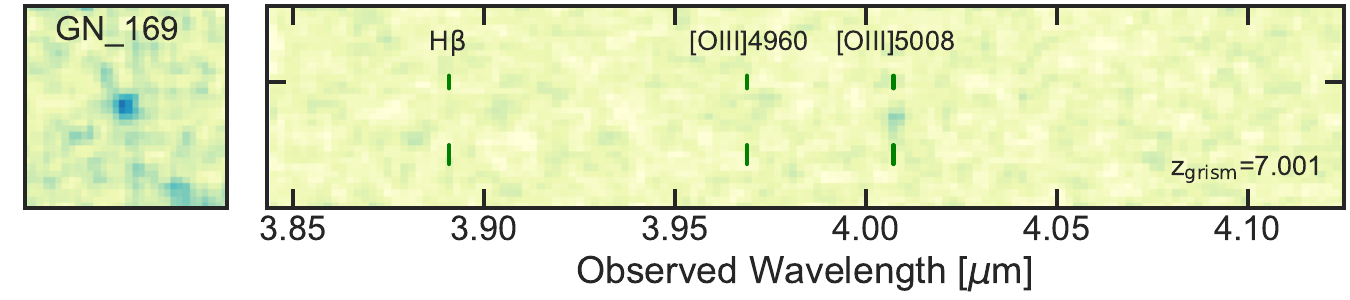}
\vspace{-0.49 cm}
\vspace*{-0.49 cm}
\includegraphics[width=1.0\textwidth]{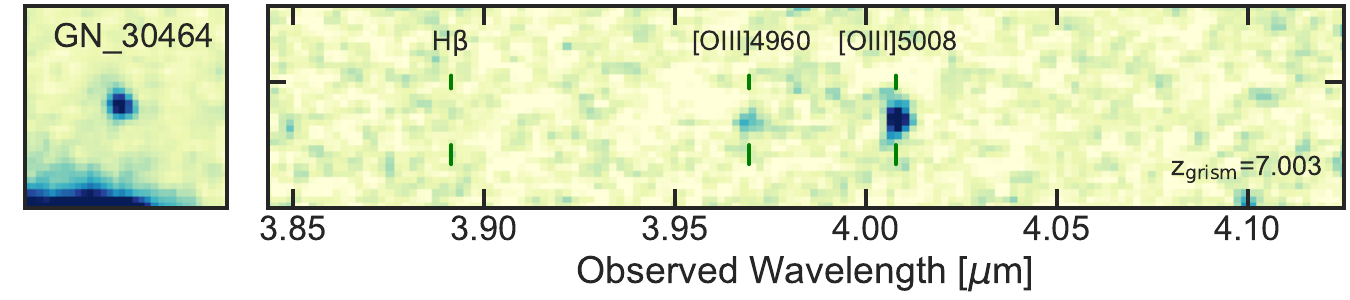}
\vspace{-0.49 cm}
\vspace*{-0.49 cm}
\includegraphics[width=1.0\textwidth]{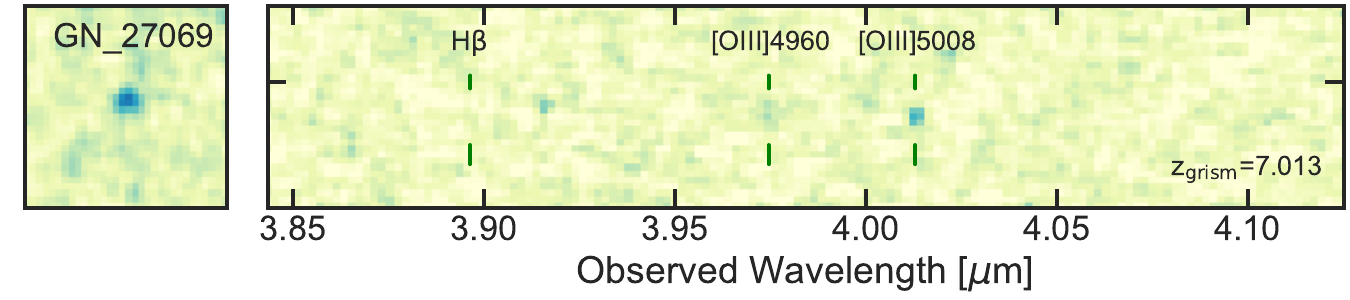}
\vspace{-0.49 cm}
\vspace*{-0.49 cm}
\includegraphics[width=1.0\textwidth]{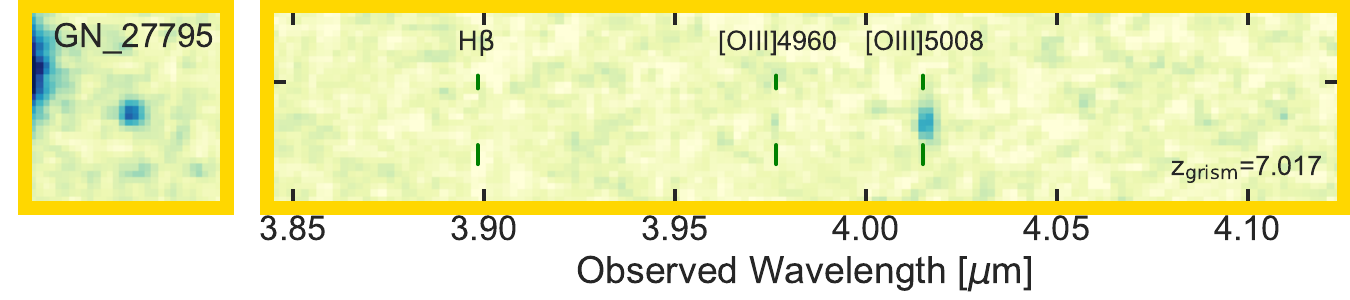}
\vspace{-0.49 cm}
\vspace*{-0.49 cm}
\includegraphics[width=1.0\textwidth]{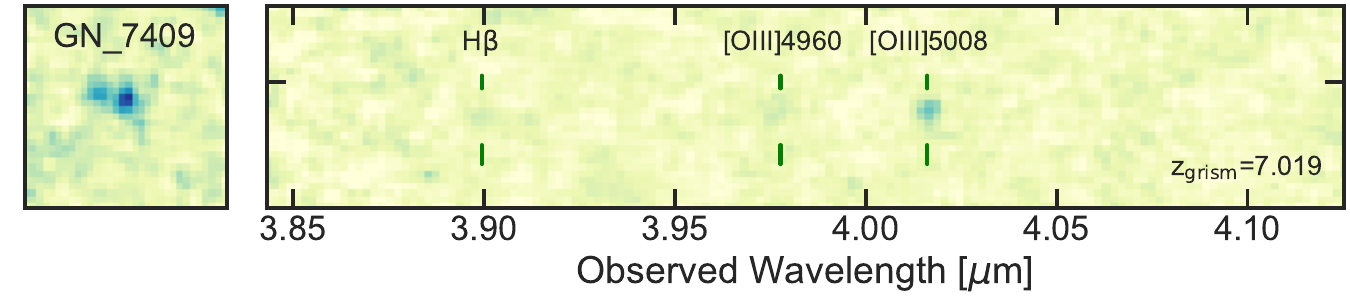}
\vspace{-0.49 cm}
\vspace*{-0.49 cm}
\includegraphics[width=1.0\textwidth]{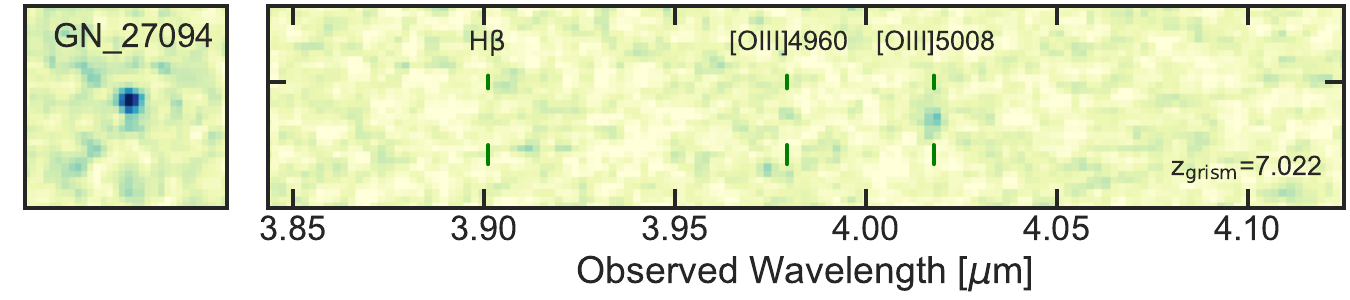}
\vspace{-0.49 cm}
\vspace*{-0.49 cm}
\includegraphics[width=1.0\textwidth]{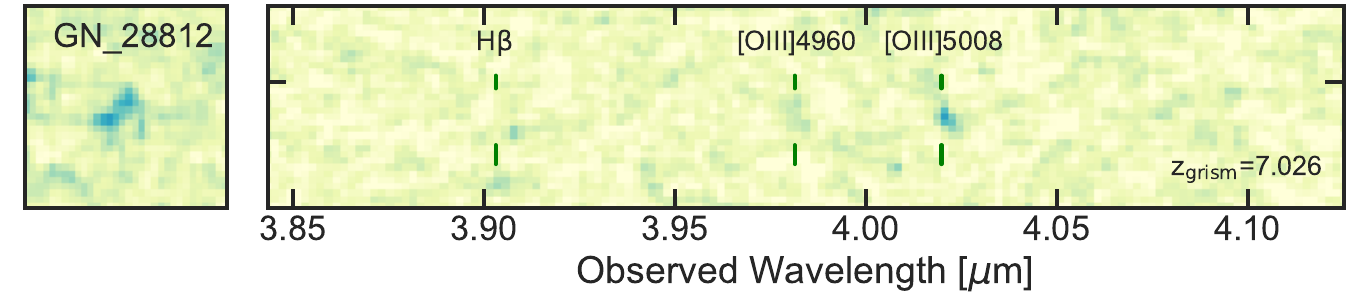}
\vspace{-0.49 cm}
\vspace*{-0.49 cm}
\includegraphics[width=1.0\textwidth]{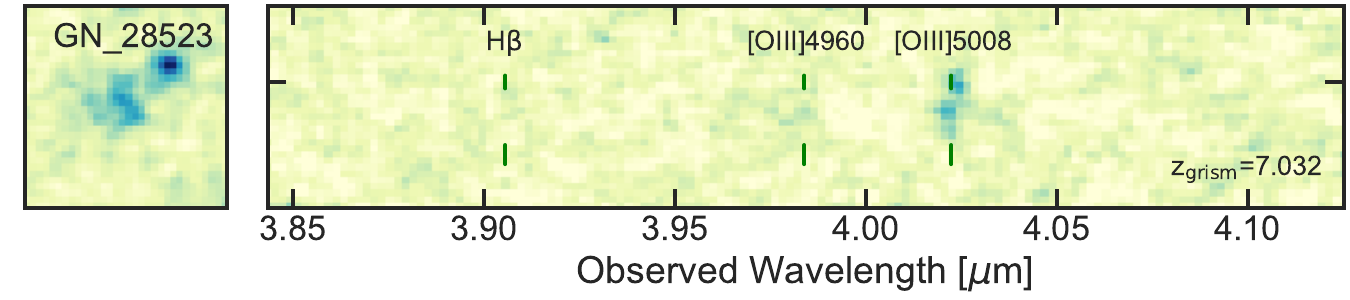}

\end{subfigure}
\hspace{0.05\textwidth}
\begin{subfigure}[t]{0.47\textwidth}
\centering
\includegraphics[width=1.0\textwidth]{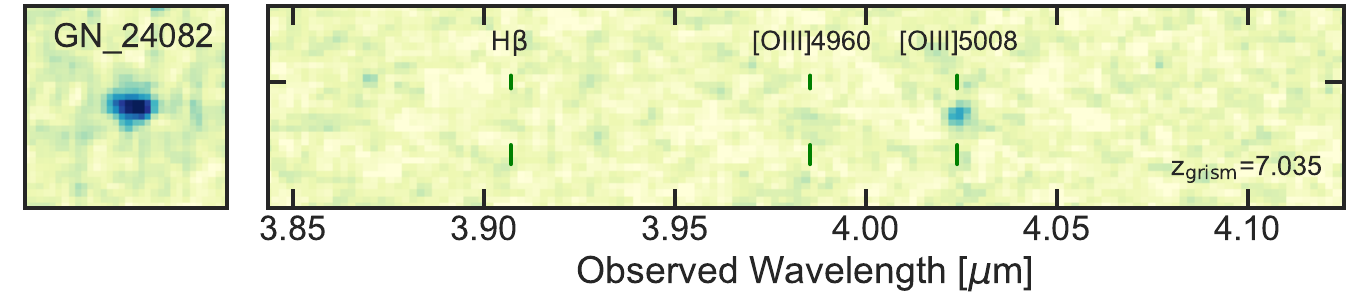}
\vspace{-0.49 cm}
\vspace*{-0.49 cm}
\includegraphics[width=1.0\textwidth]{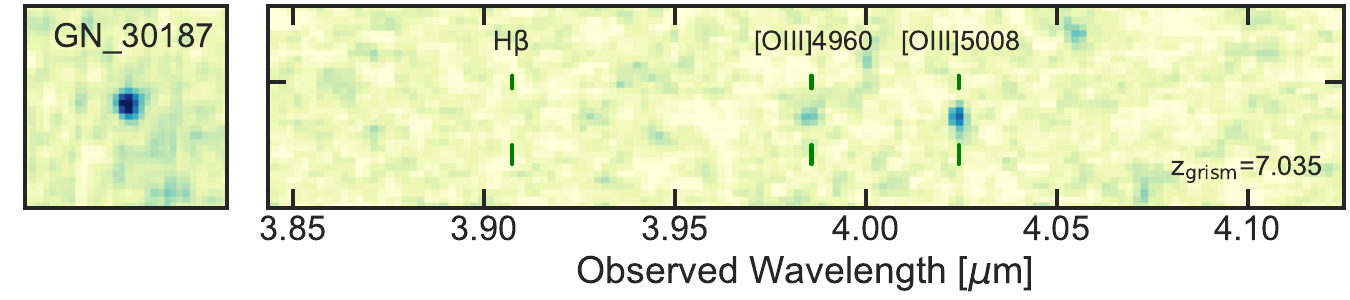}
\vspace{-0.49 cm}
\vspace*{-0.49 cm}
\includegraphics[width=1.0\textwidth]{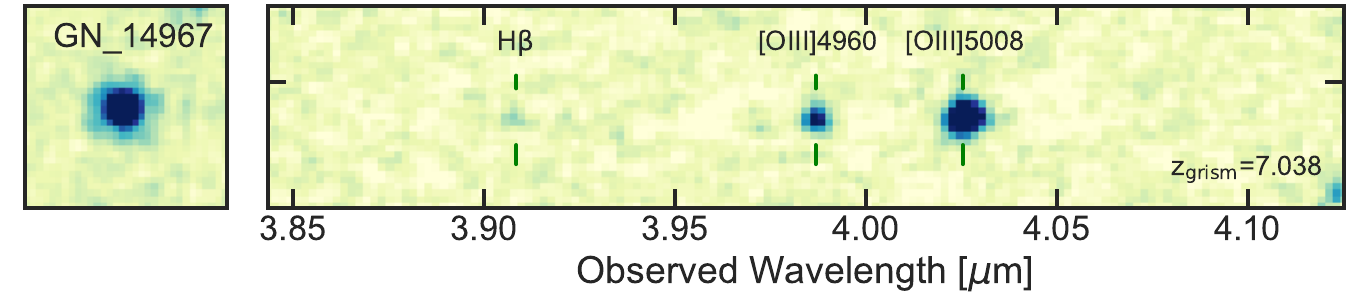}
\vspace{-0.49 cm}
\vspace*{-0.49 cm}
\includegraphics[width=1.0\textwidth]{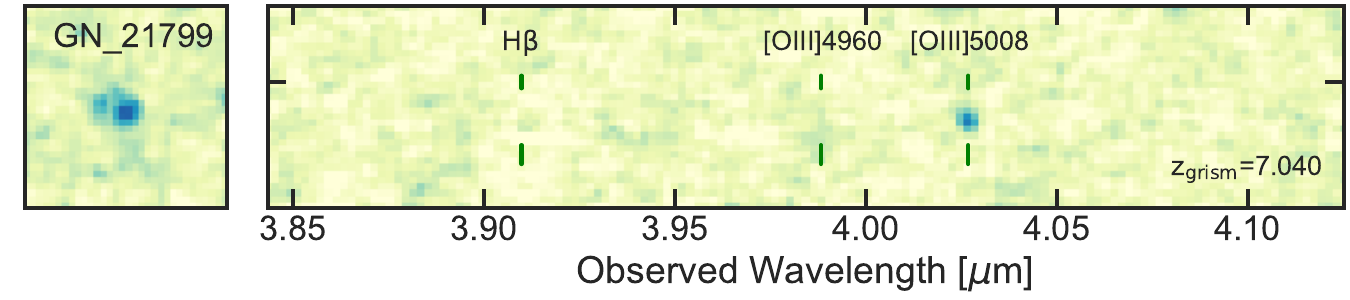}
\vspace{-0.49 cm}
\vspace*{-0.49 cm}
\includegraphics[width=1.0\textwidth]{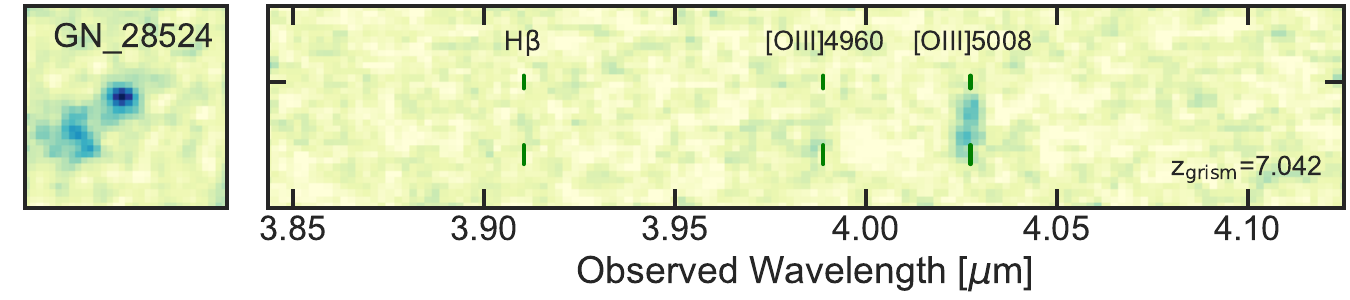}
\vspace{-0.49 cm}
\vspace*{-0.49 cm}
\includegraphics[width=1.0\textwidth]{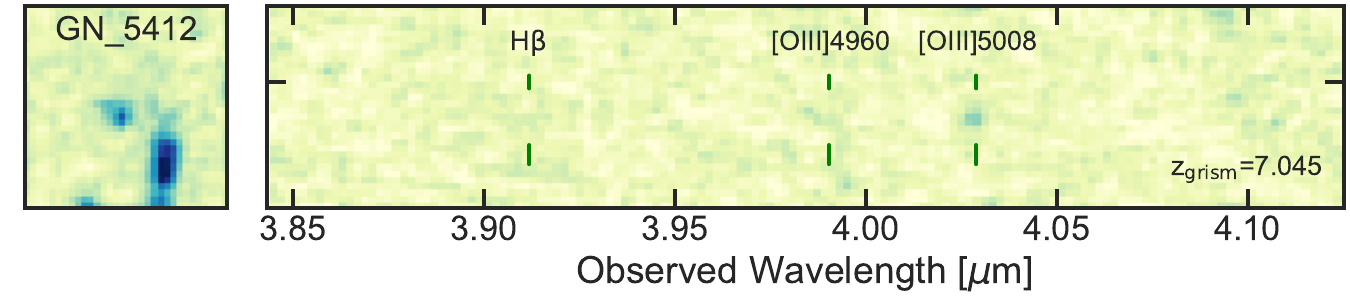}
\vspace{-0.49 cm}
\vspace*{-0.49 cm}
\includegraphics[width=1.0\textwidth]{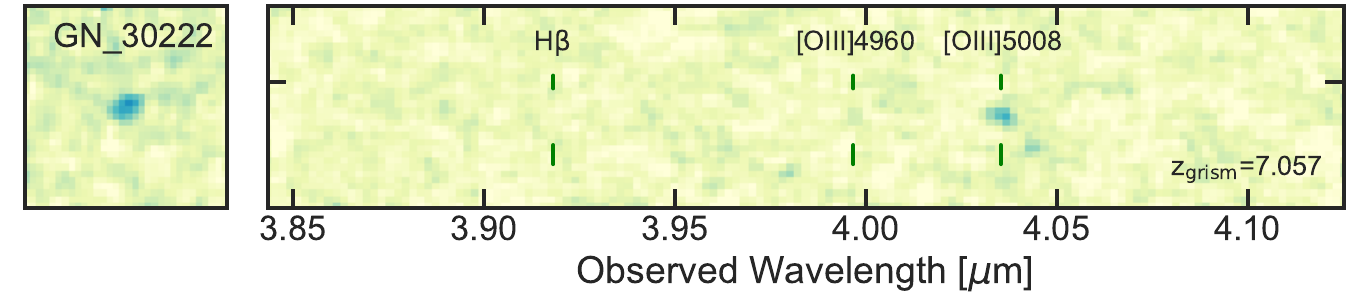}
\vspace{-0.49 cm}
\vspace*{-0.49 cm}
\includegraphics[width=1.0\textwidth]{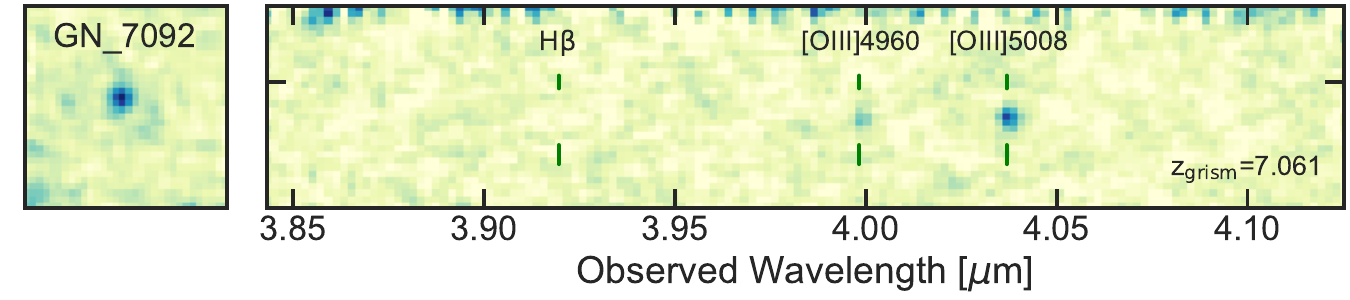}
\vspace{-0.49 cm}
\vspace*{-0.49 cm}
\includegraphics[width=1.0\textwidth]{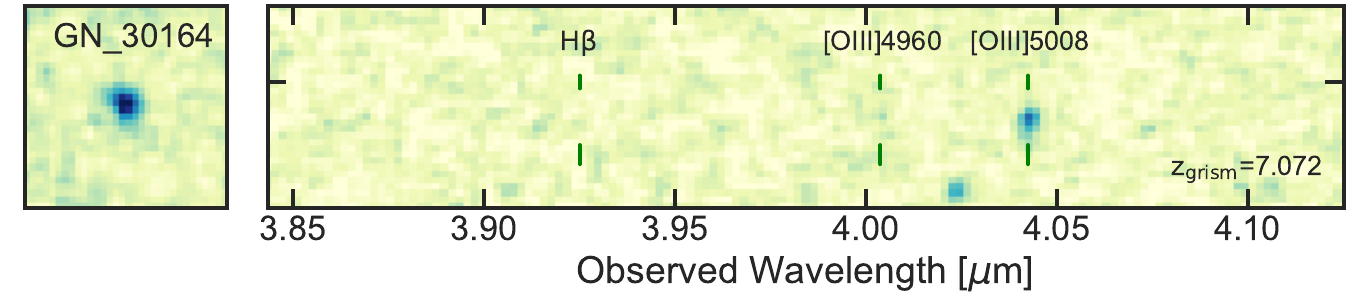}
\vspace{-0.49 cm}
\vspace*{-0.49 cm}
\includegraphics[width=1.0\textwidth]{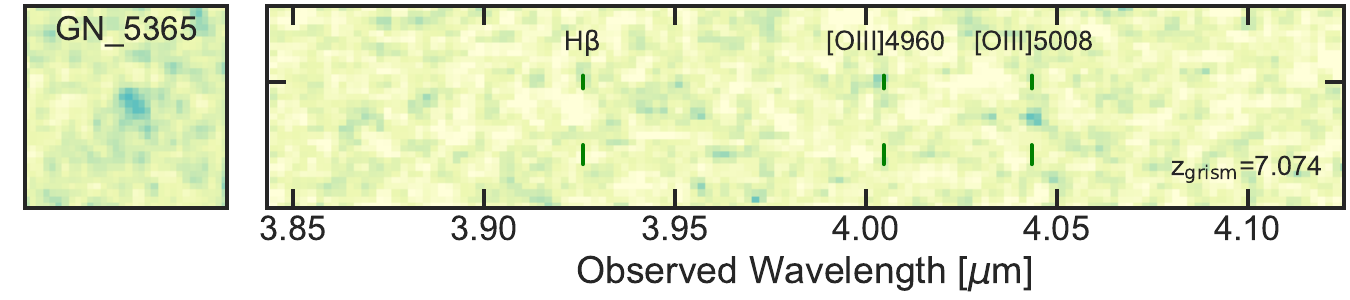}
\vspace{-0.49 cm}
\vspace*{-0.49 cm}
\includegraphics[width=1.0\textwidth]{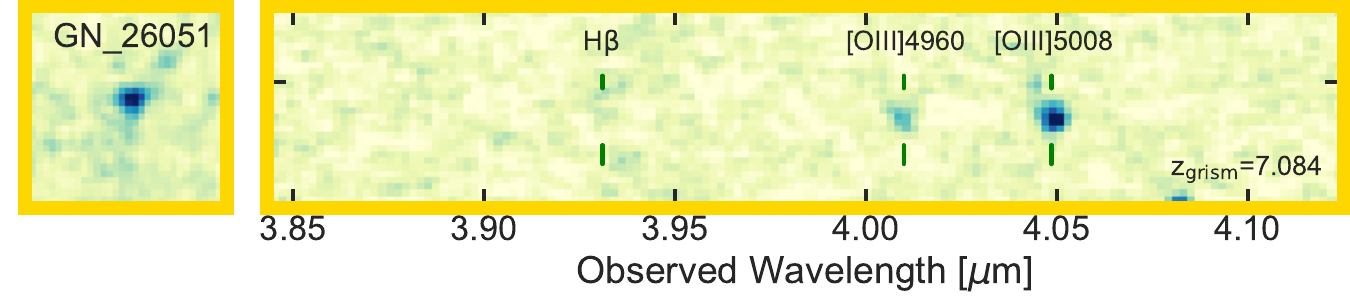}
\vspace{-0.49 cm}
\vspace*{-0.49 cm}
\includegraphics[width=1.0\textwidth]{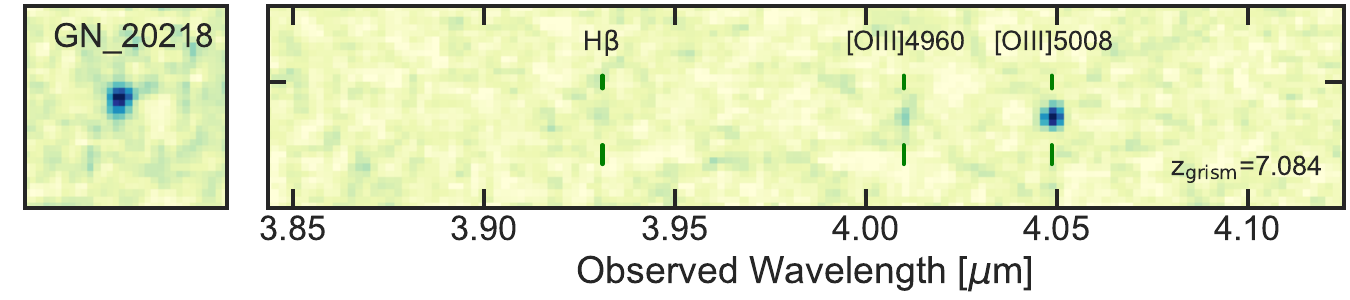}
\vspace{-0.49 cm}
\vspace*{-0.49 cm}
\includegraphics[width=1.0\textwidth]{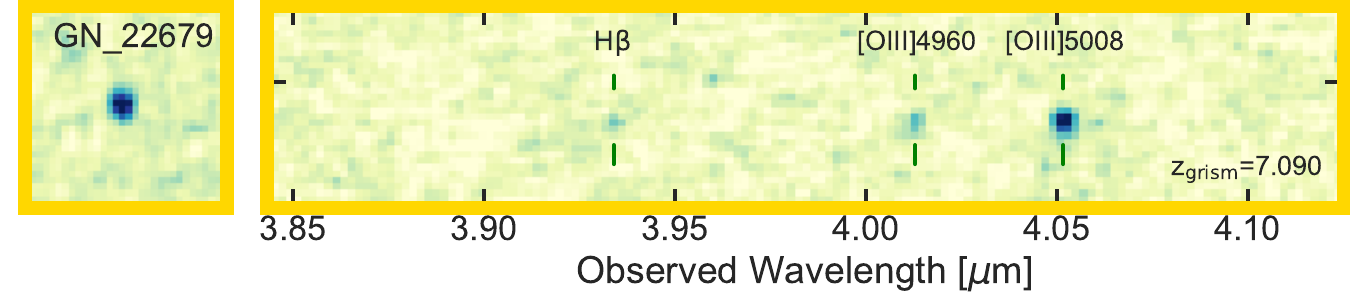}
\vspace{-0.49 cm}
\vspace*{-0.49 cm}
\includegraphics[width=1.0\textwidth]{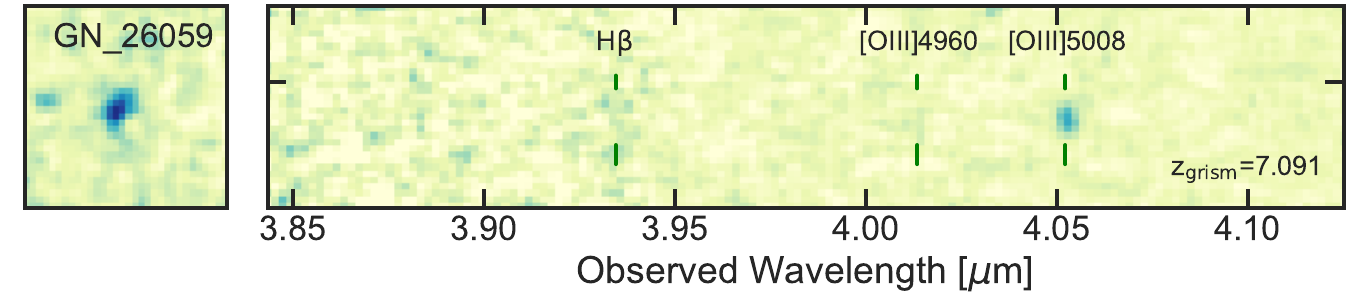}
\vspace{-0.49 cm}
\vspace*{-0.49 cm}
\includegraphics[width=1.0\textwidth]{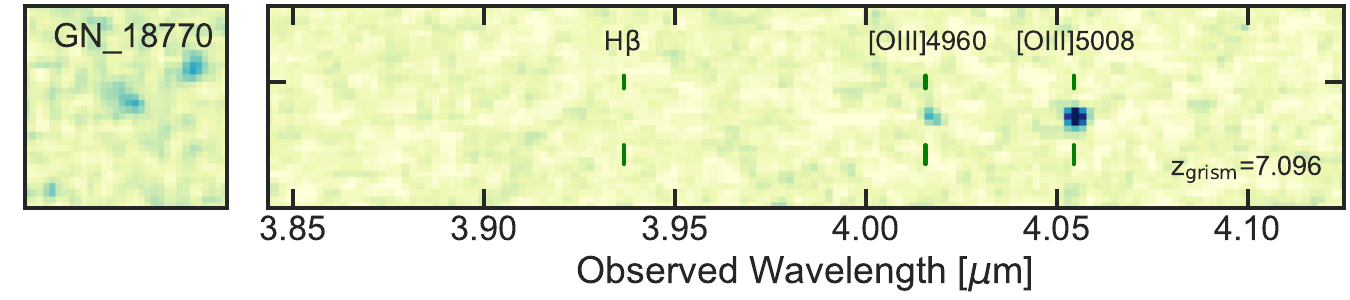}

\end{subfigure}
\caption {Top left panel: sky projection of galaxies in the GN\_z7\_0 group, where the LAE is marked as a star and its companion [OIII] emitters as circles. Top right panel: 3D spatial distribution, with the LAEs represented by stars and [OIII] emitters by circles. Axes are in proper Mpc, with redshift increasing along the x-axis. Bottom panel: 2D spectra of [OIII] emitters, with the yellow frame highlighting the Lyman-$\alpha$ emitters.}

\label{fig:group1}
\end{figure*}

%-----------------------%-----------------------
\begin{table*}
%\centering

\begin{tabular}{cccccccccc}
\hline\hline

  ID & RA & DEC & z$_\mathrm{Lya}$ & z$_\mathrm{[OIII]}$ & M$_\mathrm{{UV}}$ &  $\log_{\mathrm{M}_{\star}}$& EW$_\mathrm{{[OIII]}}$[\AA]  \\\hline\hline\\[-5pt]

\multicolumn{9}{c}{\textbf{GN\_z7\_2 group}}\\[5pt]

4394\href{#refa}{\textsuperscript{a}}  & 189.08349 & 62.20258 & 7.197 & 7.186 & -20.4 $^{+0.1}_{-0.1}$ & 9.15 $^{+0.30}_{-0.46}$ & 167 $^{+73}_{-25}$ \\
8563\href{#refa}{\textsuperscript{a}}  & 189.08405 & 62.22204 & 7.206 & 7.200 & -20.2 $^{+0.0}_{-0.0}$ & 7.81 $^{+0.12}_{-0.07}$ & 816 $^{+84}_{-74}$ \\
26844\href{#refa}{\textsuperscript{a}}  & 189.22512 & 62.28629 & 7.21 & 7.205 & -19.9 $^{+0.1}_{-0.1}$ & 7.84 $^{+0.26}_{-0.14}$ & 1306 $^{+128}_{-132}$ &\\
\hline\hline\\
5688\href{#ref1}{\textsuperscript{1}} & 189.07065 & 62.20895 & - & 7.186 & -22.8 $^{+0.0}_{-0.0}$ & 10.65 $^{+0.11}_{-0.10}$ & 7 $^{+1}_{-1}$ \\
8877 & 189.13655 & 62.22330 & - & 7.139 & -21.5 $^{+0.0}_{-0.0}$ & 8.50 $^{+0.15}_{-0.06}$ & 475 $^{+31}_{-26}$ \\
5216 & 189.07460 & 62.20674 & - & 7.188 & -21.2 $^{+0.1}_{-0.0}$ & 8.46 $^{+0.66}_{-0.18}$ & 500 $^{+53}_{-37}$ \\
14101 & 189.24983 & 62.24122 & - & 7.145 & -20.4 $^{+0.0}_{-0.0}$ & 8.62 $^{+0.34}_{-0.17}$ & 796 $^{+53}_{-49}$ \\
19551 & 189.14817 & 62.23806 & - & 7.138 & -20.4 $^{+0.1}_{-0.0}$ & 8.64 $^{+0.42}_{-0.13}$ & 437 $^{+64}_{-57}$ \\
28784 & 189.24916 & 62.29901 & - & 7.134 & -19.9 $^{+0.1}_{-0.1}$ & 8.86 $^{+0.30}_{-0.36}$ & 490 $^{+79}_{79}$ \\
256 & 189.15107 & 62.16643 & - & 7.166 & -19.6 $^{+0.2}_{-0.1}$ & 8.07 $^{+0.43}_{-0.40}$ & 340 $^{+120}_{-81}$ \\
20564 & 189.25929 & 62.23546 & - & 7.141 & -19.4 $^{+0.1}_{-0.1}$ & 7.39 $^{+0.07}_{-0.04}$ & 327 $^{+54}_{-92}$ \\
14140 & 189.25017 & 62.24139 & - & 7.150 & -19.3 $^{+0.1}_{-0.0}$ & 7.95 $^{+0.16}_{-0.15}$ & 494 $^{+68}_{-59}$ \\
12873 & 189.25888 & 62.23745 & - & 7.145 & -19.2 $^{+0.1}_{-0.1}$ & 8.14 $^{+0.08}_{-0.65}$ & 806 $^{+95}_{-91}$ \\
814 & 189.19798 & 62.17702 & - & 7.138 & -19.0 $^{+0.1}_{-0.0}$ & 7.72 $^{+0.05}_{-0.06}$ & 1789 $^{+104}_{-116}$ \\
11400 & 189.12769 & 62.23260 & - & 7.131 & -18.9 $^{+0.2}_{-0.1}$ & 8.19 $^{+0.34}_{-0.26}$ & 965 $^{+261}_{261}$ \\
13324 & 189.10994 & 62.23872 & - & 7.132 & -18.0 $^{+0.2}_{-0.2}$ & 7.64 $^{+0.23}_{-0.24}$ & 793 $^{+156}_{-141}$ \\
20960 & 189.23104 & 62.26238 & - & 7.169 & -17.3 $^{+0.6}_{-0.5}$ & 7.90 $^{+0.55}_{-0.52}$ & 1028 $^{+509}_{509}$ \\
\hline\hline
\end{tabular}

\caption{Summary of the GN$\_$z7$\_$2 group.}\label{table2}
\label{tab:2f}
\vspace{1ex}
\footnotesize{\raggedright Notes:  \textsuperscript{a} LAEs}
\footnotesize{\raggedright\par}
\footnotesize{\raggedright References:  {\textsuperscript{1}} \cite{Fujimoto2022} \par}
\end{table*}

%-----------------------%-----------------------%-----------------------

\begin{figure*}
\centering
\begin{subfigure}[t]{1\textwidth}
\centering
\includegraphics[width=0.46\textwidth]{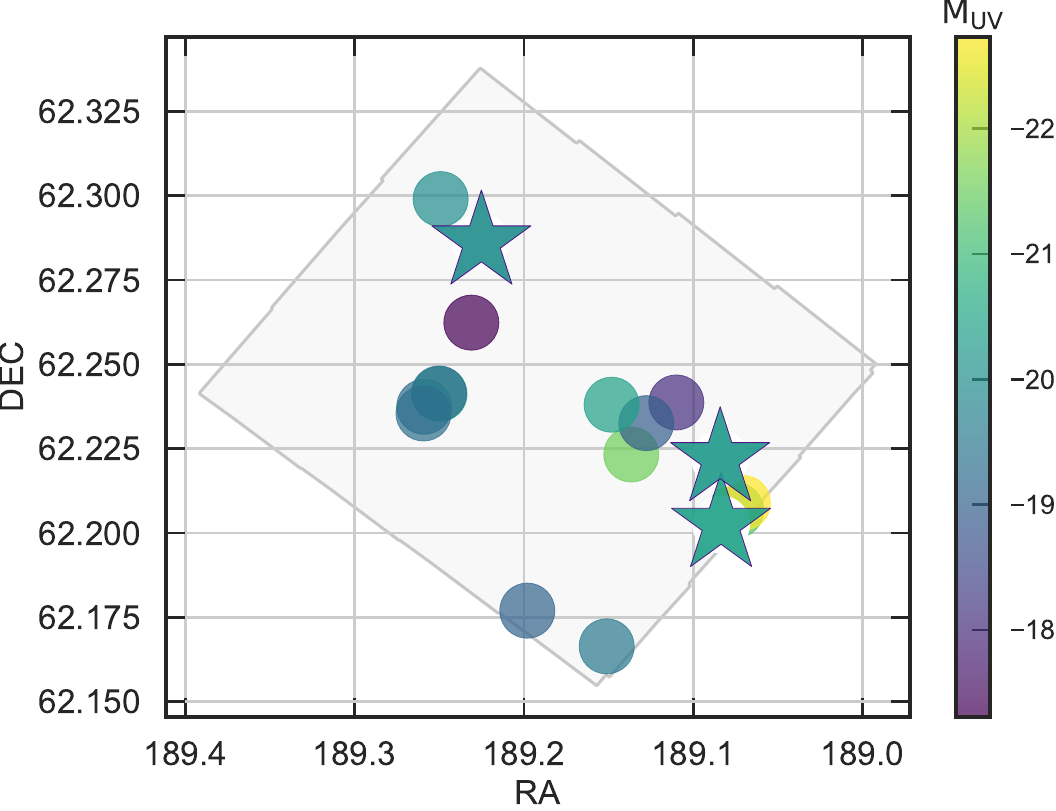}
\includegraphics[width=0.42\textwidth, height=0.42\textwidth]{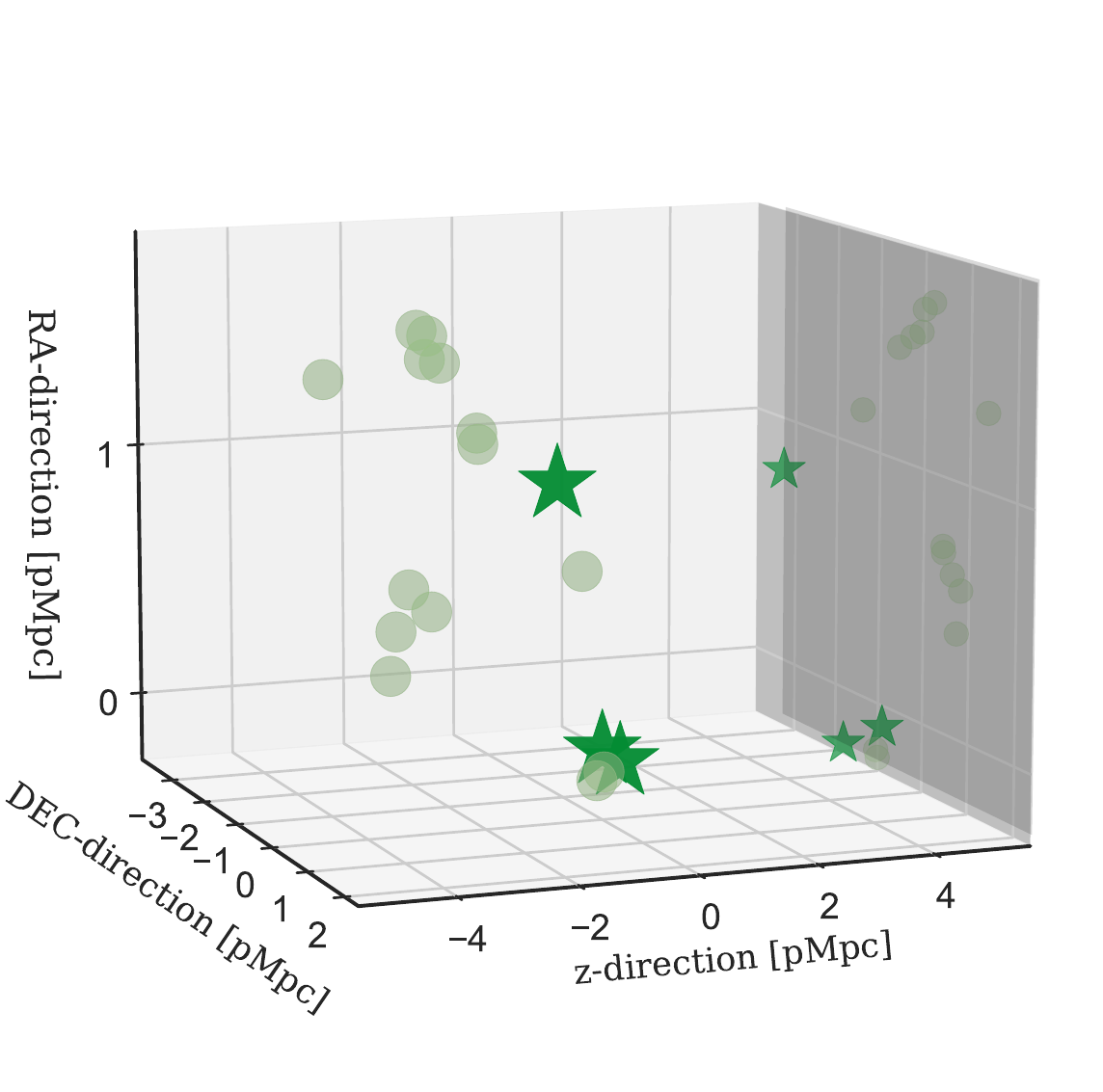}
\end{subfigure}
\begin{subfigure}[t]{0.47\textwidth}
\centering
\includegraphics[width=1.0\textwidth]{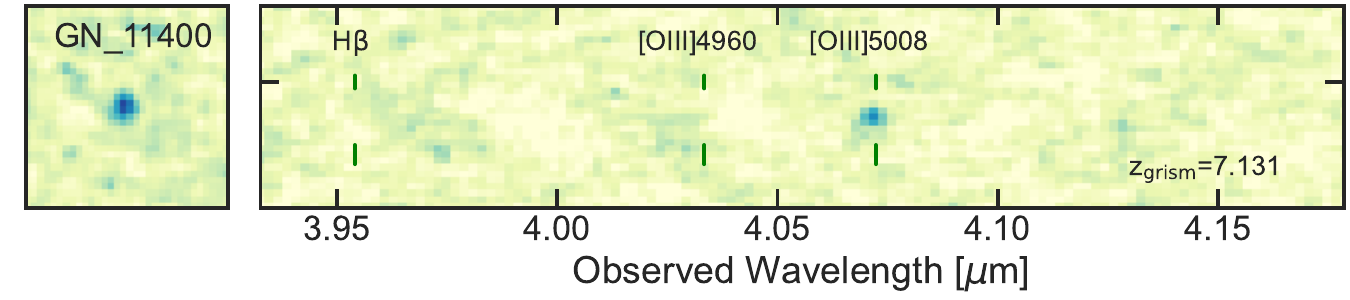}
\vspace{-0.49 cm}
\vspace*{-0.49 cm}
\includegraphics[width=1.0\textwidth]{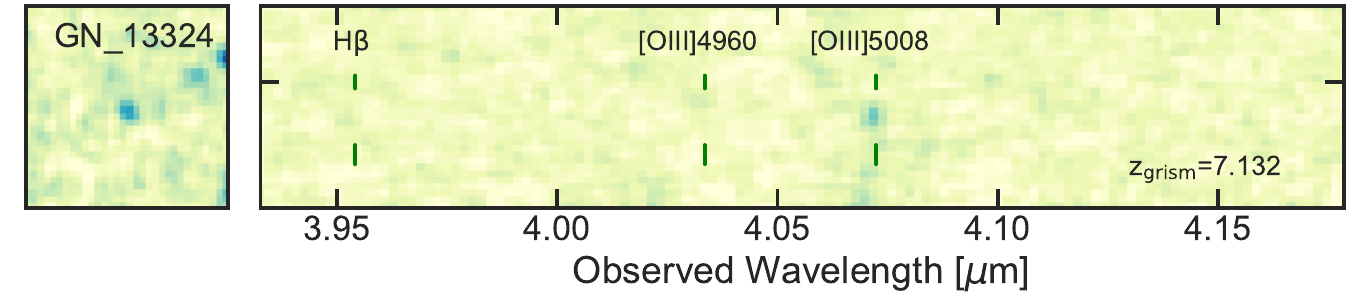}
\vspace{-0.49 cm}
\vspace*{-0.49 cm}
\includegraphics[width=1.0\textwidth]{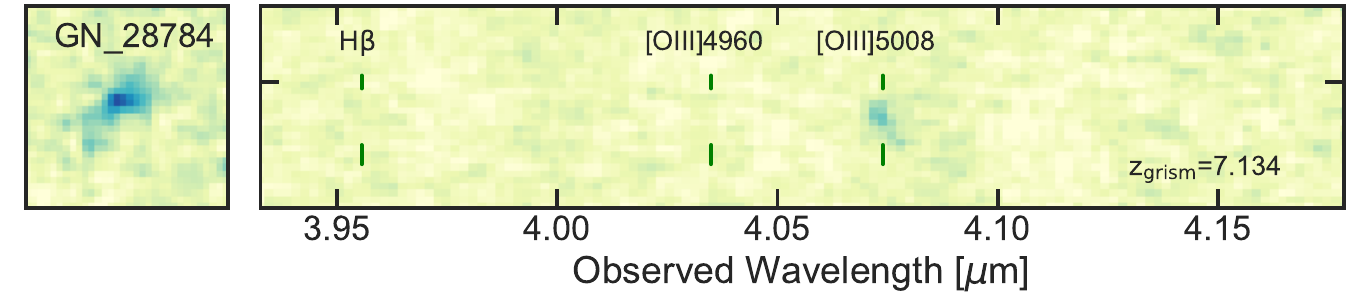}
\vspace{-0.49 cm}
\vspace*{-0.49 cm}
\includegraphics[width=1.0\textwidth]{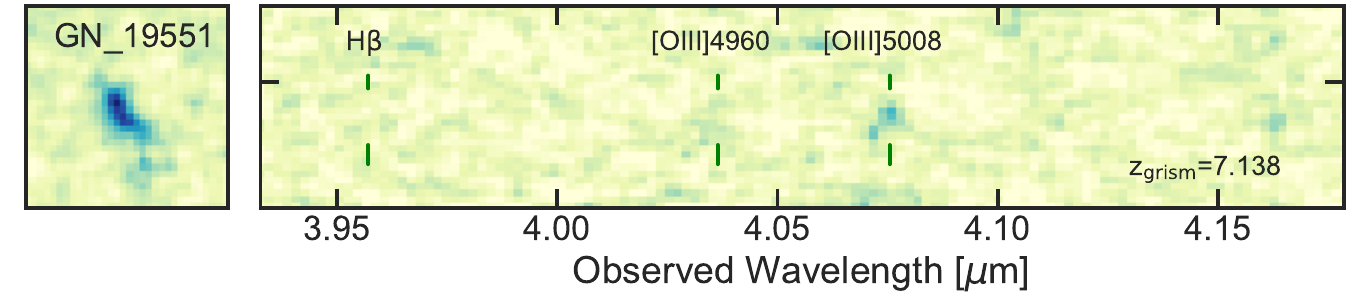}
\vspace{-0.49 cm}
\vspace*{-0.49 cm}
\includegraphics[width=1.0\textwidth]{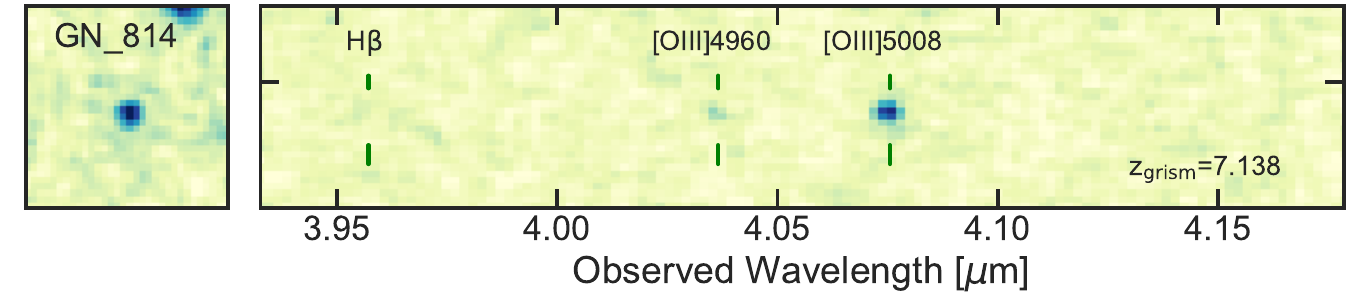}
\vspace{-0.49 cm}
\vspace*{-0.49 cm}
\includegraphics[width=1.0\textwidth]{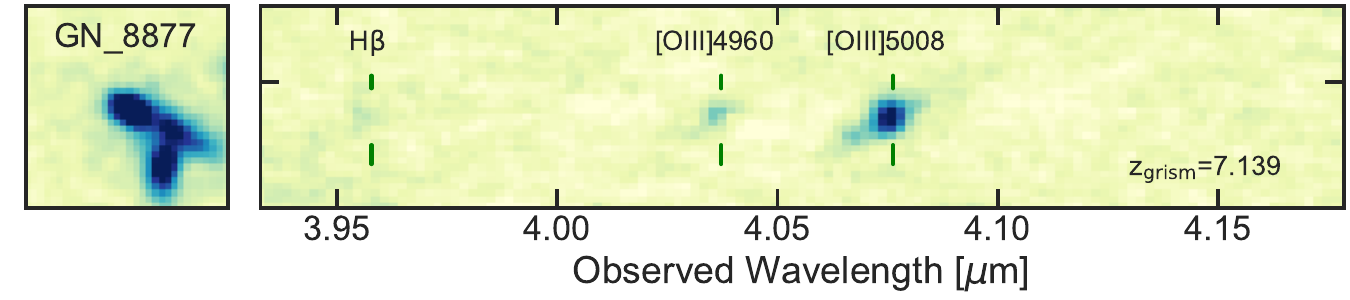}
\vspace{-0.49 cm}
\vspace*{-0.49 cm}
\includegraphics[width=1.0\textwidth]{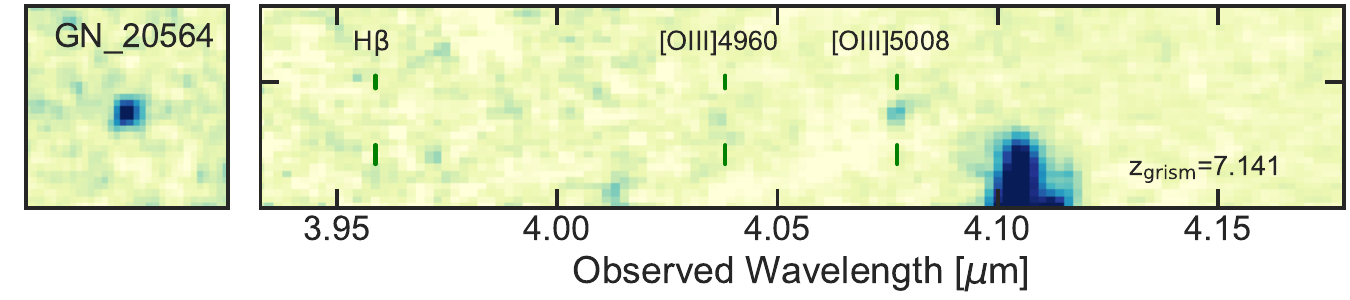}
\vspace{-0.49 cm}
\vspace*{-0.49 cm}
\includegraphics[width=1.0\textwidth]{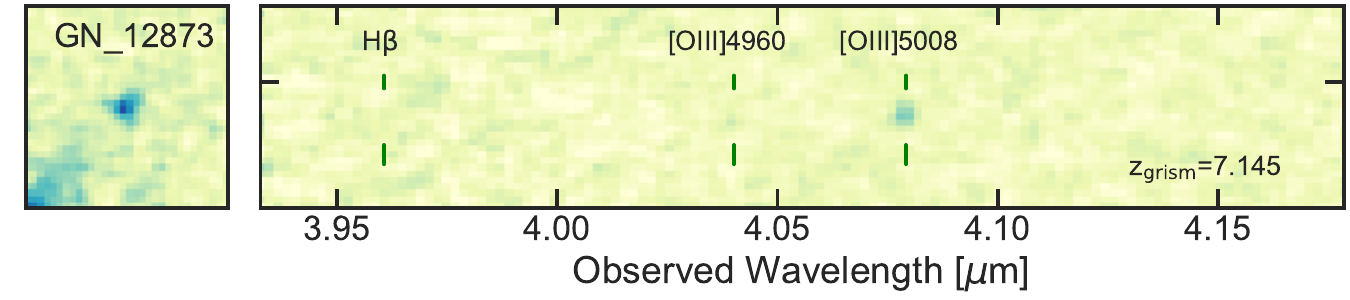}

\end{subfigure}
\hspace{0.05\textwidth}
\begin{subfigure}[t]{0.47\textwidth}
\centering
\includegraphics[width=1.0\textwidth]{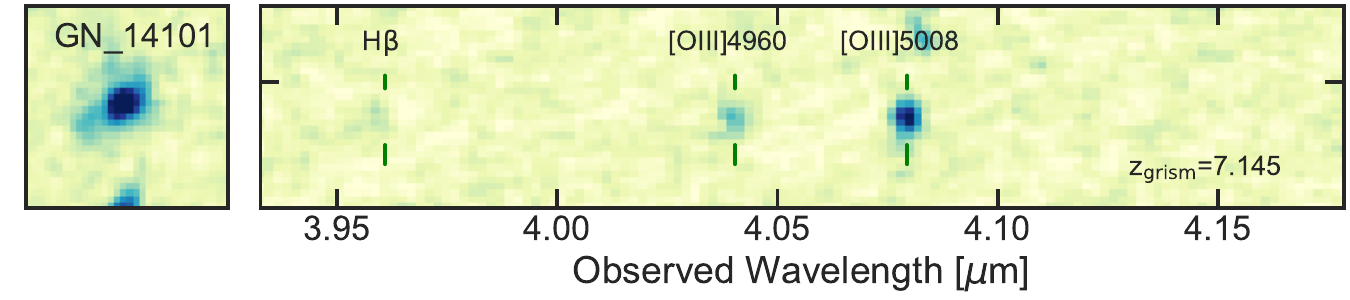}
\vspace{-0.49 cm}
\vspace*{-0.49 cm}
\includegraphics[width=1.0\textwidth]{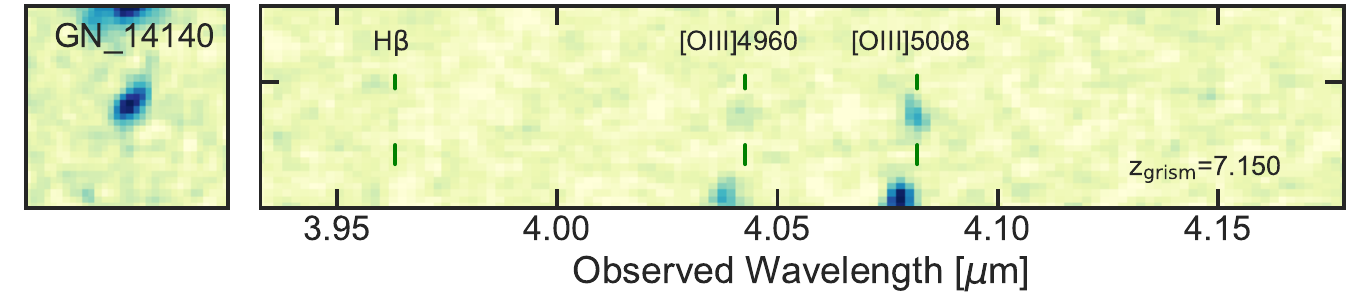}
\vspace{-0.49 cm}
\vspace*{-0.49 cm}
\includegraphics[width=1.0\textwidth]{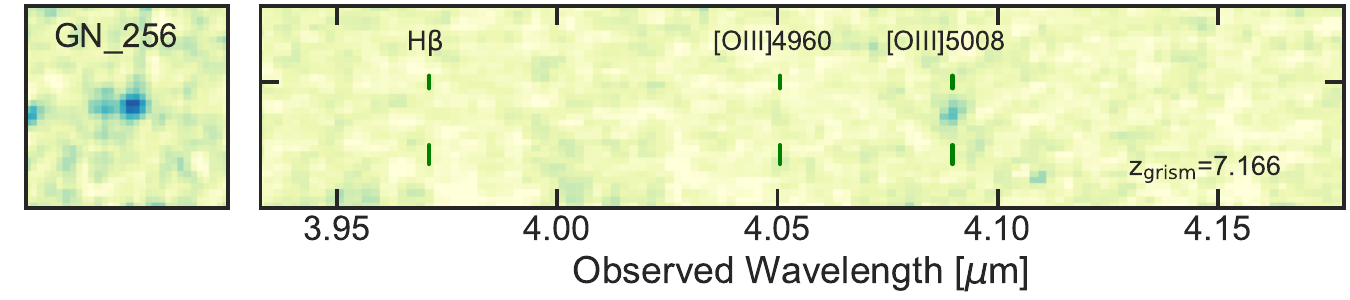}
\vspace{-0.49 cm}
\vspace*{-0.49 cm}
\includegraphics[width=1.0\textwidth]{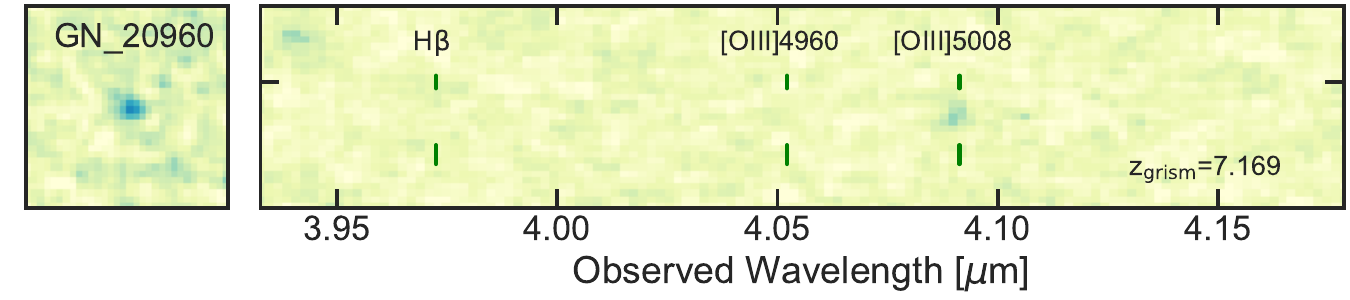}
\vspace{-0.49 cm}
\vspace*{-0.49 cm}
\includegraphics[width=1.0\textwidth]{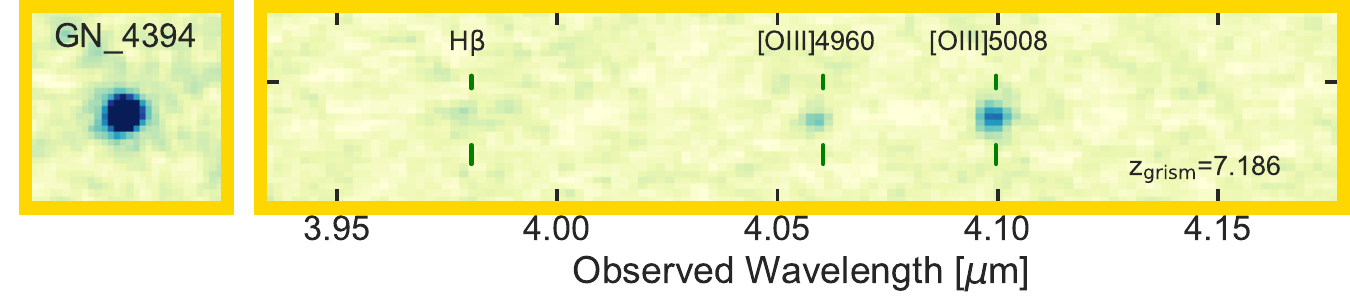}
\vspace{-0.49 cm}
\vspace*{-0.49 cm}
\includegraphics[width=1.0\textwidth]{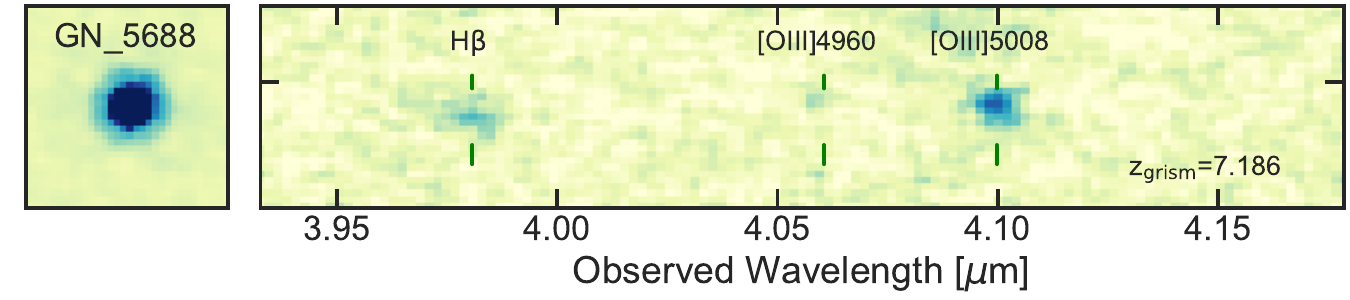}
\vspace{-0.49 cm}
\vspace*{-0.49 cm}
\includegraphics[width=1.0\textwidth]{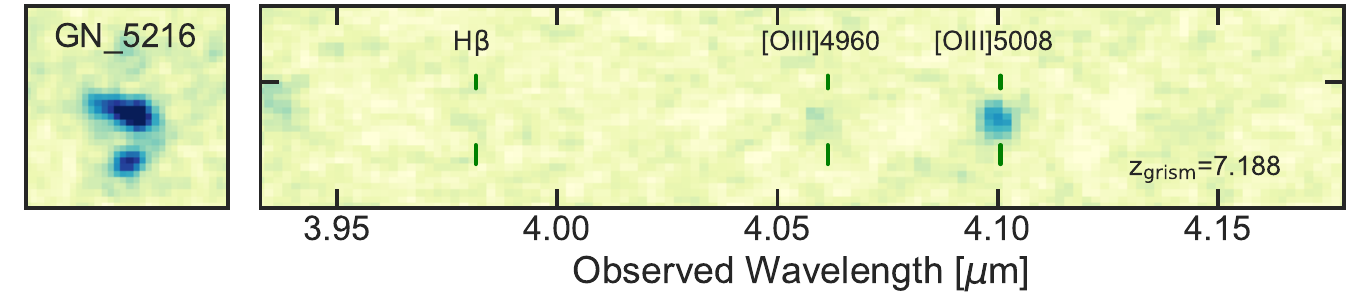}
\vspace{-0.49 cm}
\vspace*{-0.49 cm}
\includegraphics[width=1.0\textwidth]{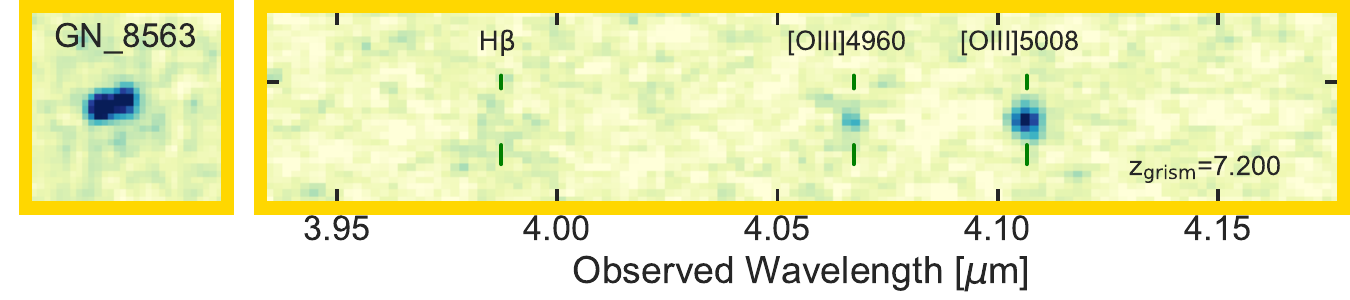}
\vspace{-0.49 cm}
\vspace*{-0.49 cm}
\includegraphics[width=1.0\textwidth]{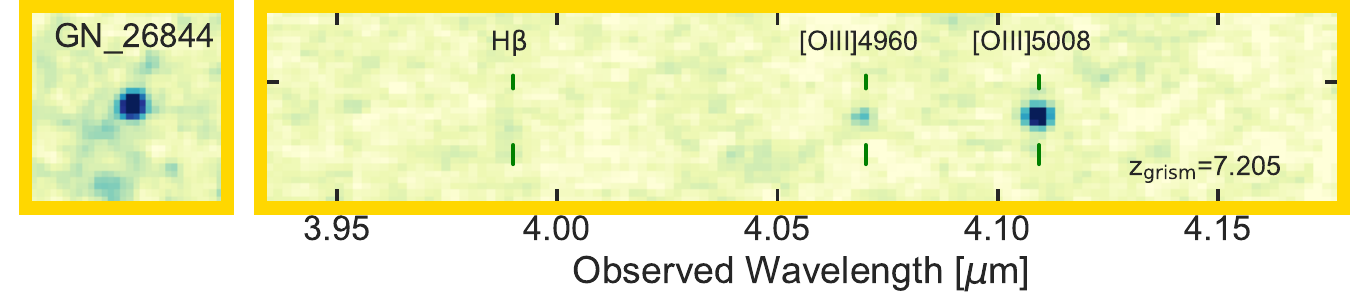}

\end{subfigure}
\caption {Top left panel: sky projection of galaxies in the GN\_z7\_2 group, where the LAE is marked as a star and its companion [OIII] emitters as circles. Top right panel: 3D spatial distribution, with the LAEs represented by stars and [OIII] emitters by circles. Axes are in proper Mpc, with redshift increasing along the x-axis. Bottom panel: 2D spectra of [OIII] emitters, with the yellow frame highlighting the Lyman-$\alpha$ emitters.}
\label{fig:group2}
\end{figure*}

%-----------------------%-----------------------
\begin{table*}
%\centering

\begin{tabular}{cccccccccc}
\hline\hline

  ID & RA & DEC & z$_\mathrm{Lya}$ & z$_\mathrm{[OIII]}$ & M$_\mathrm{{UV}}$ &  $\log_{\mathrm{M}_{\star}}$ & EW$_\mathrm{{[OIII]}}$[\AA]  \\\hline\hline\\[-5pt]

\multicolumn{9}{c}{\textbf{GN\_z7\_5 group}}\\[5pt]
29193\href{#refa}{\textsuperscript{a}} & 189.15798 & 62.30240 & 7.507 & 7.502 & -21.1 $^{+0.0}_{-0.0}$ & 8.25 $^{+0.19}_{-0.05}$ & 799 $^{+54}_{-51}$ \\
29192\href{#refa}{\textsuperscript{a}}  \href{#ref1}{\textsuperscript{1}}& 189.15779 & 62.30234 & 7.507 & 7.498 & -20.8 $^{+0.0}_{-0.0}$ & 8.22 $^{+0.26}_{-0.11}$ & 939 $^{+126}_{-96}$ &\\
\hline\hline\\
26237 & 189.26166 & 62.28327 & - & 7.427 & -19.6 $^{+0.1}_{-0.1}$ & 7.73 $^{+0.23}_{-0.12}$ & 1455 $^{+265}_{-217}$ \\
6294 & 189.26206 & 62.21209 & - & 7.423 & -19.5 $^{+0.1}_{-0.1}$ & 8.21 $^{+0.09}_{-0.09}$ & 781 $^{+89}_{-78}$ \\
29409 & 189.08771 & 62.20013 & - & 7.544 & -18.7 $^{+0.3}_{-0.3}$ & 8.43 $^{+0.39}_{-0.86}$ & 516 $^{+244}_{-151}$ \\
22186 & 189.22427 & 62.26668 & - & 7.522 & -18.7 $^{+0.2}_{-0.2}$ & 8.56 $^{+0.35}_{-0.50}$ & 413 $^{+126}_{126}$ \\
9329 & 189.26255 & 62.26825 & - & 7.552 & -18.7 $^{+0.2}_{-0.2}$ & 8.13 $^{+0.39}_{-0.49}$ & 490 $^{+168}_{-134}$ \\

\hline\hline
\end{tabular}

\caption{Summary of the GN$\_$z7$\_$5 group.}\label{table3}
\label{tab:3f}
\footnotesize{\raggedright Notes:  \textsuperscript{a} LAEs}
\footnotesize{\raggedright\par}
\footnotesize{\raggedright References:  {\textsuperscript{1}} \cite{Jung2020} \par}
\end{table*}

%-----------------------%-----------------------

\begin{figure*}
\centering
\begin{subfigure}[t]{1\textwidth}
\centering
\includegraphics[width=0.46\textwidth]{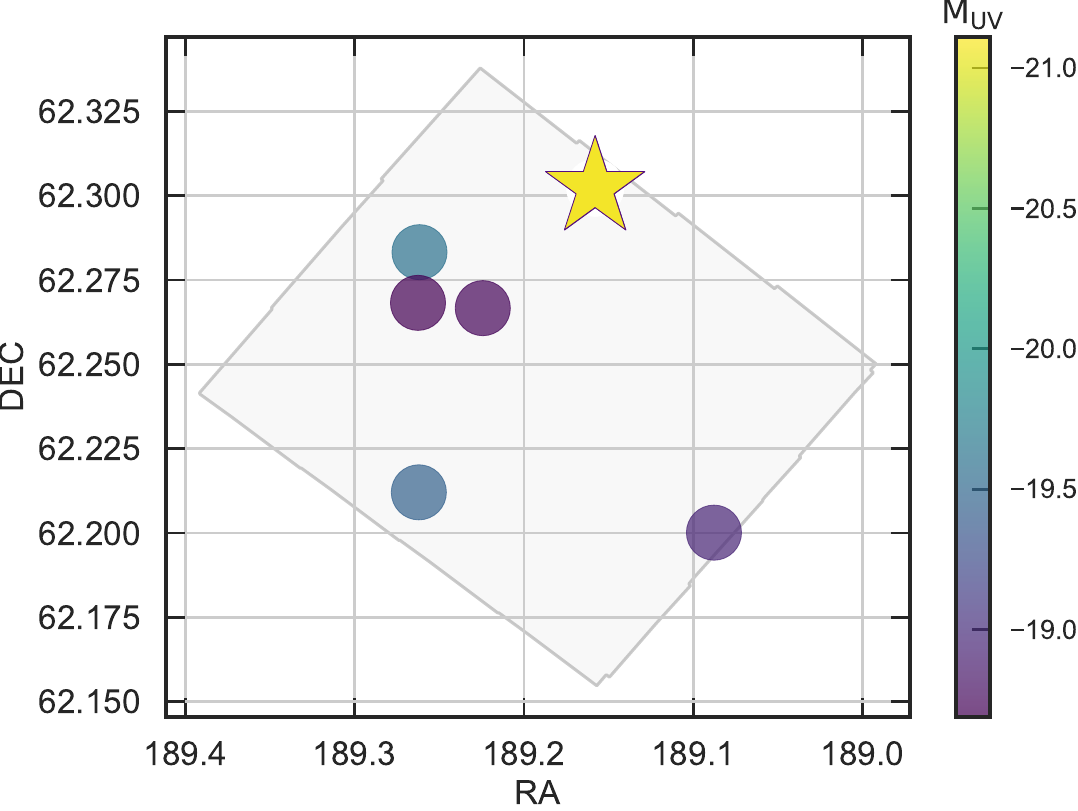}
\includegraphics[width=0.42\textwidth,height=0.4\textwidth]{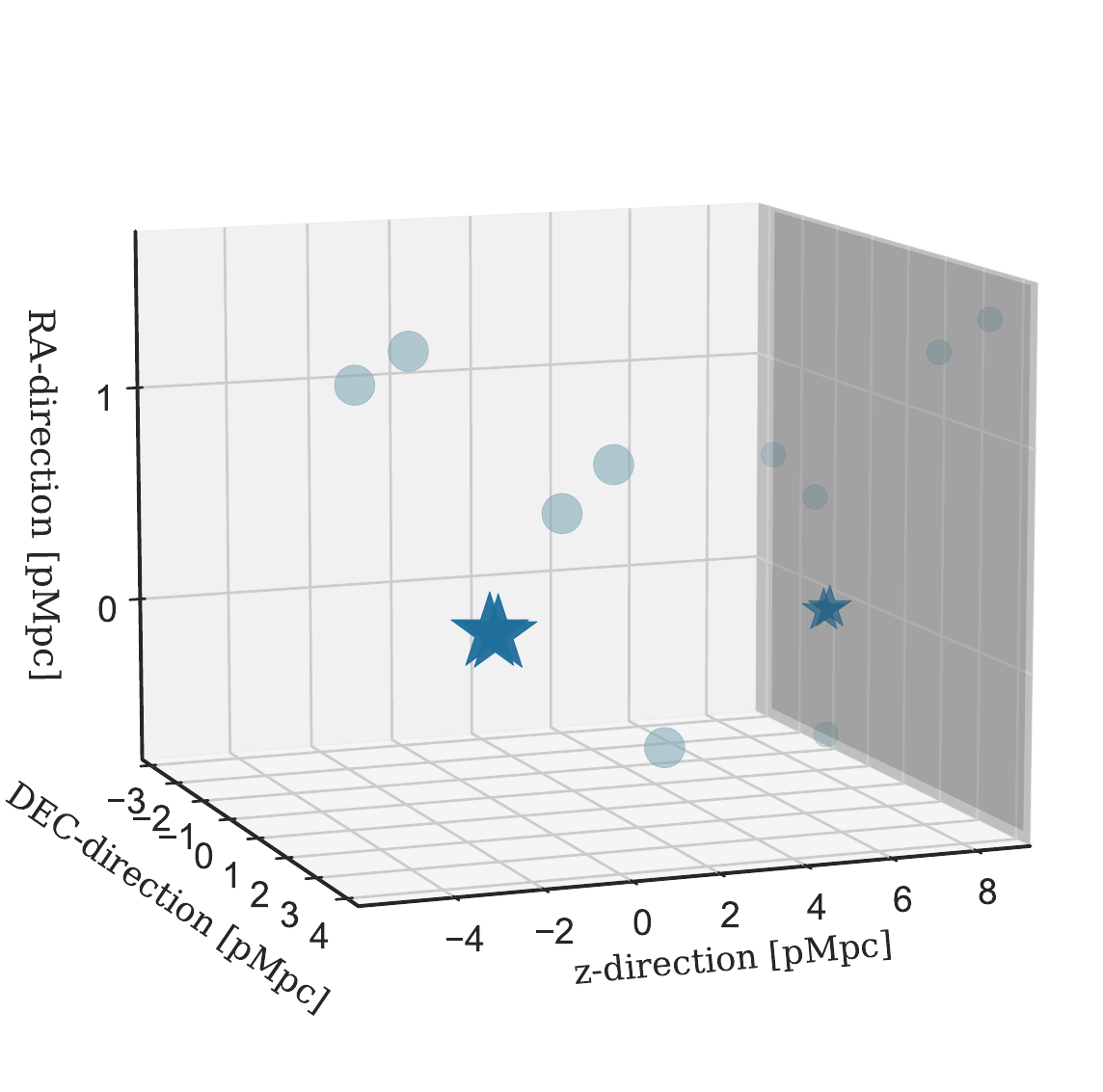}
\end{subfigure}
\begin{subfigure}[t]{0.47\textwidth}
\centering
\includegraphics[width=1.0\textwidth]{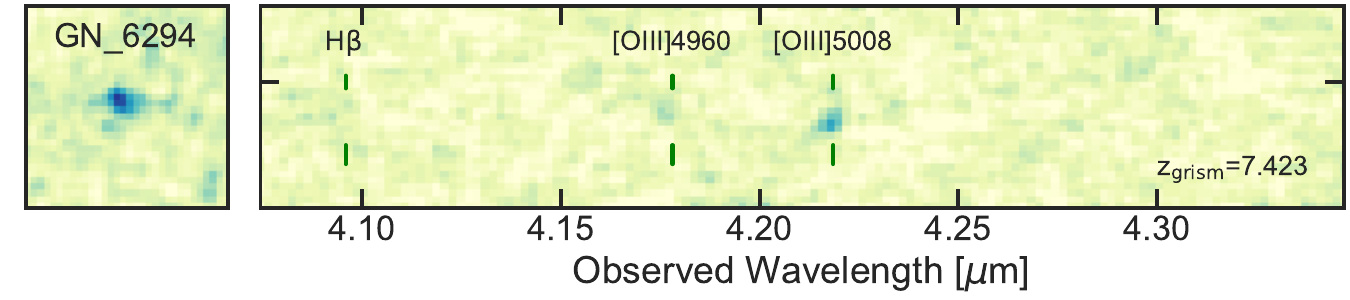}
\vspace{-0.49 cm}
\vspace*{-0.49 cm}
\includegraphics[width=1.0\textwidth]{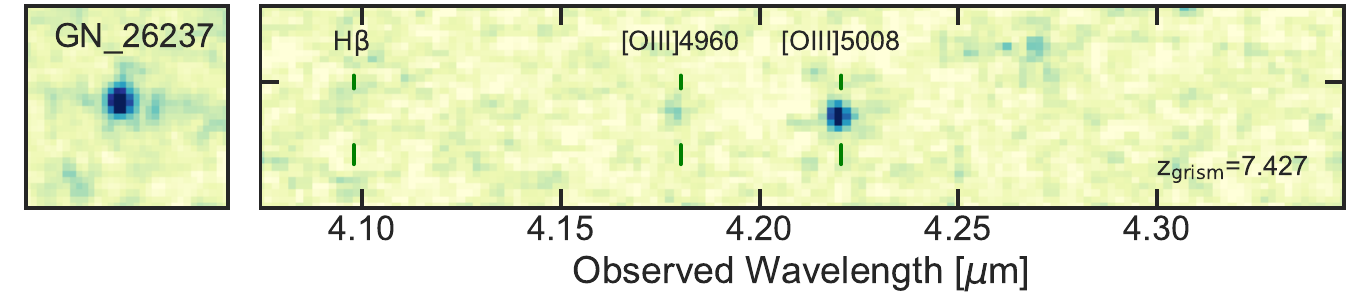}
\vspace{-0.49 cm}
\vspace*{-0.49 cm}
\includegraphics[width=1.0\textwidth]{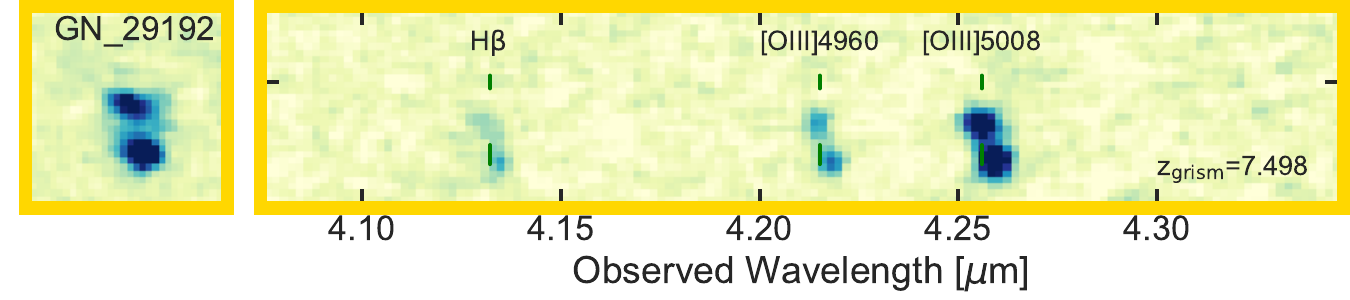}
\vspace{-0.49 cm}
\vspace*{-0.49 cm}
\includegraphics[width=1.0\textwidth]{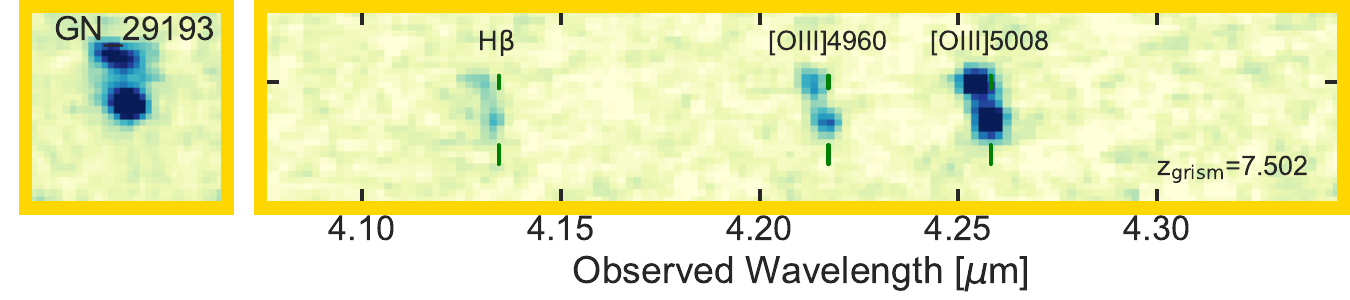}

\end{subfigure}
\hspace{0.05\textwidth}
\begin{subfigure}[t]{0.47\textwidth}
\centering
\includegraphics[width=1.0\textwidth]{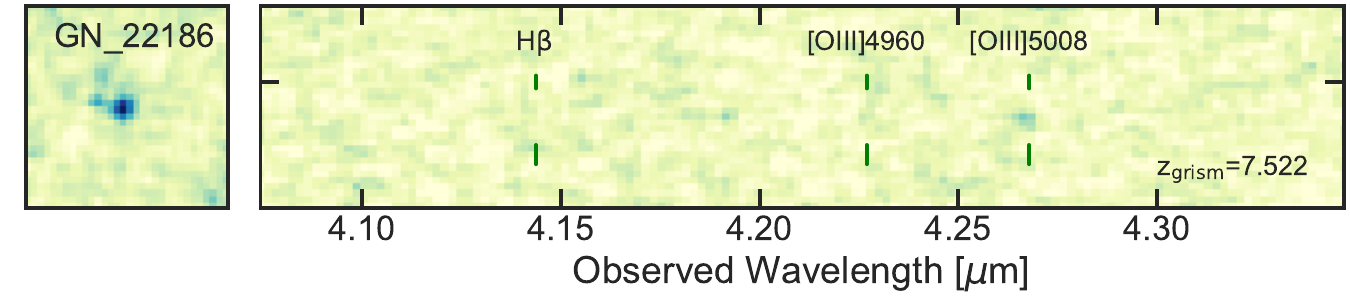}
\vspace{-0.49 cm}
\vspace*{-0.49 cm}
\includegraphics[width=1.0\textwidth]{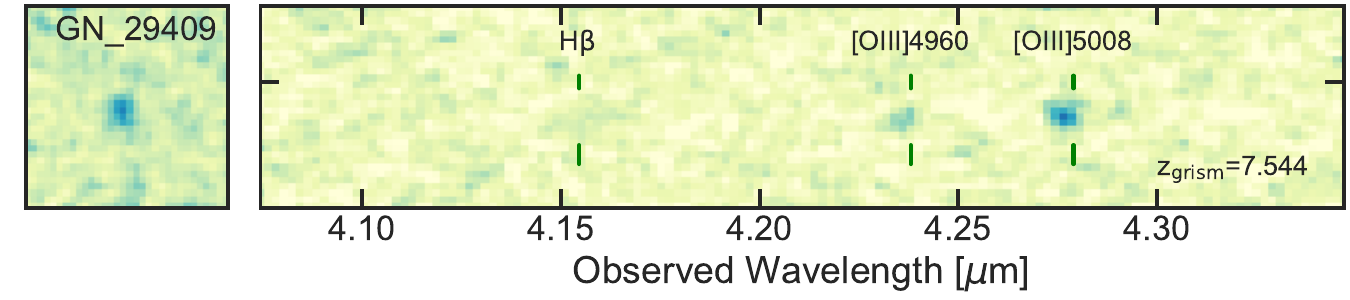}
\vspace{-0.49 cm}
\vspace*{-0.49 cm}
\includegraphics[width=1.0\textwidth]{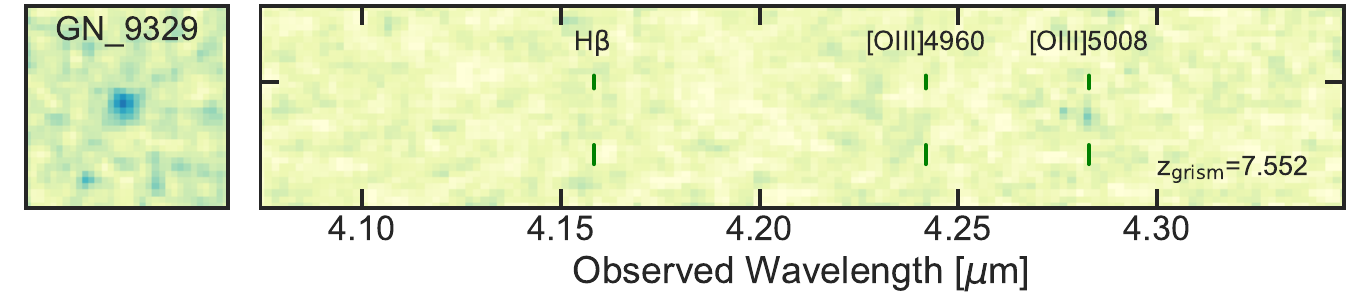}

\end{subfigure}
\caption {Top left panel: sky projection of galaxies in the GN\_z7\_5 group, where the LAE is marked as a star and its companion [OIII] emitters as circles. Top right panel: 3D spatial distribution, with the LAEs represented by stars and [OIII] emitters by circles. Axes are in proper Mpc, with redshift increasing along the x-axis. Bottom panel: 2D spectra of [OIII] emitters, with the yellow frame highlighting the Lyman-$\alpha$ emitters.}
\label{fig:group3}
\end{figure*}

\begin{table*}
%\centering

\begin{tabular}{cccccccccc}
\hline\hline

  ID & RA & DEC & z$_\mathrm{Lya}$ & z$_\mathrm{[OIII]}$ & M$_\mathrm{{UV}}$ &  $\log_{\mathrm{M}_{\star}}$ & EW$_\mathrm{{[OIII]}}$[\AA]  \\\hline\hline\\[-5pt]

\multicolumn{9}{c}{\textbf{GN\_z7\_6 group}}\\[5pt]

19441\href{#refa}{\textsuperscript{a}}\href{#ref1}{\textsuperscript{1}} & 189.33307 & 62.25722 & 7.6 & 7.595 & -21.2 $^{+0.0}_{-0.0}$ & 8.33 $^{+0.26}_{-0.08}$ & 1930 $^{+151}_{-168}$ \\
14535\href{#refa}{\textsuperscript{a}} & 189.20307 & 62.24249 & 7.657 & 7.648 & -20.4 $^{+0.1}_{-0.1}$ & 8.61 $^{+0.37}_{-0.40}$ & 856 $^{+159}_{159}$ \\
25604\href{#refa}{\textsuperscript{a}} & 189.30014 & 62.28034 & 7.7 & 7.694 & -19.9 $^{+0.1}_{-0.1}$ & 8.74 $^{+0.40}_{-0.26}$ & 439 $^{+75}_{75}$ \\
9680\href{#refa}{\textsuperscript{a}} & 189.28867 & 62.22651 & 7.635 & 7.629 & -19.9 $^{+0.1}_{-0.0}$ & 8.15 $^{+0.07}_{-0.19}$ & 1028 $^{+130}_{-161}$  &\\
\hline\hline\\
26281 & 189.20129 & 62.21439 & - & 7.647 & -21.1 $^{+0.1}_{-0.1}$ & 9.01 $^{+0.19}_{-0.24}$ & 141 $^{+17}_{-23}$ \\
27240 & 189.27002 & 62.28855 & - & 7.629 & -20.7 $^{+0.1}_{-0.0}$ & 8.82 $^{+0.14}_{-0.16}$ & 1214 $^{+111}_{-135}$ \\
30094 & 189.22439 & 62.31137 & - & 7.620 & -20.3 $^{+0.1}_{-0.1}$ & 8.10 $^{+0.56}_{-0.08}$ & 819 $^{+92}_{-65}$ \\
30469 & 189.22201 & 62.31576 & - & 7.600 & -20.1 $^{+0.1}_{-0.1}$ & 7.87 $^{+0.26}_{-0.21}$ & 539 $^{+87}_{-81}$ \\
16884 & 189.03149 & 62.24946 & - & 7.603 & -19.7 $^{+0.2}_{-0.1}$ & 8.36 $^{+0.55}_{-0.43}$ & 447 $^{+103}_{103}$ \\
21941 & 189.06594 & 62.26573 & - & 7.607 & -19.5 $^{+0.1}_{-0.1}$ & 8.16 $^{+0.14}_{-0.15}$ & 412 $^{+115}_{115}$ \\
16560 & 189.35297 & 62.24850 & - & 7.586 & -19.4 $^{+0.1}_{-0.2}$ & 8.77 $^{+0.34}_{-0.36}$ & 341 $^{+86}_{86}$ \\
11294 & 189.33119 & 62.26185 & - & 7.595 & -19.4 $^{+0.1}_{-0.1}$ & 8.31 $^{+0.61}_{-0.42}$ & 474 $^{+115}_{-121}$ \\
14534 & 189.20319 & 62.24245 & - & 7.633 & -19.2 $^{+0.1}_{-0.1}$ & 8.72 $^{+0.28}_{-0.28}$ & 655 $^{+152}_{152}$ \\
18605 & 189.04517 & 62.25453 & - & 7.651 & -19.2 $^{+0.2}_{-0.2}$ & 8.05 $^{+0.52}_{-0.24}$ & 580 $^{+198}_{-163}$ \\
28698 & 189.26083 & 62.29841 & - & 7.704 & -18.6 $^{+0.2}_{-0.2}$ & 7.90 $^{+0.47}_{-0.30}$ & 747 $^{+256}_{256}$ \\
8328 & 189.17372 & 62.27178 & - & 7.613 & -17.7 $^{+0.5}_{-0.5}$ & 7.84 $^{+0.44}_{-0.44}$ & 761 $^{+266}_{-242}$ \\
\hline\hline
\end{tabular}

\caption{Summary of the GN$\_$z7$\_$6 group.}\label{table4}
\label{tab:4f}
\vspace{1ex}
\footnotesize{\raggedright Notes:  \textsuperscript{a} LAEs}
\footnotesize{\raggedright\par}
\footnotesize{\raggedright References:  {\textsuperscript{1}} \cite{Jung2020} \par}
\end{table*}

%-----------------------%-----------------------
\begin{figure*}
\centering
\begin{subfigure}[t]{1\textwidth}
\centering
\includegraphics[width=0.46\textwidth]{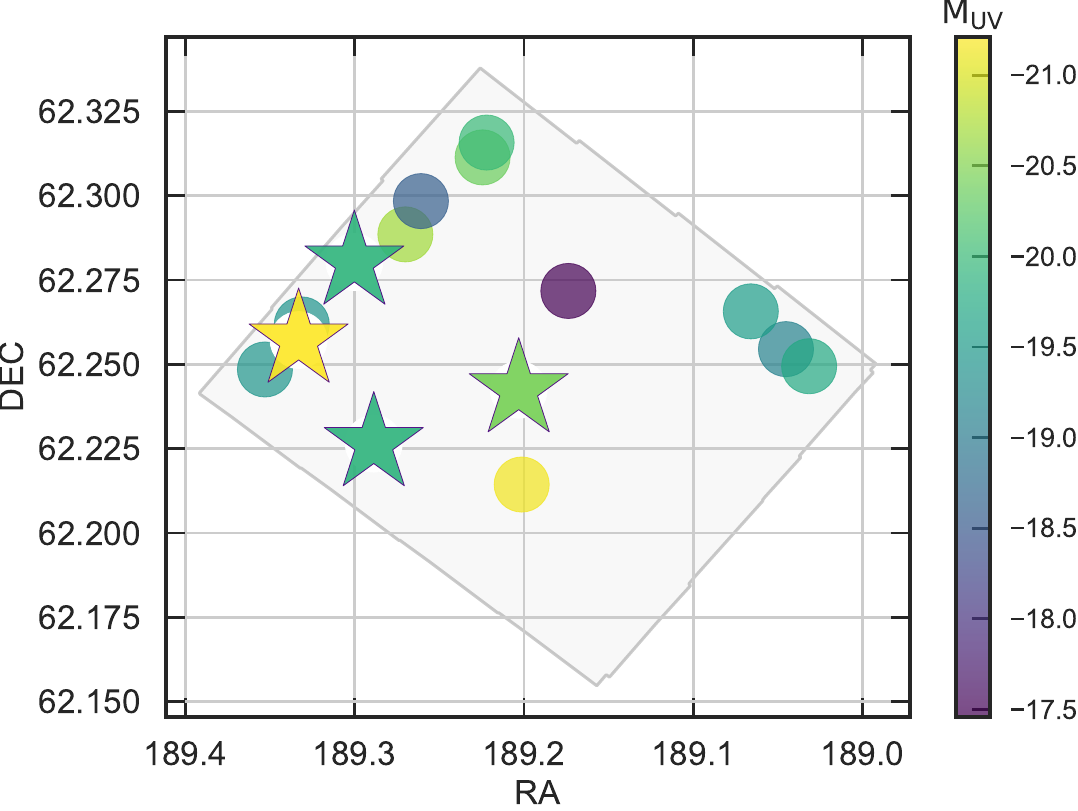}
\includegraphics[width=0.42\textwidth, height=0.43\textwidth]{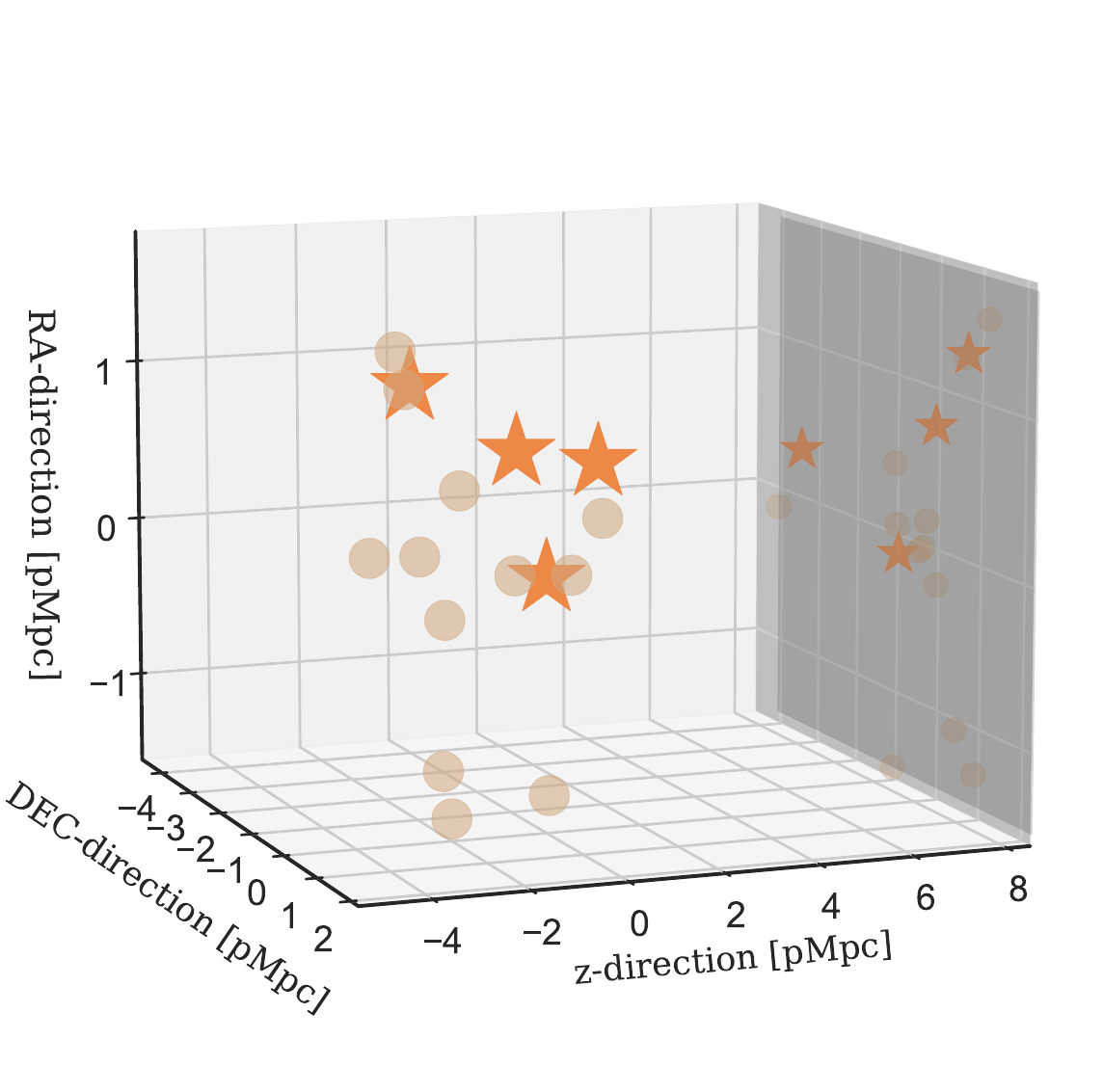}
\end{subfigure}
\begin{subfigure}[t]{0.47\textwidth}
\centering
\includegraphics[width=1.0\textwidth]{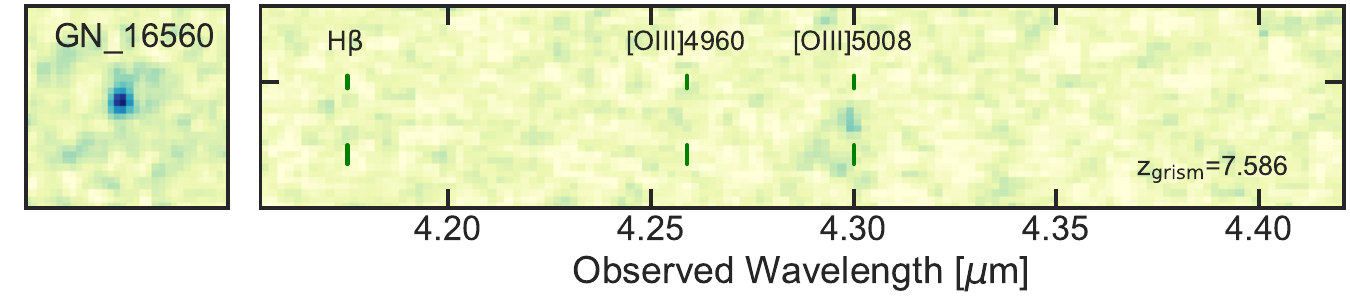}
\vspace{-0.49 cm}
\vspace*{-0.49 cm}
\includegraphics[width=1.0\textwidth]{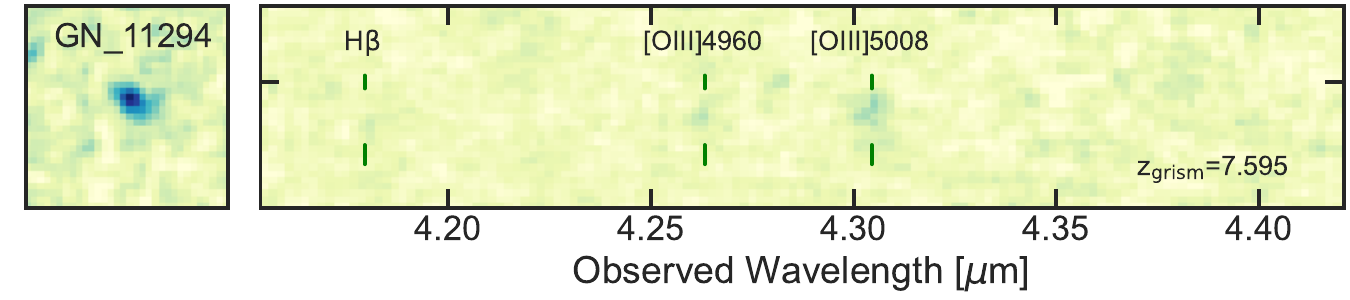}
\vspace{-0.49 cm}
\vspace*{-0.49 cm}
\includegraphics[width=1.0\textwidth]{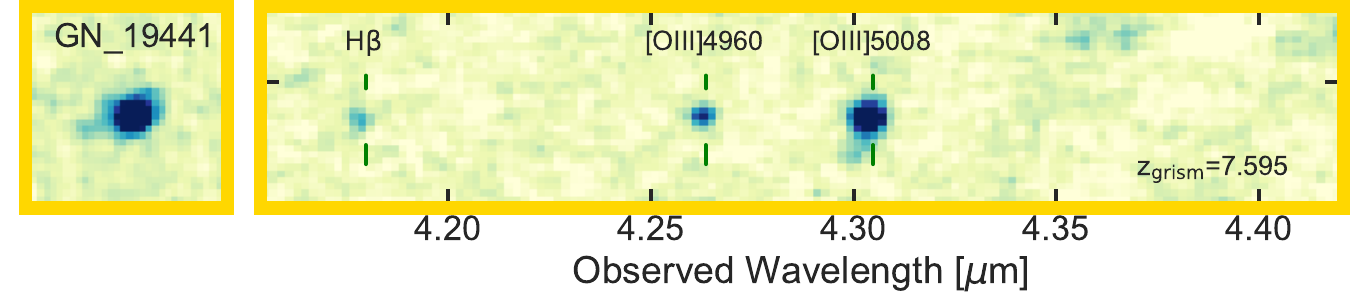}
\vspace{-0.49 cm}
\vspace*{-0.49 cm}
\includegraphics[width=1.0\textwidth]{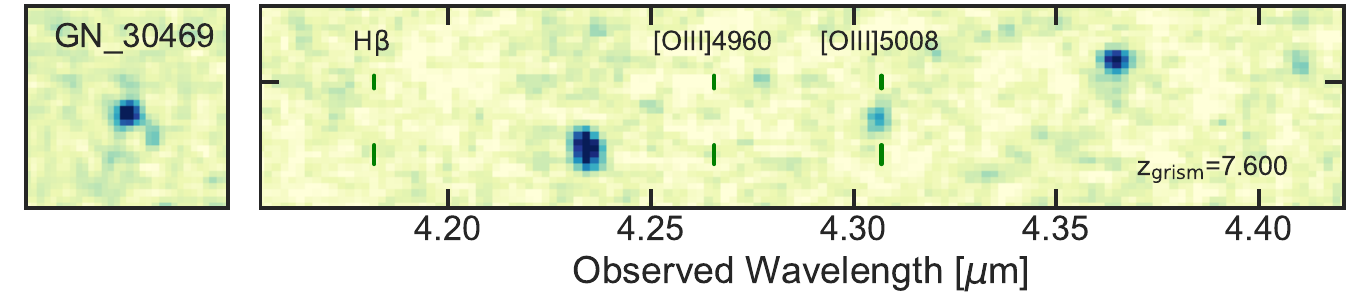}
\vspace{-0.49 cm}
\vspace*{-0.49 cm}
\includegraphics[width=1.0\textwidth]{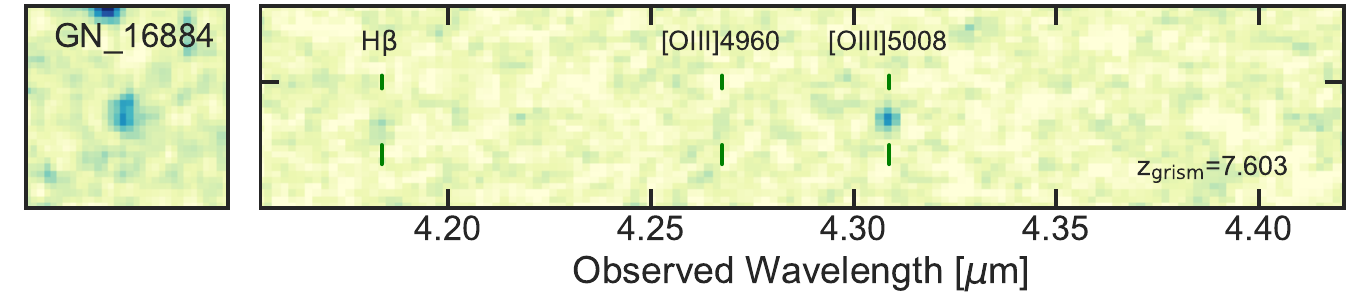}
\vspace{-0.49 cm}
\vspace*{-0.49 cm}
\includegraphics[width=1.0\textwidth]{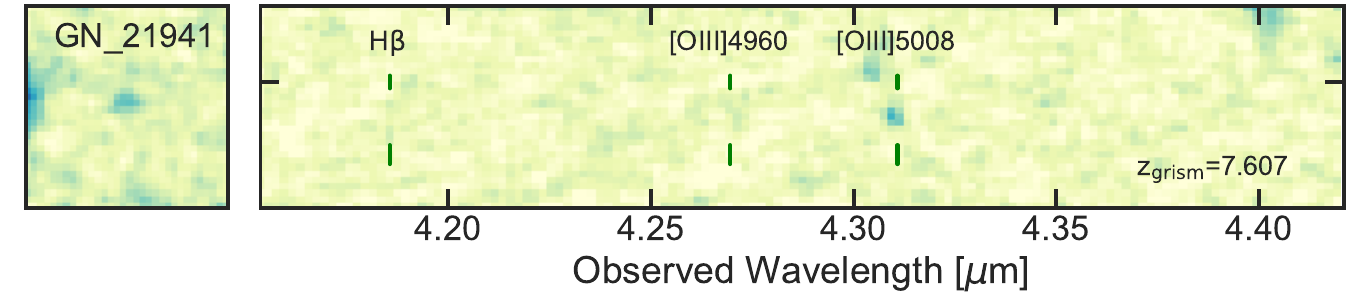}
\vspace{-0.49 cm}
\vspace*{-0.49 cm}
\includegraphics[width=1.0\textwidth]{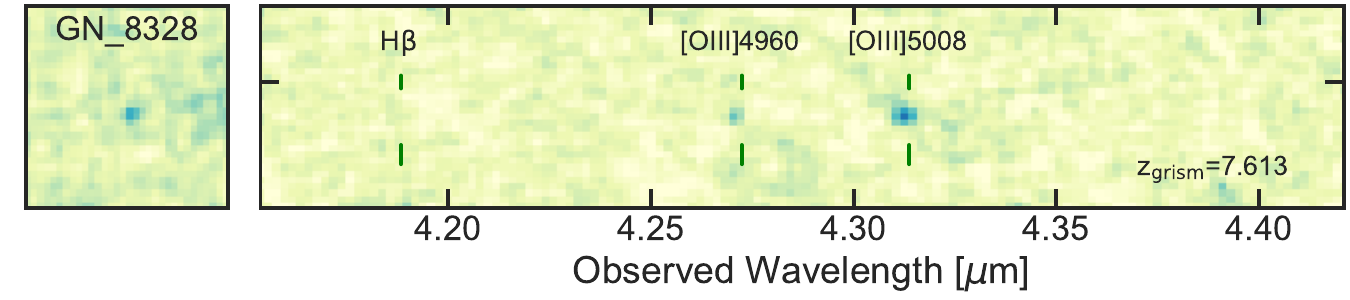}
\vspace{-0.49 cm}
\vspace*{-0.49 cm}
\includegraphics[width=1.0\textwidth]{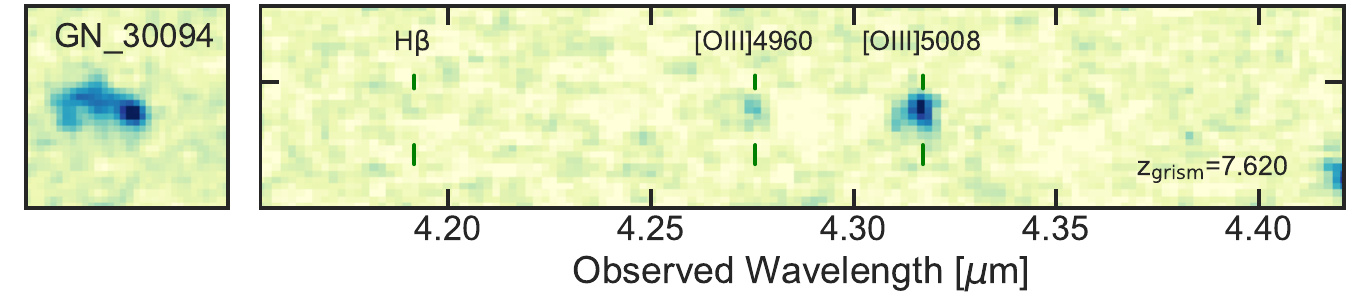}

\end{subfigure}
\hspace{0.05\textwidth}
\begin{subfigure}[t]{0.47\textwidth}
\centering
\includegraphics[width=1.0\textwidth]{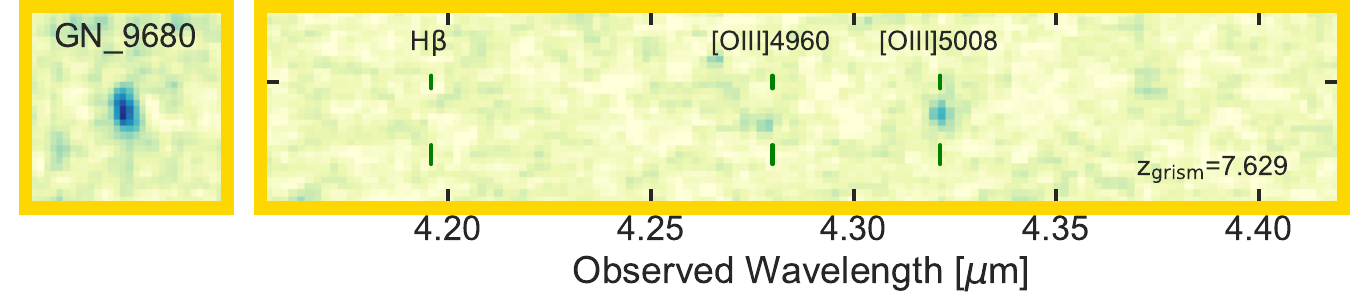}
\vspace{-0.49 cm}
\vspace*{-0.49 cm}
\includegraphics[width=1.0\textwidth]{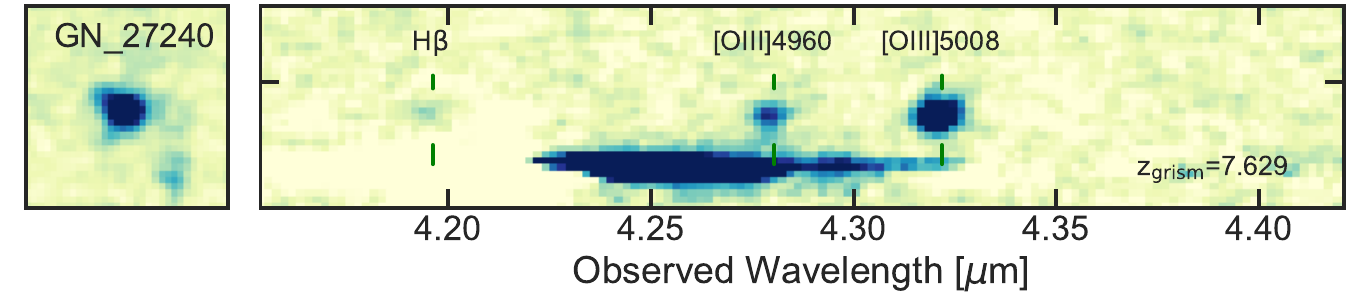}
\vspace{-0.49 cm}
\vspace*{-0.49 cm}
\includegraphics[width=1.0\textwidth]{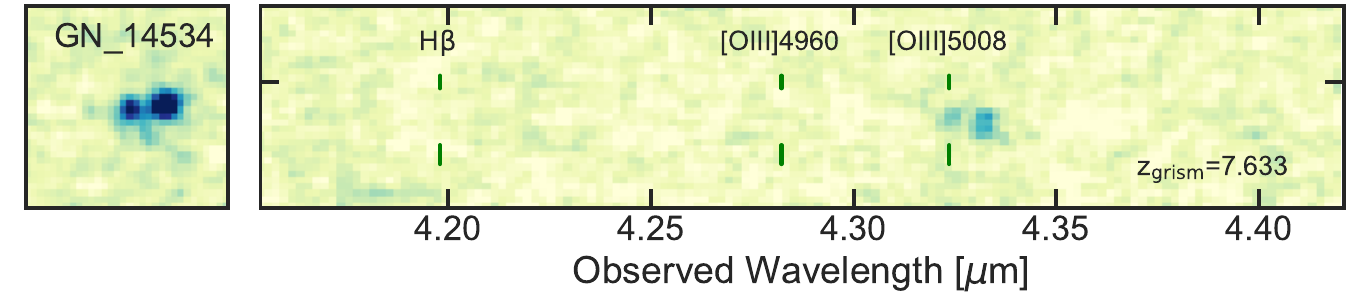}
\vspace{-0.49 cm}
\vspace*{-0.49 cm}
\includegraphics[width=1.0\textwidth]{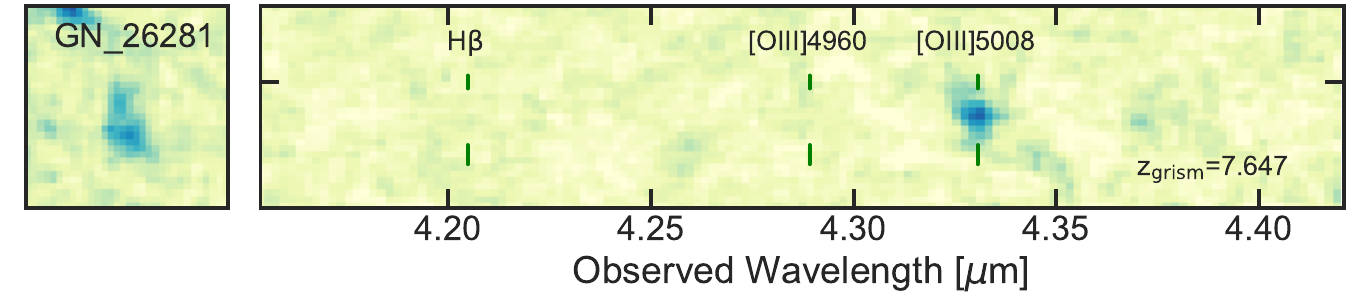}
\vspace{-0.49 cm}
\vspace*{-0.49 cm}
\includegraphics[width=1.0\textwidth]{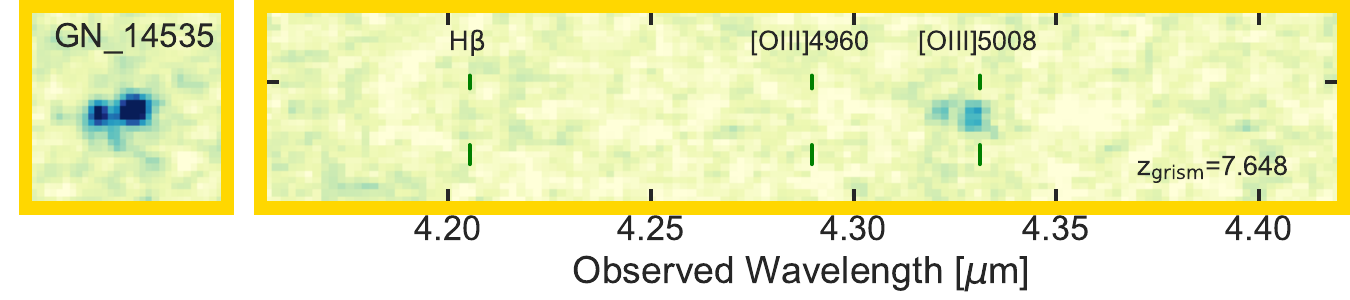}
\vspace{-0.49 cm}
\vspace*{-0.49 cm}
\includegraphics[width=1.0\textwidth]{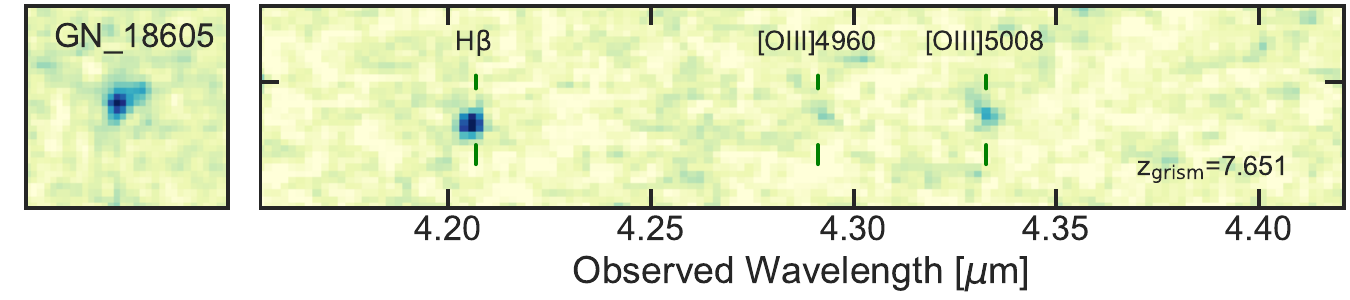}
\vspace{-0.49 cm}
\vspace*{-0.49 cm}
\includegraphics[width=1.0\textwidth]{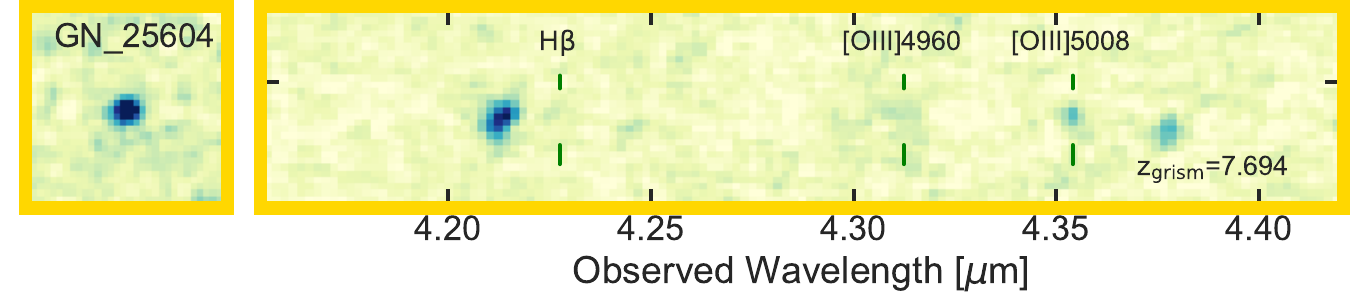}
\vspace{-0.49 cm}
\vspace*{-0.49 cm}
\includegraphics[width=1.0\textwidth]{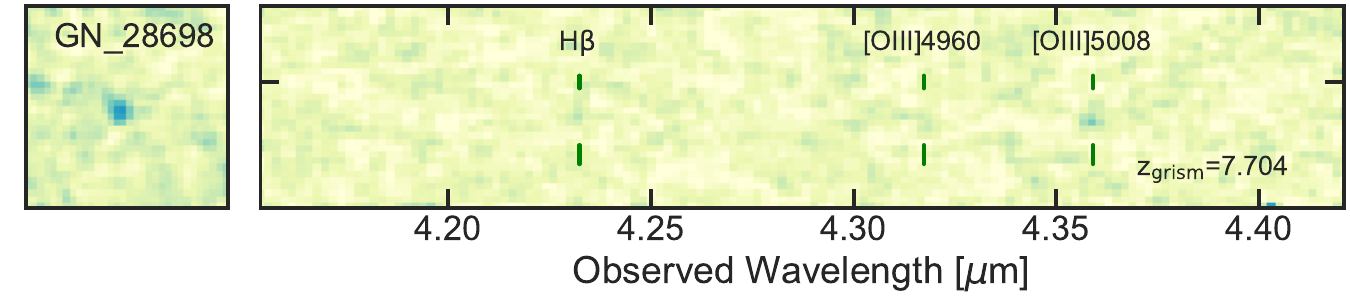}

\end{subfigure}
\caption {Top left panel: sky projection of galaxies in the GN\_z7\_6 group, where the LAE is marked as a star and its companion [OIII] emitters as circles. Top right panel: 3D spatial distribution, with the LAE represented by stars and [OIII] emitters by circles. Axes are in proper Mpc, with redshift increasing along the x-axis. Bottom panel: 2D spectra of [OIII] emitters, with the yellow frame highlighting the Lyman-$\alpha$ emitters.}
\label{fig:group4}
\end{figure*}

\begin{table*}
%\centering

\begin{tabular}{cccccccccc}
\hline\hline

 ID & RA & DEC & z$_\mathrm{Lya}$ & z$_\mathrm{[OIII]}$ & M$_\mathrm{{UV}}$ &  $\log_{\mathrm{M}_{\star}}$ & EW$_\mathrm{{[OIII]}}$[\AA]  \\\hline\hline\\[-5pt]

\multicolumn{8}{c}{\textbf{GS\_z7\_2 group}}\\[5pt]

1744\href{#refa}{\textsuperscript{a}} & 53.16958 & -27.73806 & 7.242 & 7.242 & -20.3 $^{+ 0.0}_{-0.0}$ & 8.21 $^{+ 0.08}_{-0.13}$ & 305 $^{+ 75}_{-49}$\\
JADES-13682\href{#refa}{\textsuperscript{a}}\href{#ref1}{\textsuperscript{1}}\href{#ref2}{\textsuperscript{2}} & 53.16746 & -27.7720&7.275\href{#ref2}{\textsuperscript{2}}&-&-16.7 &-&-&\\ \hline \hline

23162 & 53.19540 & -27.83548 & - & 7.276 & -20.4 $^{+ 0.1}_{-0.1}$ & 8.69 $^{+ 0.46}_{-0.16}$ & 427 $^{+ 39}_{-57}$ \\
12667 & 53.16481 & -27.78826 & - & 7.231 & -20.3 $^{+ 0.0}_{-0.0}$ & 8.12 $^{+ 0.28}_{-0.14}$ & 794 $^{+ 80}_{-74}$ \\
12061 & 53.16171 & -27.78539 & - & 7.236 & -20.3 $^{+ 0.1}_{-0.0}$ & 9.36 $^{+ 0.05}_{-0.09}$ & 150 $^{+ 29}_{-29}$ \\
23161 & 53.19541 & -27.83543 & - & 7.277 & -20.2 $^{+ 0.1}_{-0.0}$ & 7.90 $^{+ 0.23}_{-0.26}$ & 885 $^{+ 129}_{-116}$ \\
14008 & 53.18373 & -27.79390 & - & 7.261 & -19.8 $^{+ 0.1}_{-0.1}$ & 8.36 $^{+ 0.26}_{-0.17}$ & 339 $^{+ 46}_{-46}$ \\
12669 & 53.16469 & -27.78823 & - & 7.249 & -19.8 $^{+ 0.0}_{-0.0}$ & 8.29 $^{+ 0.19}_{-0.19}$ & 997 $^{+ 74}_{-95}$ \\
13902 & 53.11354 & -27.79349 & - & 7.168 & -19.7 $^{+ 0.0}_{-0.0}$ & 7.80 $^{+ 0.13}_{-0.05}$ & 2332 $^{+ 158}_{-177}$ \\
27493 & 53.07543 & -27.85520 & - & 7.215 & -19.6 $^{+ 0.1}_{-0.1}$ & 8.42 $^{+ 0.04}_{-0.03}$ & 1173 $^{+ 139}_{-139}$ \\
5947 & 53.15086 & -27.75734 & - & 7.255 & -19.4 $^{+ 0.1}_{-0.1}$ & 8.65 $^{+ 0.12}_{-0.34}$ & 192 $^{+ 29}_{-37}$ \\
9779 & 53.17975 & -27.77465 & - & 7.275 & -19.3 $^{+ 0.0}_{-0.0}$ & 8.29 $^{+ 0.15}_{-0.20}$ & 916 $^{+ 66}_{-66}$ \\
4357 & 53.10019 & -27.75030 & - & 7.221 & -19.0 $^{+ 0.2}_{-0.2}$ & 8.74 $^{+ 0.38}_{-0.83}$ & 321 $^{+ 257}_{-65}$ \\

\hline\hline
\end{tabular}
\caption{Summary of the GS$\_$z7$\_$2 group.}
\label{tab:group5} 
\footnotesize{\raggedright Notes:  \textsuperscript{a} LAEs. JADES-13682 was not detected in FRESCO due to the sensitivity of the program.}
\footnotesize{\raggedright\par}
\footnotesize{\raggedright References:  {\textsuperscript{1}} \cite{Saxena23}, {\textsuperscript{2}} \cite{Tang2024} \par}

\end{table*}

\begin{figure*}
\centering
\begin{subfigure}[t]{1\textwidth}
\centering
\includegraphics[width=0.46\textwidth]{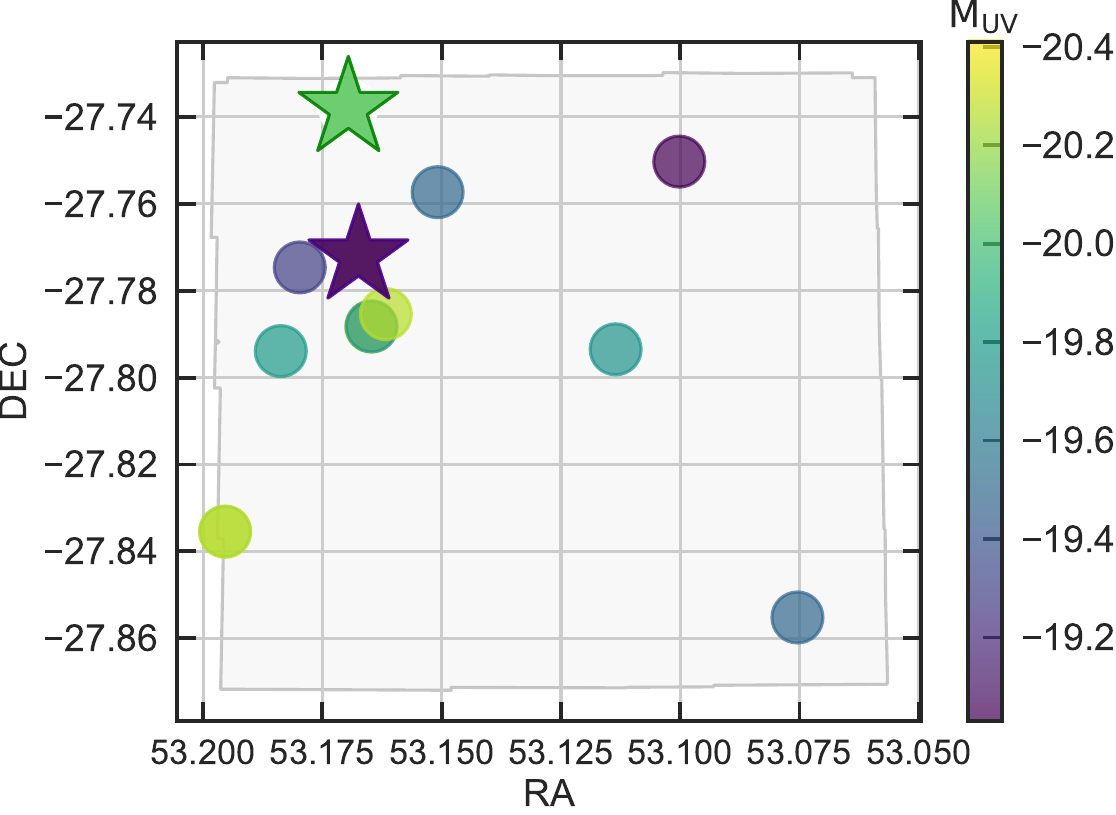}
\includegraphics[width=0.42\textwidth]{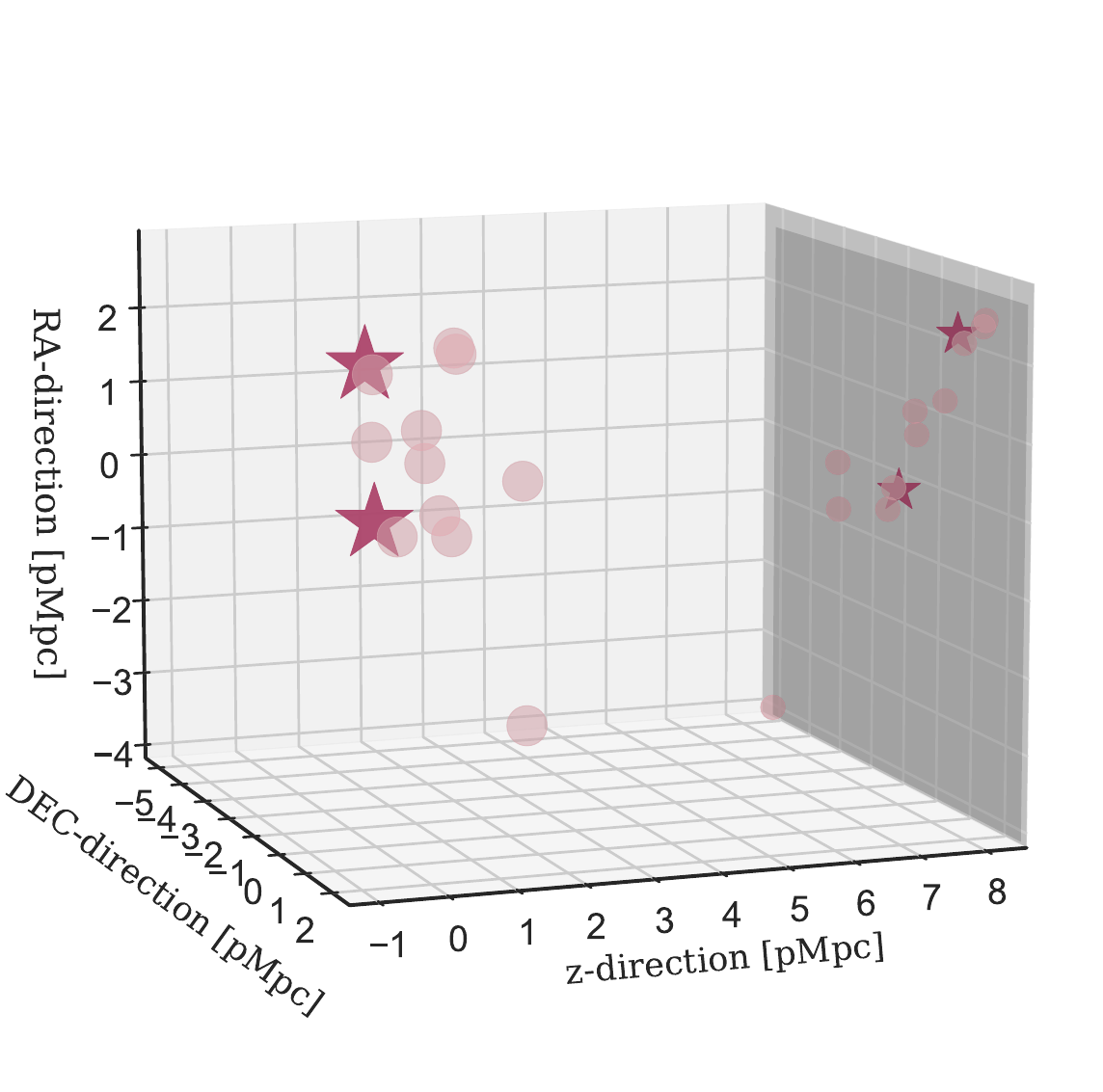}
\end{subfigure}
\begin{subfigure}[t]{0.47\textwidth}
\centering
\includegraphics[width=1.0\textwidth]{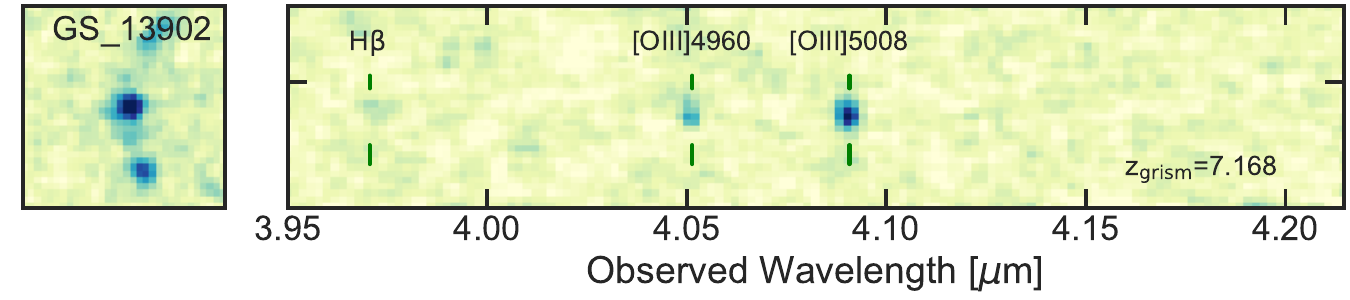}
\vspace{-0.49 cm}
\vspace*{-0.49 cm}
\includegraphics[width=1.0\textwidth]{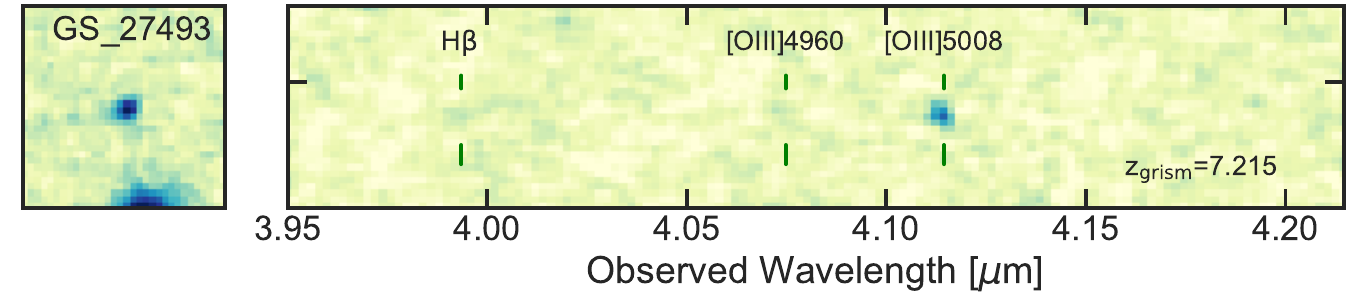}
\vspace{-0.49 cm}
\vspace*{-0.49 cm}

\includegraphics[width=1.0\textwidth]{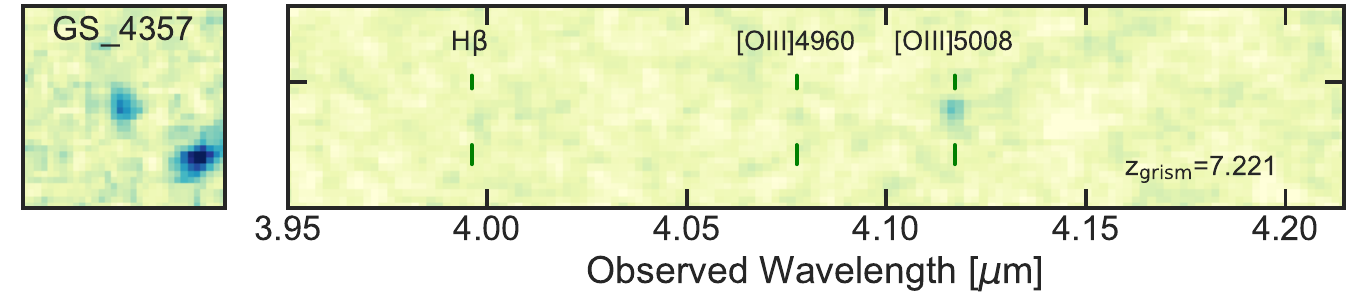}
\vspace{-0.49 cm}
\vspace*{-0.49 cm}
\includegraphics[width=1.0\textwidth]{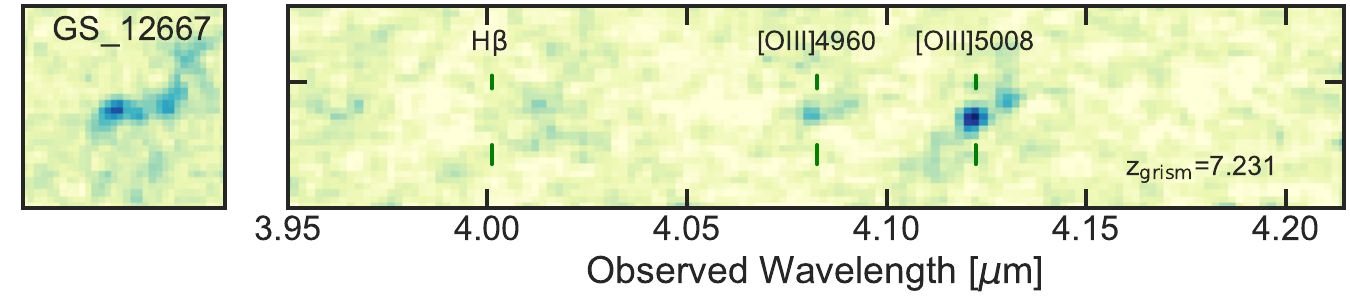}
\vspace{-0.49 cm}
\vspace*{-0.49 cm}
\includegraphics[width=1.0\textwidth]{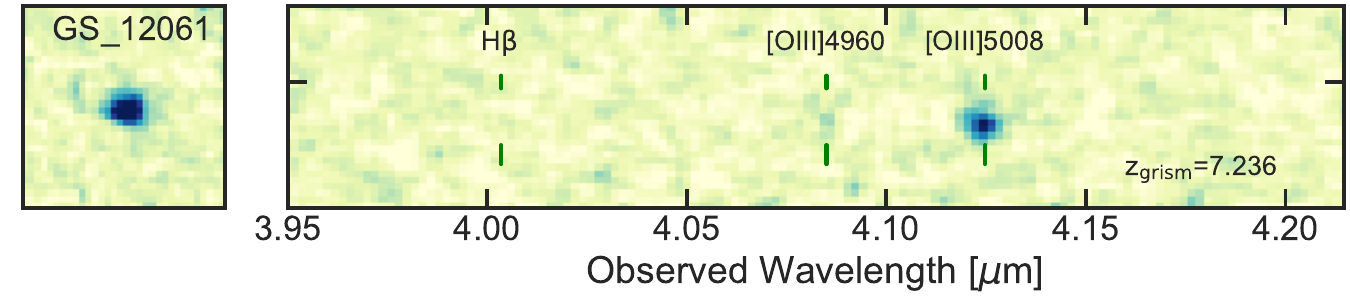}
\vspace{-0.46 cm}
\vspace*{-0.46 cm}
\includegraphics[width=1.0\textwidth]{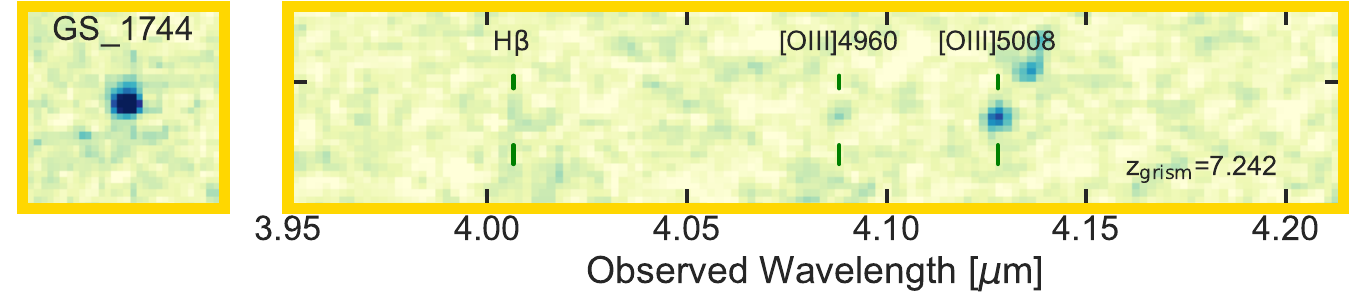}

\end{subfigure}
\hspace{0.05\textwidth}
\begin{subfigure}[t]{0.47\textwidth}
\centering
\includegraphics[width=1.0\textwidth]{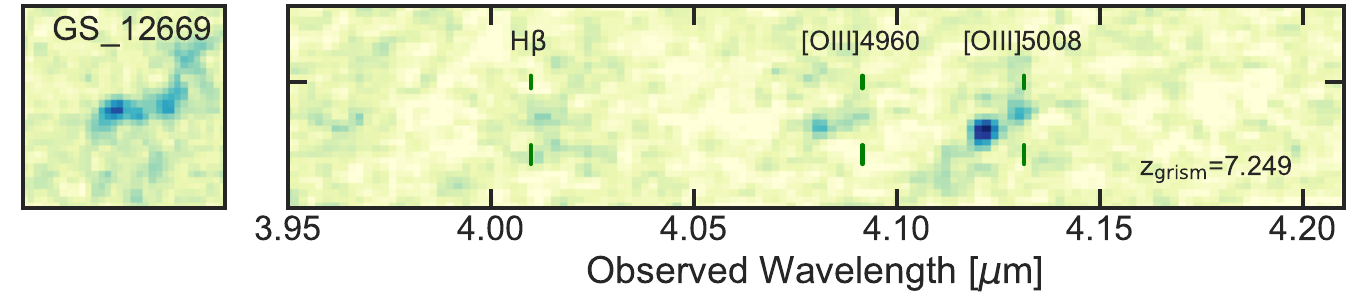}
\vspace{-0.49 cm}
\vspace*{-0.49 cm}
\includegraphics[width=1.0\textwidth]{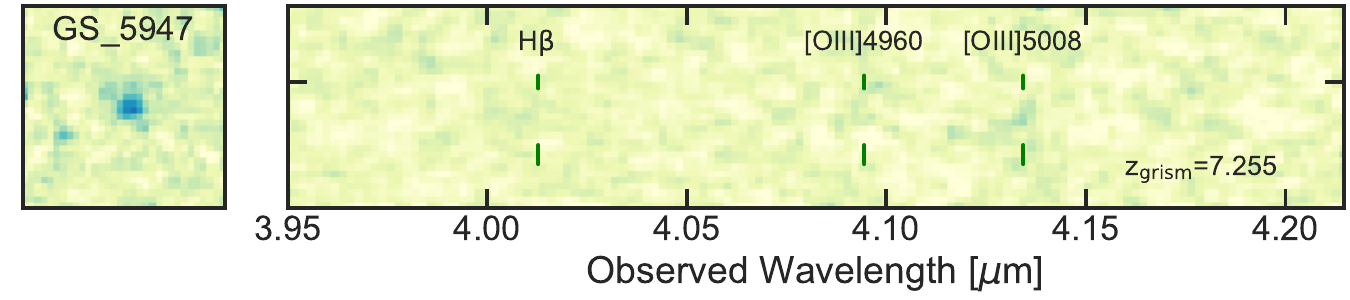}
\vspace{-0.49 cm}
\vspace*{-0.49 cm}
\includegraphics[width=1.0\textwidth]{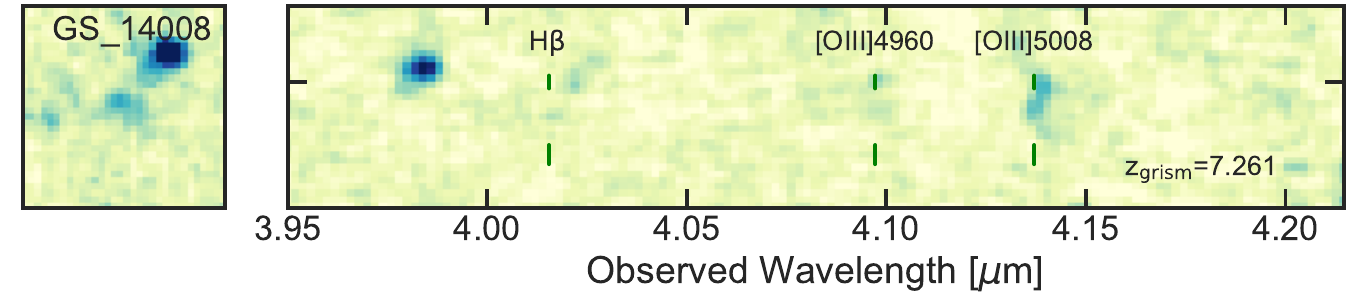}
\vspace{-0.49 cm}
\vspace*{-0.49 cm}
\includegraphics[width=1.0\textwidth]{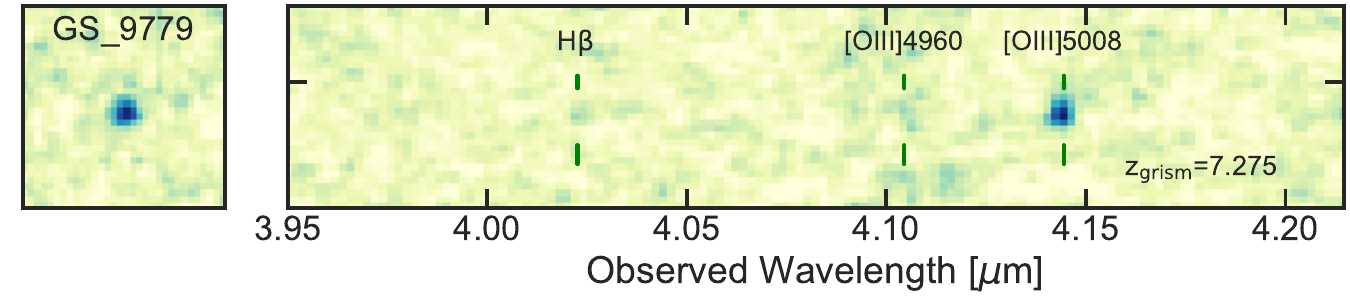}
\vspace{-0.49 cm}
\vspace*{-0.49 cm}
\includegraphics[width=1.0\textwidth]{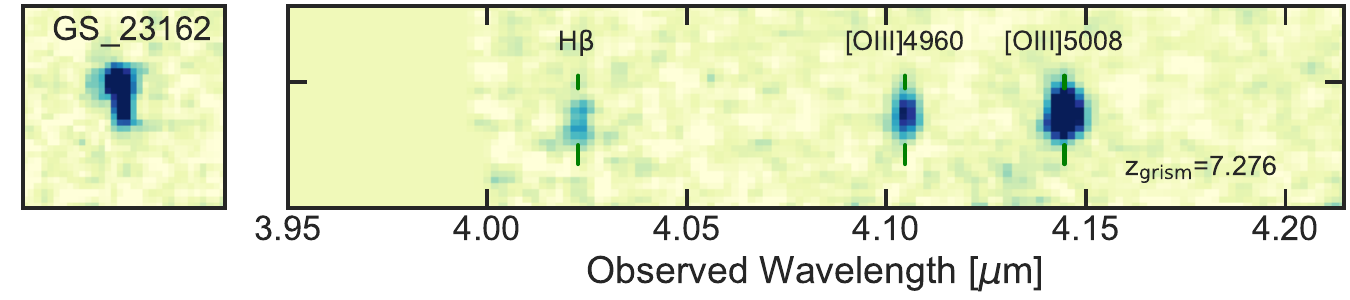}
\vspace{-0.49 cm}
\vspace*{-0.49 cm}
\includegraphics[width=1.0\textwidth]{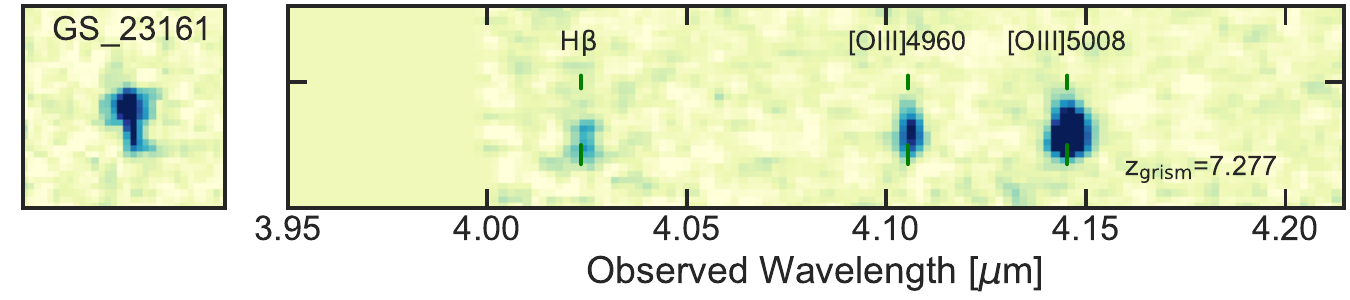}

\end{subfigure}
\caption {Top left panel: sky projection of galaxies in the GS\_z7\_2 group, where the LAE is marked as a star and its companion [OIII] emitters as circles. Top right panel: 3D spatial distribution, with the LAEs represented by stars and [OIII] emitters by circles. Axes are in proper Mpc, with redshift increasing along the x-axis. Bottom panel: 2D spectra of [OIII] emitters, with the yellow frame highlighting the Lyman-$\alpha$ emitters.}
\label{fig:group5}
\end{figure*}

%---------------------------------------
\begin{table*}
%\centering

\begin{tabular}{cccccccccc}
\hline\hline

  ID & RA & DEC & z$_\mathrm{Lya}$ & z$_\mathrm{[OIII]}$ & M$_\mathrm{{UV}}$ &  $\log_{\mathrm{M}_{\star}}$ & EW$_\mathrm{{[OIII]}}$[\AA]    \\\hline\hline\\[-5pt]

\multicolumn{9}{c}{\textbf{GS\_z7\_7 group}}\\[5pt]
6644\href{#refa}{\textsuperscript{a}}\href{#ref1}{\textsuperscript{1}}\href{#ref2}{\textsuperscript{2}} & 53.13346 & -27.76038 & 7.660 & 7.661 & -20.8 $^{+ 0.1}_{-0.1}$ & 9.17 $^{+ 0.48}_{-0.34}$ & 262 $^{+ 175}_{-51}$  & \\ \hline \hline

27780 & 53.12001 & -27.85644 & - & 7.652 & -20.4 $^{+ 0.0}_{-0.1}$ & 8.12 $^{+ 0.40}_{-0.37}$ & 723 $^{+ 71}_{-96}$ \\
23584 & 53.15709 & -27.83707 & - & 7.671 & -20.3 $^{+ 0.1}_{-0.1}$ & 7.52 $^{+ 0.28}_{-0.06}$ & 1247 $^{+ 148}_{-116}$ \\
17172 & 53.05892 & -27.80809 & - & 7.681 & -20.2 $^{+ 0.1}_{-0.1}$ & 9.43 $^{+ 0.26}_{-0.30}$ & 308 $^{+ 64}_{-31}$ \\
20315 & 53.08631 & -27.82396 & - & 7.6 & -20.1 $^{+ 0.0}_{-0.1}$ & 7.66 $^{+ 0.07}_{-0.05}$ & 1367 $^{+ 128}_{-94}$ \\
27775 & 53.12021 & -27.85633 & - & 7.649 & -19.9 $^{+ 0.1}_{-0.1}$ & 8.41 $^{+ 0.20}_{-0.43}$ & 885 $^{+ 248}_{-138}$ \\
6645 & 53.13349 & -27.76046 & - & 7.661 & -19.5 $^{+ 0.0}_{-0.0}$ & 8.86 $^{+ 0.09}_{-0.13}$ & 383 $^{+ 26}_{-26}$ \\
\hline\hline
\end{tabular}
\caption{Table of the GS$\_$z7$\_$7 group.}\label{table1}
\label{tab:group6}

\footnotesize{\raggedright Notes:  \textsuperscript{a} LAE}
\footnotesize{\raggedright\par}
\footnotesize{\raggedright References:  {\textsuperscript{1}} \cite{Song16}, {\textsuperscript{2}} \cite{Tang2024} \par}

\end{table*}
%---------------------------------------
\begin{figure*}
\centering
\begin{subfigure}[t]{1\textwidth}
\centering
\includegraphics[width=0.46\textwidth]{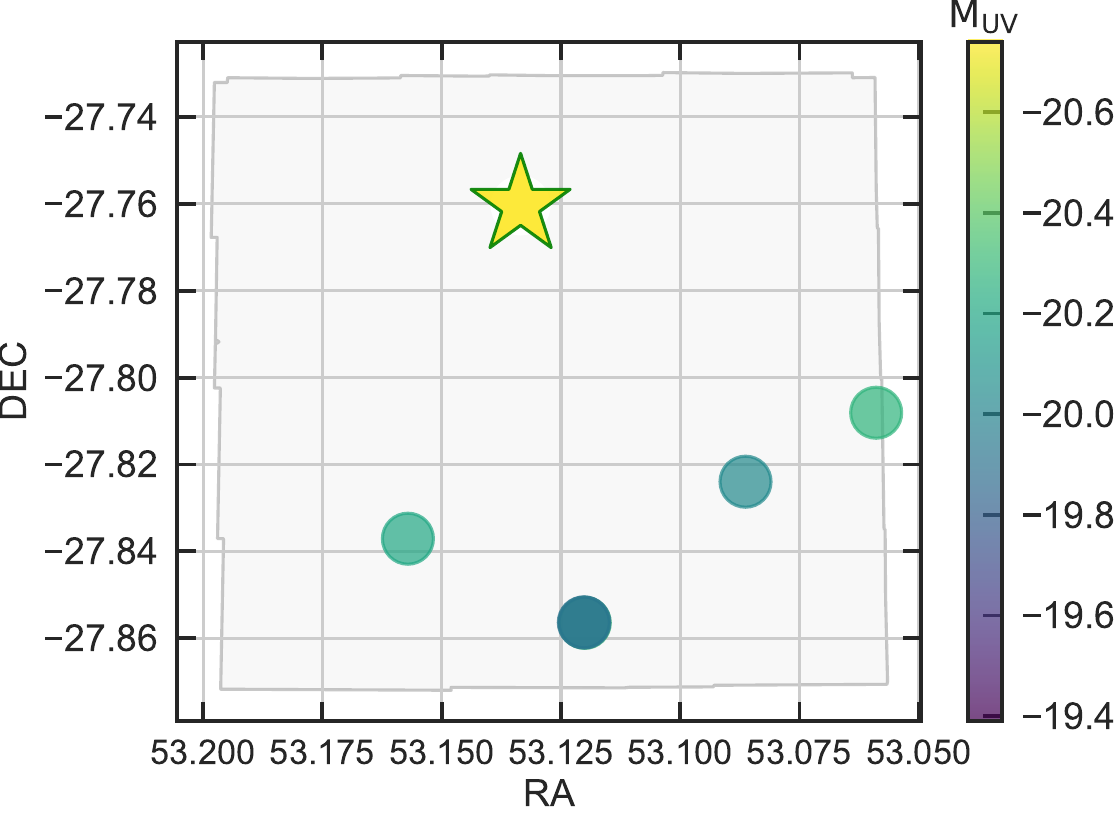}
\includegraphics[width=0.42\textwidth,height=0.4\textwidth]{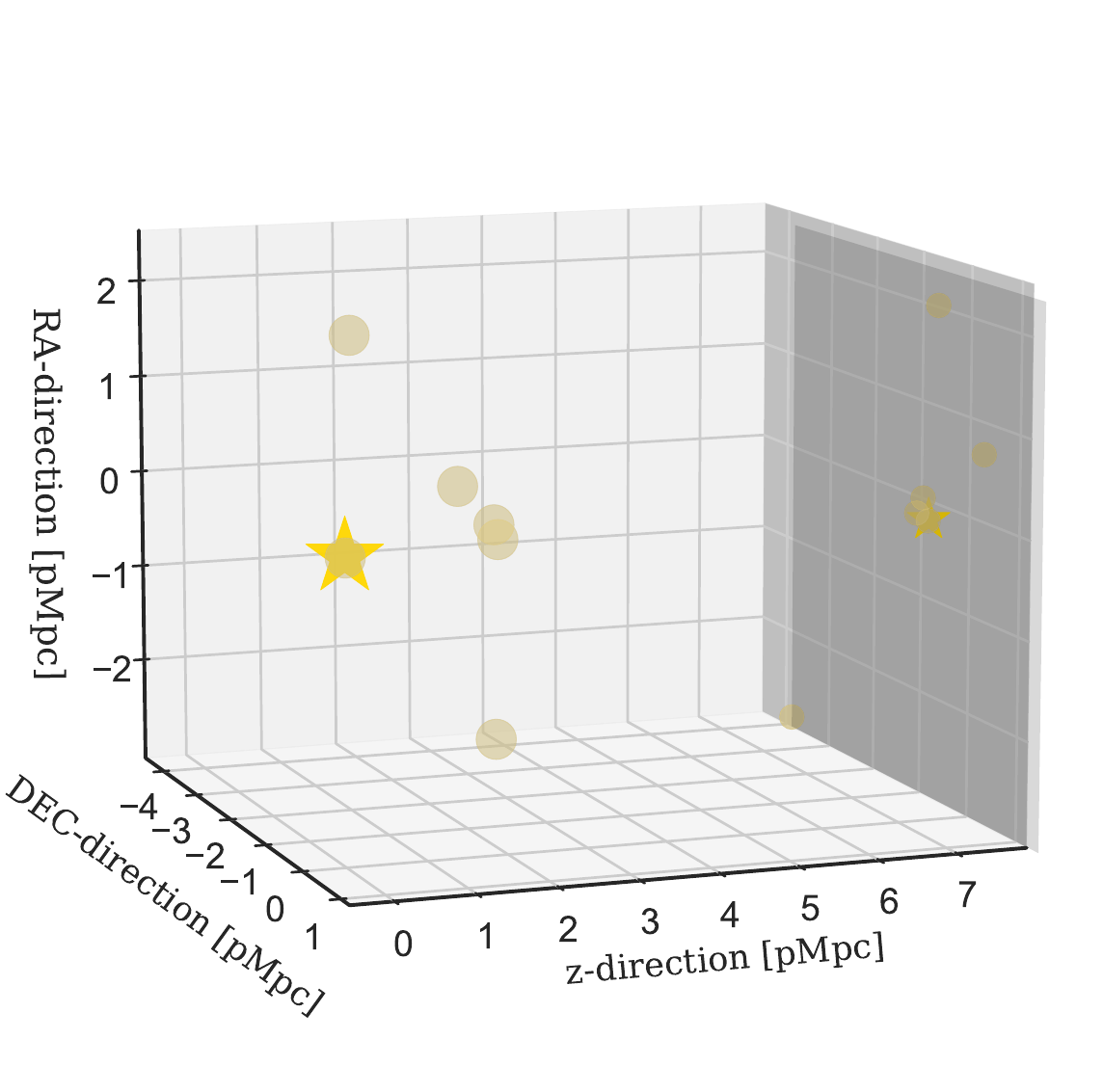}
\end{subfigure}
\begin{subfigure}[t]{0.47\textwidth}
\centering
\includegraphics[width=1.0\textwidth]{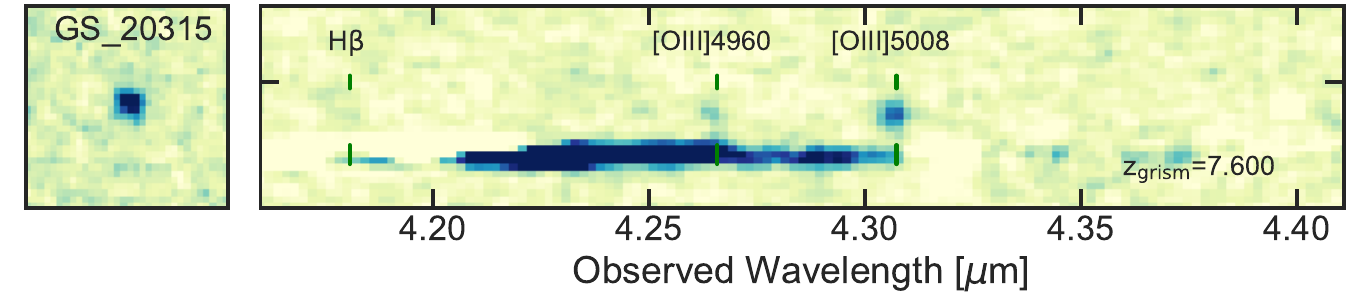}
\vspace{-0.49 cm}
\vspace*{-0.49 cm}
\includegraphics[width=1.0\textwidth]{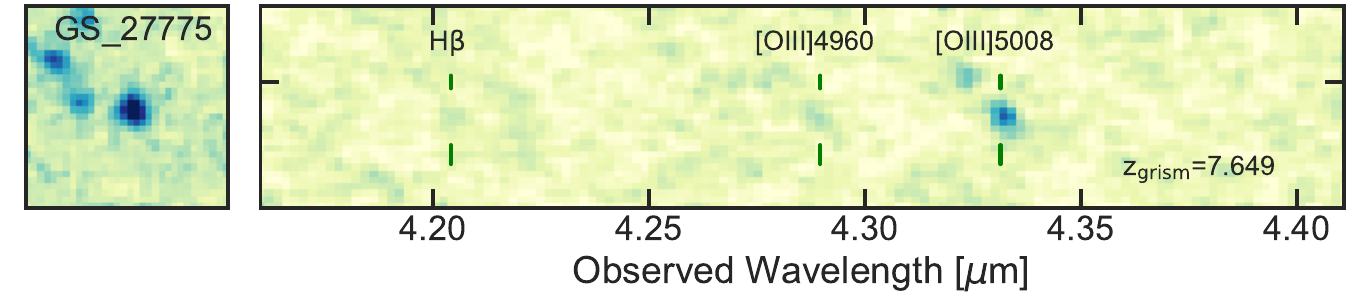}
\vspace{-0.49 cm}
\vspace*{-0.49 cm}
\includegraphics[width=1.0\textwidth]{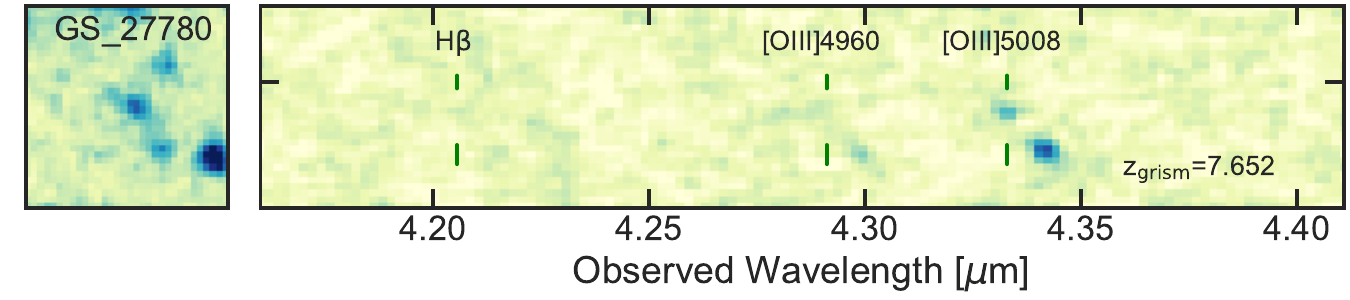}
\vspace{-0.49 cm}
\vspace*{-0.49 cm}
\includegraphics[width=1.0\textwidth]{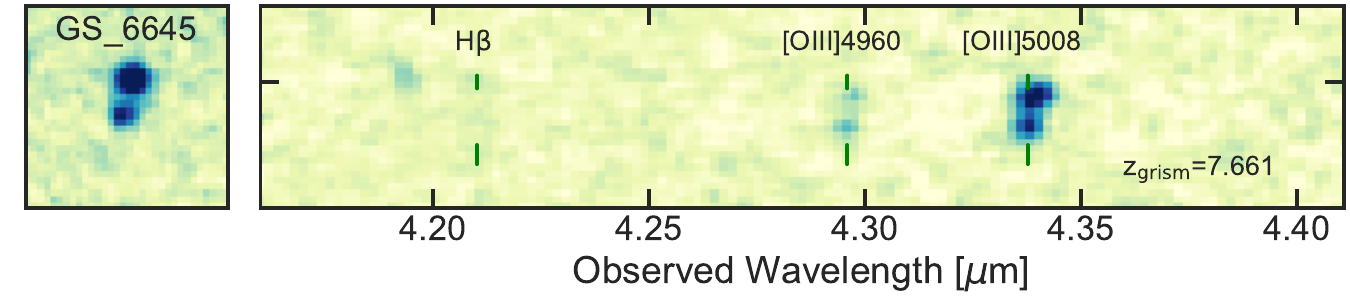}

\end{subfigure}
\hspace{0.05\textwidth}
\begin{subfigure}[t]{0.47\textwidth}
\centering
\includegraphics[width=1.0\textwidth]{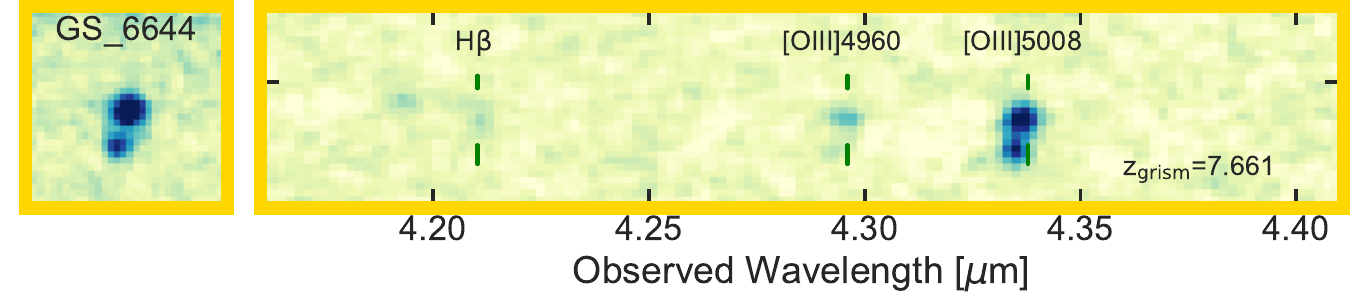}
\vspace{-0.49 cm}
\vspace*{-0.49 cm}
\includegraphics[width=1.0\textwidth]{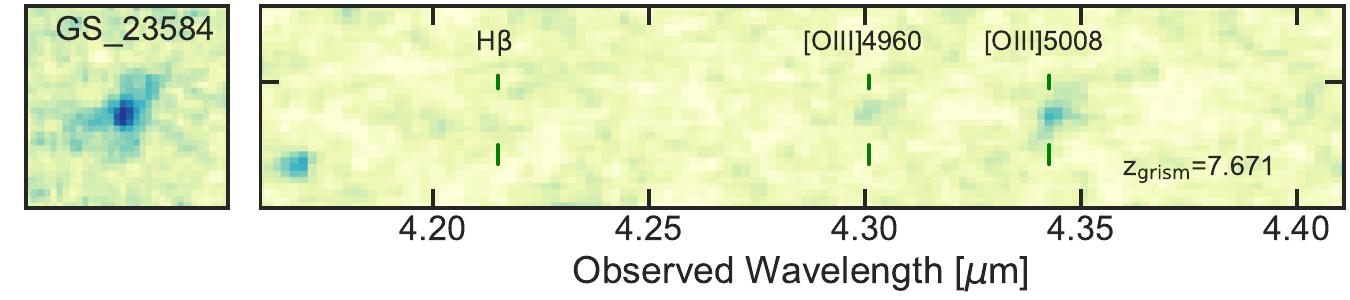}
\vspace{-0.49 cm}
\vspace*{-0.49 cm}
\includegraphics[width=1.0\textwidth]{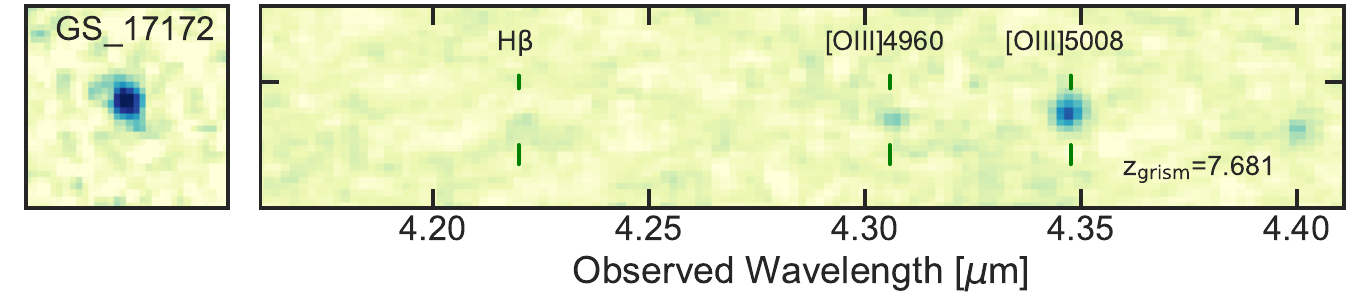}

\end{subfigure}
\caption {Top left panel: sky projection of galaxies in the GS\_z7\_7 group, where the LAE is marked as a star and its companion [OIII] emitters as circles. Top right panel: 3D spatial distribution, with the LAE represented by a star and [OIII] emitters by circles. Axes are in proper Mpc, with redshift increasing along the x-axis. Bottom panel: 2D spectra of [OIII] emitters, with the yellow frame highlighting the Lyman-$\alpha$ emitters.}
\label{fig:group6}
\end{figure*}

%---------------------------------------
\begin{table*}
%\centering

\begin{tabular}{ccccccccc}
\hline\hline

  ID & RA & DEC & z$_\mathrm{Lya}$ & z$_\mathrm{[OIII]}$ & M$_\mathrm{{UV}}$ &  $\log_{\mathrm{M}_{\star}}$ & EW$_\mathrm{{[OIII]}}$[\AA]  \\\hline\hline\\[-5pt]

\multicolumn{9}{c}{\textbf{GS\_z8 group}}\\[5pt]

28631\href{#refa}{\textsuperscript{a}}\href{#ref1}{\textsuperscript{1}} & 53.08738 & -27.86031 & 7.962 & 7.958 & -20.4 $^{+ 0.0}_{-0.0}$ & 8.59 $^{+ 0.17}_{-0.06}$ & 605 $^{+ 48}_{-52}$ &  \\ \hline \hline

284 & 53.06882 & -27.73139 & - & 7.967 & -21.3 $^{+ 0.0}_{-0.1}$ & 7.54 $^{+ 0.20}_{-0.06}$ & 742 $^{+ 110}_{-133}$ \\
29419 & 53.06028 & -27.86353 & - & 7.885 & -19.7 $^{+ 0.1}_{-0.1}$ & 8.34 $^{+ 0.37}_{-0.09}$ & 789 $^{+ 76}_{-63}$ \\
28440 & 53.08649 & -27.85919 & - & 7.953 & -19.7 $^{+ 0.1}_{-0.0}$ & 8.40 $^{+ 0.14}_{-0.18}$ & 416 $^{+ 75}_{-70}$ \\
4864 & 53.19211 & -27.75252 & - & 7.991 & -18.9 $^{+ 0.7}_{-0.6}$ & 8.49 $^{+ 0.39}_{-0.64}$ & 705 $^{+ 475}_{-205}$ \\
8102 & 53.15683 & -27.76717 & - & 7.982 & -18.8 $^{+ 0.1}_{-0.1}$ & 7.52 $^{+ 0.36}_{-0.26}$ & 823 $^{+ 204}_{-154}$ \\
5402 & 53.10354 & -27.75481 & - & 7.908 & -17.9 $^{+ 0.1}_{-0.1}$ & 9.40 $^{+ 0.18}_{-0.21}$ & 415 $^{+ 61}_{-43}$ \\

\hline\hline
\end{tabular}
\caption{Table of the GN$\_$z8 group.}
\label{tab:group7}

\footnotesize{\raggedright Notes:  \textsuperscript{a} LAE}
\footnotesize{\raggedright\par}
\footnotesize{\raggedright References:  {\textsuperscript{1}} \cite{Roberts-Borsani2023}\par}

\end{table*}
%---------------------------------------
\begin{figure*}
\centering
\begin{subfigure}[t]{1\textwidth}
\centering
\includegraphics[width=0.46\textwidth]{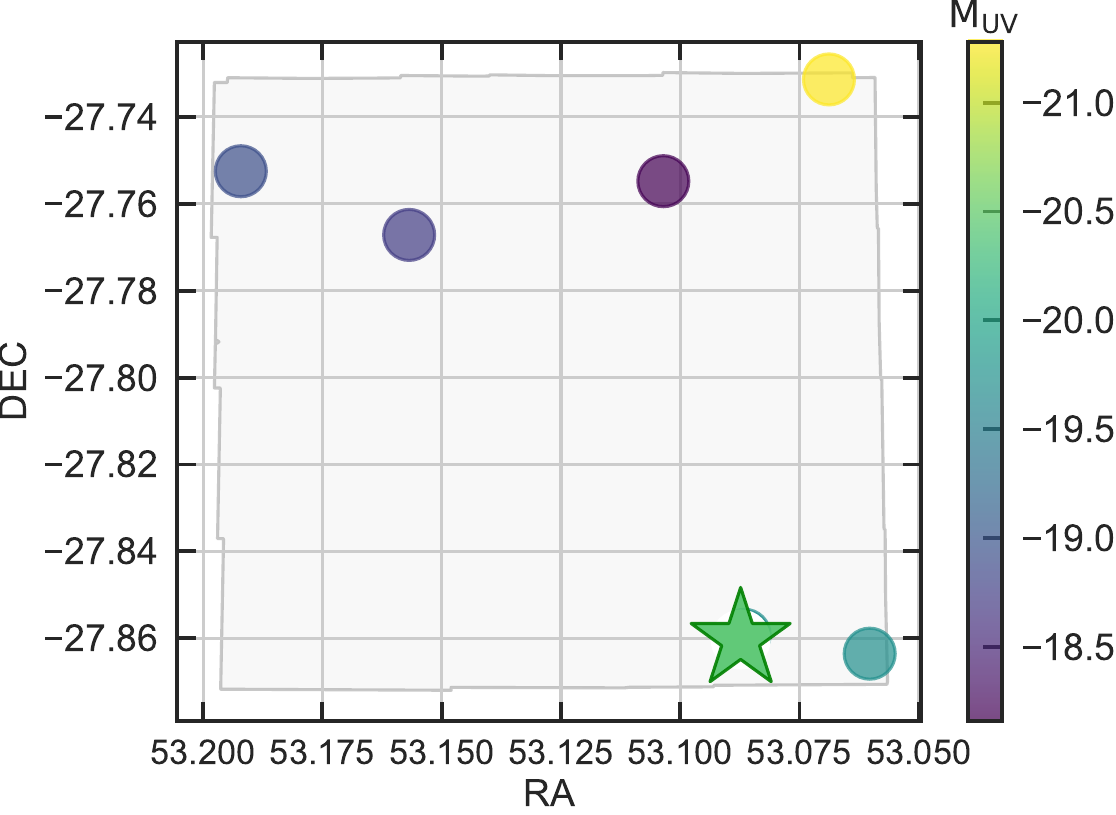}
\includegraphics[width=0.42\textwidth,height=0.4\textwidth]{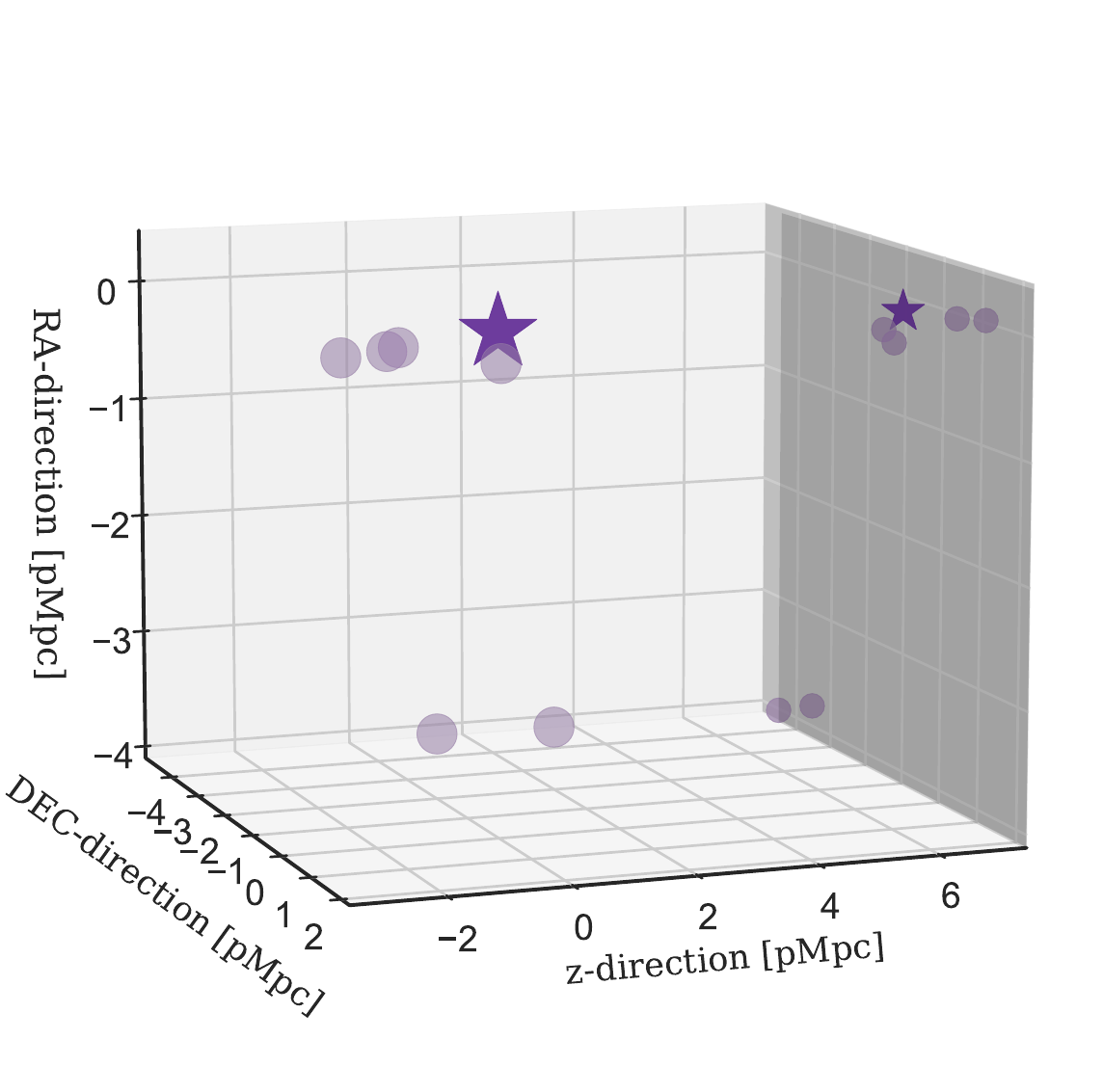}
\end{subfigure}
\begin{subfigure}[t]{0.47\textwidth}
\centering
\includegraphics[width=1.0\textwidth]{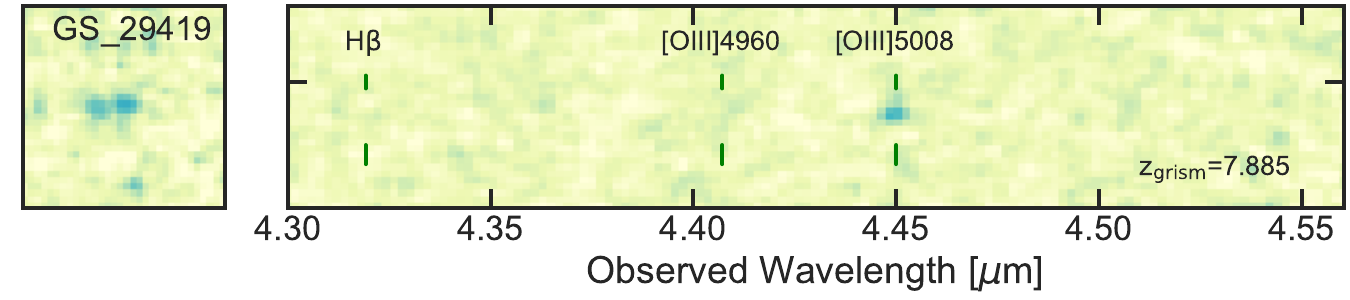}
\vspace{-0.49 cm}
\vspace*{-0.49 cm}
\includegraphics[width=1.0\textwidth]{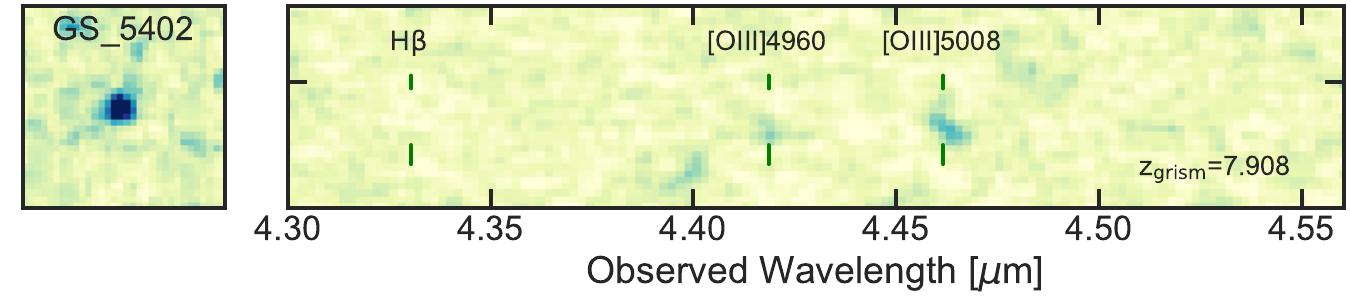}
\vspace{-0.49 cm}
\vspace*{-0.49 cm}
\includegraphics[width=1.0\textwidth]{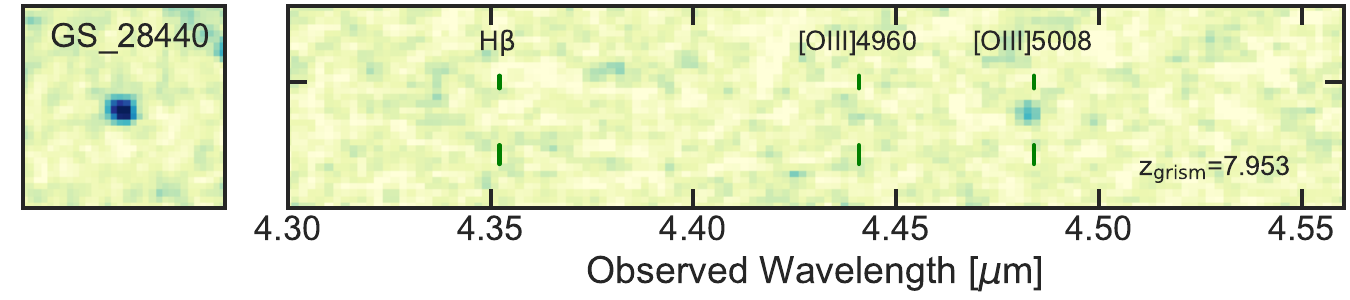}
\vspace{-0.49 cm}
\vspace*{-0.49 cm}
\includegraphics[width=1.0\textwidth]{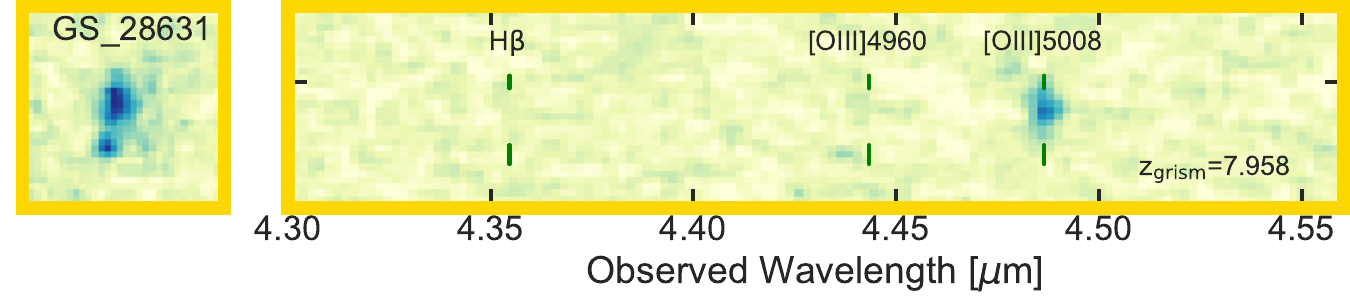}

\end{subfigure}
\hspace{0.05\textwidth}
\begin{subfigure}[t]{0.47\textwidth}
\centering
\includegraphics[width=1.0\textwidth]{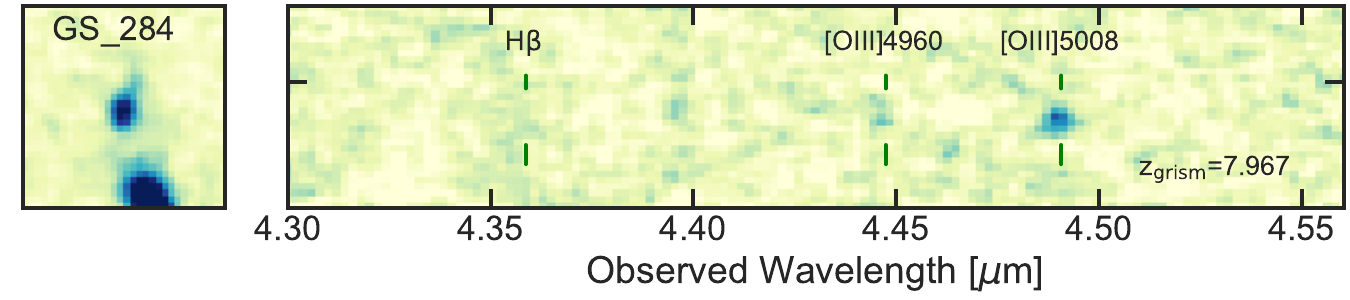}
\vspace{-0.49 cm}
\vspace*{-0.49 cm}
\includegraphics[width=1.0\textwidth]{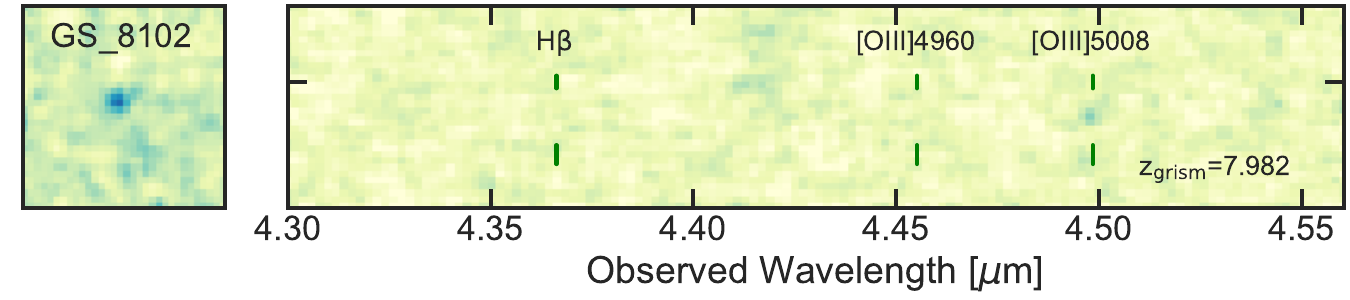}
\vspace{-0.49 cm}
\vspace*{-0.49 cm}
\includegraphics[width=1.0\textwidth]{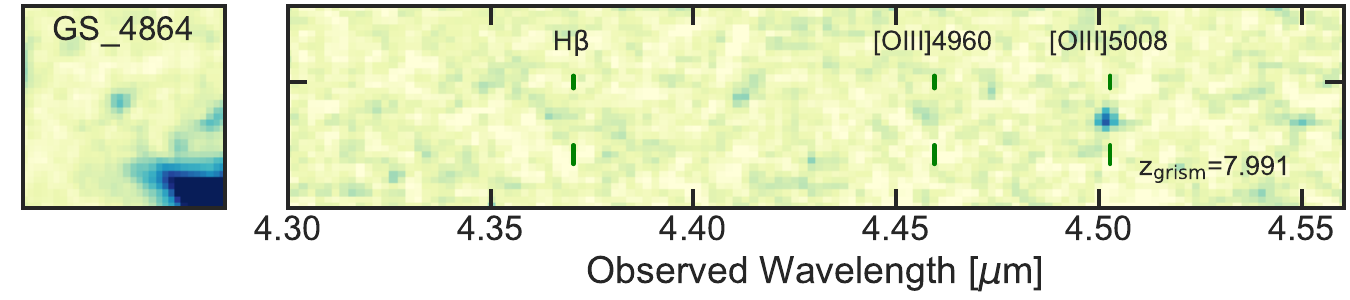}

\end{subfigure}
\caption {Top left panel: sky projection of galaxies in the GS\_z7\_8 group, where the LAE is marked as a star and its companion [OIII] emitters as circles. Top right panel: 3D spatial distribution, with the LAE represented by a star and [OIII] emitters by circles. Axes are in proper Mpc, with redshift increasing along the x-axis. Bottom panel: 2D spectra of [OIII] emitters, with the yellow frame highlighting the Lyman-$\alpha$ emitters.}
\label{fig:group7}
\end{figure*}

%%%%%%%%%%%%%%%%%%%%%%%%%%%%%%%%%%%%%%%%%%%%%%%%%%

\bsp	% typesetting comment
\label{lastpage}
\end{document}